\pdfoutput=1
\documentclass[cernpreprint, texlive=2023, UKenglish, orcidlogo, texmf, orcidlogo]{atlasdoc}
 
\usepackage[subfigure=true]{atlaspackage}
\usepackage{atlasbiblatex}
 
\usepackage{atlasphysics}
 
\graphicspath{{logos/}{figures/}}
 
\usepackage{ANA-HDBS-2018-40-PAPER-defs}
 
\usepackage{pifont}
\usepackage{tikz} 
\usetikzlibrary{tikzmark, positioning, fit, decorations.pathreplacing, calc}
\usepackage{tikz-feynman}
\tikzfeynmanset{compat=1.1.0}
 
\usepackage{multirow}

\AtlasTitle{Search for resonant and non-resonant Higgs boson pair production in the $b\bar b\tau^+\tau^-$ decay channel using 13~TeV $pp$ collision data from the ATLAS detector}
 
\AtlasJournal{JHEP}

\AtlasAbstract{
A search for Higgs boson pair production in events with two $b$-jets and two $\tau$-leptons is presented, using a proton--proton collision dataset with an integrated luminosity of 139~fb$^{-1}$ collected at $\sqrt{s}=13$~TeV by the ATLAS experiment at the LHC. Higgs boson pairs produced non-resonantly or in the decay of a narrow scalar resonance in the mass range from 251 to 1600~GeV are targeted. Events in which at least one $\tau$-lepton decays hadronically are considered, and multivariate discriminants are used to reject the backgrounds. No significant excess of events above the expected background is observed in the non-resonant search. The largest excess in the resonant search is observed at a resonance mass of 1~TeV, with a local (global) significance of $3.1\sigma$ ($2.0\sigma$). Observed (expected) 95\% confidence-level upper limits are set on the non-resonant Higgs boson pair-production cross-section at 4.7 (3.9) times the Standard Model prediction, assuming Standard Model kinematics, and on the resonant Higgs boson pair-production cross-section at between 21 and 900~fb (12 and 840~fb), depending on the mass of the narrow scalar resonance.
}
 
\author{The ATLAS Collaboration}
 
\PreprintIdNumber{CERN-EP-2022-109}

\AtlasJournalRef{JHEP 07 (2023) 040}
 
\AtlasDOI{10.1007/JHEP07(2023)040}

\hypersetup{pdftitle={ATLAS document},pdfauthor={The ATLAS Collaboration}}
 
\begin{document}
 
\maketitle

\section{Introduction}
\label{sec:intro}

\begin{figure}[!htbp]
\centering
\subfigure[]{\includegraphics[width=0.49\textwidth]{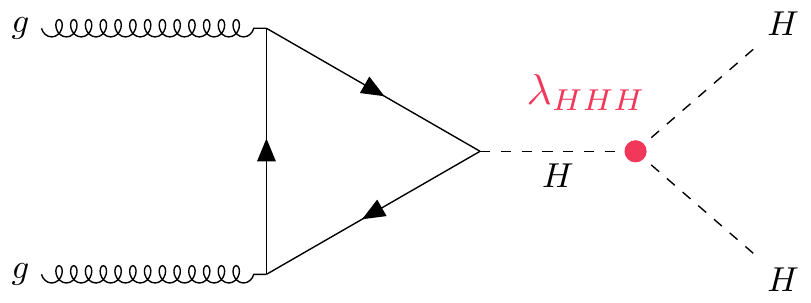}}
\subfigure[]{\includegraphics[width=0.49\textwidth]{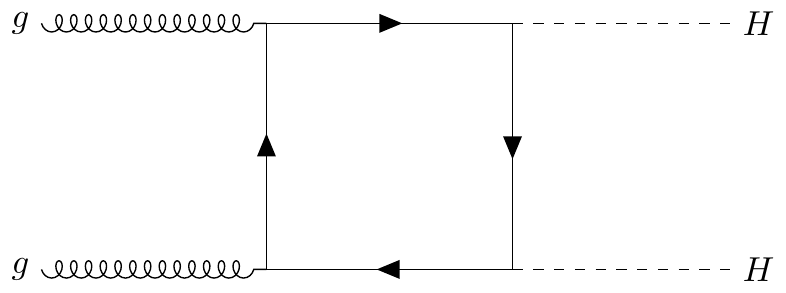}}
\caption{Leading-order Feynman diagrams for ggF non-resonant production of Higgs boson pairs: (a) the `triangle diagram' and (b) the `box diagram'.
The Higgs boson trilinear self-coupling is denoted by $\lambda_{HHH}$.
}
\label{fig:intro:triangleAndBox}
\end{figure}

 
The discovery of a Higgs boson ($H$) with a mass of about 125~\GeV~\cite{HIGG-2012-27,CMS-HIG-12-028} has led to a comprehensive programme of measurements and searches by the ATLAS~\cite{PERF-2007-01} and CMS~\cite{CMS-TDR-08-001} collaborations at the Large Hadron Collider (LHC)~\cite{Evans:2008zzb} using proton--proton ($pp$) collision data. To date, all of the measured properties of the Higgs boson are found to be consistent with their
Standard Model (SM) predictions~\cite{HIGG-2013-17-witherratum,HIGG-2019-01,CMS-HIG-12-041,CMS-HIG-19-009,HIGG-2017-06,CMS-HIG-18-002,ATLAS:2022vkf,CMS:2022dwd},
and no unexpected particles or Higgs boson decay modes have been observed. The SM predicts non-resonant $HH$ production, with approximately 90\% of the total cross-section being due to the gluon--gluon fusion (ggF) process. The leading-order (LO) contributions to  ggF $HH$ production are the `triangle diagram', which includes a Higgs boson self-coupling vertex, and the heavy-quark `box diagram', which has two fermion--fermion--Higgs vertices, as shown in Figure~\ref{fig:intro:triangleAndBox}(a) and Figure~\ref{fig:intro:triangleAndBox}(b), respectively. These diagrams interfere destructively, leading to a small SM ggF non-resonant $HH$ cross-section, which is predicted to be $31.1^{+2.1}_{-7.2}$~fb at next-to-next-to-leading (NNLO) order in $\alphas$, including an approximation of finite top-quark-mass effects, for $m_H=125$~\GeV\ and $\sqrt{s}=13$~\TeV~\cite{Dawson:1998py,Borowka:2016ehy,Baglio:2018lrj,deFlorian:2013jea,Shao:2013bz,deFlorian:2015moa,Grazzini:2018bsd,Baglio:2020wgt}. The vector-boson fusion (VBF) process provides a sub-leading source of $HH$ production in the SM, and has a cross-section of $1.73\pm 0.04$~fb at N$^{3}$LO accuracy in QCD, for $m_H=125$~\GeV\ and $\sqrt{s}=13$~\TeV~\cite{Baglio:2012np,Frederix:2014hta,Ling:2014sne,Dreyer:2018rfu,Dreyer:2018qbw}. Due to these small cross-sections, an observation of SM non-resonant $HH$ production is not expected with the currently available LHC dataset, although significant non-resonant and resonant enhancements to the $HH$ cross-section are predicted in many beyond-the-SM (BSM) theories.
 
Due to the diagram shown in Figure~\ref{fig:intro:triangleAndBox}(a) and its interference with the diagram shown in Figure~\ref{fig:intro:triangleAndBox}(b), non-resonant $HH$ production is a sensitive probe of the Higgs boson trilinear self-coupling and the shape of the Higgs field potential, which have important implications for the stability of the electroweak vacuum~\cite{Branchina:2008pc,Buttazzo:2013uya} and for baryogenesis~\cite{Reichert:2017puo} and  inflation~\cite{Bezrukov:2007ep,Masina:2011aa}. Modifications to the non-resonant $HH$ cross-section occur in BSM scenarios with new, light, coloured scalars~\cite{Kribs:2012kz}, composite Higgs models~\cite{Grober:2010yv}, theoretical scenarios with couplings between pairs of top quarks and pairs of Higgs bosons~\cite{Contino:2012xk}, as well as models with a modified coupling of the Higgs boson to the top quark.
 
Previous searches for non-resonant $HH$ production were performed by ATLAS and CMS in the
$b\bar b\tau^+\tau^-$~\cite{HIGG-2016-16-witherratum,CMS-HIG-17-002},
$b\bar b\gamma\gamma$~\cite{HIGG-2016-15,CMS-HIG-17-008},
$b\bar bb\bar b$~\cite{EXOT-2016-31,CMS-HIG-17-017} decay channels,
by ATLAS in the $b\bar bqq\ell\nu$~\cite{HIGG-2016-27}, $WW^*\gamma\gamma$~\cite{HIGG-2016-20}
and $WW^*WW^*$~\cite{HIGG-2016-24} decay channels,
and by CMS in the $b\bar b\ell^+\nu\ell^-\nu$~\cite{CMS-HIG-17-006} decay channel.
In the $b\bar b\tau^+\tau^-$ decay channel, ATLAS and CMS set observed (expected) upper limits on the
non-resonant $HH$ production cross-section at 12.7 (14.8)~\cite{HIGG-2016-16-witherratum} and 30 (25)~\cite{CMS-HIG-17-002} times
the SM expectation using 36.1~\ifb\ and 35.9~\ifb\ of 13~\TeV\ $pp$ collision data, respectively.
Using these datasets, ATLAS and CMS each combined their results from several search channels,
improving these observed (expected) limits to 6.9 (10)~\cite{HDBS-2018-58} and 22.2 (12.8)~\cite{CMS-HIG-17-030} times
the SM expectation, respectively.
ATLAS and CMS recently updated many of their searches for $HH$ production using their full Run~2 $pp$ collision datasets.
Both experiments have published their results in the $b\bar{b} b\bar{b}$~\cite{HDBS-2018-18,CMS-HIG-20-005_edited,CMS:2022nmn} and
$b\bar{b}\gamma\gamma$~\cite{ATLAS:2021ifb,CMS-HIG-19-018} decay modes.
CMS also performed the search in the $b\bar b\tau^+\tau^-$~\cite{CMS:2022hgz},
$b\bar b\ell^+\ell^-\ell^+\ell^-$~\cite{CMS:2022omp}
and $WW^*WW^*/WW^*\tau^+\tau^-/\tau^+\tau^-\tau^+\tau^-$~\cite{CMS:2022kdx} decay channels.
ATLAS also searched for the $HH$ production in the $b\bar b\ell^+\nu\ell^-\nu$~\cite{HDBS-2018-33} final state.
For the search in the $b\bar b\tau^+\tau^-$ decay channel, CMS set an observed (expected) upper limit on the
SM non-resonant $HH$ cross-section of 3.3 (5.2) times the SM expectation by using 138~\ifb\ of 13~\TeV\ $pp$ collision data.
ATLAS and CMS have performed combinations of $HH$ searches using their full Run~2 datasets, which set observed (expected) upper limits on the SM non-resonant $HH$ cross-section at 2.4 (2.9) and 3.4 (2.5) times the SM expectation, respectively~\cite{HDBS-2022-03,CMS:2022dwd}.
 
Various BSM scenarios predict heavy resonances that can decay into pairs of Higgs bosons. These BSM resonances include heavy Higgs bosons from extended Higgs sectors such as those in two-Higgs-doublet models (2HDMs)~\cite{Branco:2011iw}, the minimal supersymmetric extension of the SM~\cite{HABER198575,DIMOPOULOS1981150}, twin Higgs models~\cite{Chacko:2005vw}, and composite Higgs models~\cite{Grober:2010yv,Mrazek:2011iu}. Heavy resonances that decay into pairs of Higgs bosons also include spin-0 radions and spin-2 gravitons from the Randall--Sundrum model~\cite{Randall:1999ee,Tang:2012pv,Cheung:2000rw}, and stoponium states in supersymmetric models~\cite{Kumar:2014bca}. Searches for resonant $HH$ production have been performed in many final states by ATLAS and CMS, and no significant excesses have been observed~\cite{HIGG-2016-16-witherratum,CMS-HIG-17-002,ATLAS:2021ifb,CMS-HIG-17-008,ATLAS:2022hwc,HIGG-2016-27,CMS-HIG-17-006,HIGG-2016-20,HIGG-2016-24,HDBS-2018-58,CMS-HIG-17-030,CMS-HIG-20-014,CMS-HIG-17-009,HIGG-2016-15,EXOT-2016-31,HDBS-2018-18,CMS-HIG-18-013,CMS:2022kdx,ATLAS:2020azv}.
 
This paper describes a search for non-resonant and resonant $HH$ production in the final state with two $\tau$-leptons and two jets containing $b$-hadrons ($b$-jets). The sizeable fraction of all possible SM decays that result in this final state, $\mathcal{B}(HH\to b\bar b\tau^+\tau^-)=7.3\%$~\cite{deFlorian:2016spz_edited,Djouadi:1997yw}, and relatively low backgrounds make this one of the most sensitive $HH$ search signatures. In the search for non-resonant $HH$ production, the signal kinematics are assumed to follow the SM prediction, and the search is optimised for maximum sensitivity to the cross-section rather than the Higgs boson self-coupling. Additionally, only the ggF and VBF non-resonant $HH$ production modes are considered, because other production modes are not expected to contribute significant additional sensitivity in this search. A narrow CP-even scalar particle ($X$) with a mass between 251 and 1600~\GeV\ is used as the benchmark model for the resonant signal. Decay modes in which both $\tau$-leptons decay hadronically ($\tau_\text{had}$), or in which one decays hadronically and the other leptonically ($\tau_\text{lep}$), are considered; these are referred to as $\tau_\text{had}\tau_\text{had}$ and $\tau_\text{lep}\tau_\text{had}$, respectively. The presence of $\tau_\text{had}$ are determined by detector signatures compatible with the expected visible decay products ($\tau_\text{had-vis}$). Events are categorised according to the type of trigger that accepted the event, and are required to contain two oppositely charged $\tau_\text{had-vis}$ and two $b$-tagged jets in the $\tau_\text{had}\tau_\text{had}$ final state, or an electron or muon and an oppositely charged $\tau_\text{had-vis}$ and two $b$-tagged jets in the $\tau_\text{lep}\tau_\text{had}$ final state. Signal events also contain neutrinos from the decay of $\tau$-leptons and $b$-hadrons, which manifest themselves as missing momentum transverse to the beamline ($\mathbf{p}_\text{T}^\text{miss}$). Backgrounds in this search include the production of top-quark pairs ($t\bar t$), single top quarks, $W$ and $Z$~bosons in association with jets, dibosons $(WW, WZ, ZZ)$, single Higgs bosons, and multi-jet events. In some background events, quark- or gluon-initiated jets are misidentified as $\tau_\text{had-vis}$. The selected events are tested for the presence of the signal by performing profile-likelihood fits to multivariate discriminant distributions. The search is performed using a dataset obtained from $pp$ collisions delivered by the LHC during Run~2, between 2015 and 2018, at a centre-of-mass energy of $\sqrt{s}=13$~\TeV. The data were collected by the ATLAS detector~\cite{PERF-2007-01} and correspond to an integrated luminosity of 139~\ifb. Compared to the previous ATLAS search~\cite{HIGG-2016-16-witherratum}, in addition to the greater integrated luminosity, this search profits from improved $\tau_\text{had-vis}$ and $b$-jet reconstruction and identification algorithms, more sophisticated multivariate techniques used to target the resonant signal hypotheses, and new background estimation techniques. In particular, the combined reconstruction and identification efficiencies for $\tau_\text{had-vis}$ and $b$-jets increased by around 25\%--38\% and 10\%, respectively.

The rest of this paper is organised as follows. A brief description of the ATLAS detector is given in Section~\ref{sec:detector}, and the data and simulation samples used are outlined in Section~\ref{sec:datamc}. Overviews of the reconstruction of physics objects and the event selection are presented in Sections~\ref{sec:objects} and \ref{sec:selection}, respectively. The background modelling strategy is described in Section~\ref{sec:modelling:background}. The systematic uncertainties relevant to this search and the statistical interpretation are described in Sections~\ref{sec:systematics} and \ref{sec:stats}, respectively. The results of the search are given in Section~\ref{sec:results}. Finally, Section~\ref{sec:conclusions} presents the conclusions.


\section{ATLAS detector}
\label{sec:detector}

\newcommand{\AtlasCoordFootnote}{
ATLAS uses a right-handed coordinate system with its origin at the nominal interaction point (IP)
in the centre of the detector and the $z$-axis along the beam pipe.
The $x$-axis points from the IP to the centre of the LHC ring,
and the $y$-axis points upwards.
Cylindrical coordinates $(r,\phi)$ are used in the transverse plane,
$\phi$ being the azimuthal angle around the $z$-axis.
The pseudorapidity is defined in terms of the polar angle $\theta$ as $\eta = -\ln \tan(\theta/2)$.
The distance in ($\eta,\phi$) coordinates, $\Delta R = \sqrt{(\Delta\eta)^{2}+(\Delta\phi)^{2}}$,
is also used to define cone sizes.
Transverse momentum and energy are defined as $p_{\text{T}}=p \sin{\theta}$ and
$E_{\text{T}}=E\sin{\theta}$, respectively.
}

ATLAS~\cite{PERF-2007-01} is a general-purpose particle detector covering nearly the entire
solid angle\footnote{\AtlasCoordFootnote} around the collision point.
It is composed of an inner tracking detector system surrounded by a thin superconducting solenoid,
electromagnetic and hadronic calorimeters, and a muon spectrometer incorporating
three large superconducting toroidal magnets.
 
The inner detector, located within a 2~T axial magnetic field generated by the superconducting solenoid,
is used to measure the trajectories and momenta of charged particles.
The inner layers, consisting of high-granularity silicon pixel detectors,
instrument a pseudorapidity range $\vert\eta\vert<2.5$.
An innermost silicon pixel layer, the insertable B-layer~\cite{ATLAS-TDR-19,PIX-2018-001},
was added to the detector between Run 1 and Run 2.
The insertable B-layer improves the ability to identify displaced vertices, which significantly improves
the $b$-jet tagging performance~\cite{ATL-PHYS-PUB-2015-022}.
Silicon strip detectors, which cover the range $\vert\eta\vert<2.5$, surround the pixel detectors.
Outside the strip detectors and covering $\vert\eta\vert<2.0$, there are straw-tube tracking detectors,
which also provide measurements of transition radiation that are used in electron identification.
 
The calorimeter system covers the range $\vert\eta\vert<4.9$.
Within the region $\vert\eta\vert <3.2$, electromagnetic calorimetry is provided by
barrel ($\vert\eta\vert <1.475$) and endcap ($1.375 < \vert\eta\vert <3.2$)
highly segmented lead/liquid-argon (LAr) electromagnetic calorimeters,
with an additional thin LAr presampler covering $\vert\eta\vert <1.8$
to correct for energy loss in material upstream of the calorimeters.
Hadronic calorimetry is provided by a steel/scintillator-tile calorimeter,
segmented into three barrel structures within $\vert\eta\vert <1.7$,
and two copper/LAr hadronic endcap calorimeters extend the coverage to $\vert\eta\vert=3.2$.
The region of $3.2  < \vert\eta\vert < 4.9$ is instrumented with
copper/LAr and tungsten/LAr calorimeter modules optimised for electromagnetic and
hadronic measurements, respectively.
 
The outermost part of the detector is the muon spectrometer, which measures the
curved trajectories of muons in the field of three large air-core toroidal magnets.
High-precision tracking is performed within the range $\vert\eta\vert <2.7$, and
there are chambers for fast triggering within the range $\vert\eta\vert <2.4$.
 
The ATLAS detector has a two-level trigger system~\cite{TRIG-2016-01} to select events of interest.
The first-level (L1) trigger is implemented in custom electronics and, using a subset of the information
from the detector, it accepts events from the 40~MHz LHC proton
bunch crossings at a rate of about 100~kHz.
This is followed by a software-based high-level trigger (HLT) that reduces the accepted event rate to approximately 1~kHz.
 
An extensive software suite~\cite{ATL-SOFT-PUB-2021-001} is used in the reconstruction and analysis of real and simulated data, in detector operations, and in the trigger and data acquisition systems of the experiment.


\section{Data and simulation samples}
\label{sec:datamc}

\subsection{Data samples}
\label{sec:samples:data}
 
The data used in this search were collected at a centre-of-mass energy
of 13~\TeV\ between 2015 and 2018, using triggers to select events with at least one lepton
(where a lepton is defined as an electron or a muon),
at least one $\tau_\text{had-vis}$,
at least one lepton and one $\tau_\text{had-vis}$,
or at least two $\tau_\text{had-vis}$.
Details about these triggers are discussed in Section~\ref{sec:selection:srs}.
Events are selected for analysis only if they are of good quality and
if all the relevant detector components are known to be in good operating condition~\cite{DAPR-2018-01}.
The total integrated luminosity of the data, after meeting the good-quality criteria,
is $139.0 \pm 2.4$~fb$^{-1}$~\cite{ATLAS-CONF-2019-021, ATLAS-Lumi2}.
The recorded events contain an average of 34 simultaneous inelastic $pp$ collisions per bunch-crossing (pile-up).
 
\subsection{Simulated event samples}
\label{sec:samples:simulation}
Monte Carlo (MC) simulated events are used to model SM background production,
SM non-resonant $HH$ signal production, and BSM resonant $HH$ signal production.
The events were passed through the full ATLAS detector simulation~\cite{SOFT-2010-01}
based on \GEANT~\cite{Agostinelli:2002hh}, with the exception of the BSM resonant $HH$ signal events,
which were passed through a fast simulation in which the response of the calorimeters
is parameterised~\cite{ATL-PHYS-PUB-2010-013,ATL-SOFT-PUB-2018-002}
rather than fully simulated.
The effects of pile-up in the same and neighbouring bunch crossings
were modelled by overlaying each hard-scatter event with minimum-bias events, simulated using the soft quantum chromodynamics (QCD) processes
of \Pythia[8.186]~\cite{Sjostrand:2007gs} with a set of tuned parameters called the A3 tune~\cite{ATL-PHYS-PUB-2016-017}
and the \NNPDF[2.3lo]~\cite{Ball:2012cx} parton distribution functions (PDFs).
The \EVTGEN program~\cite{Lange:2001uf} was used to model the decays of bottom and charm hadrons
in all samples of simulated events, except those generated using \Sherpa~\cite{Bothmann:2019yzt}.
The samples generated with \Sherpa used the bottom- and charm-hadron decay model
implemented within the generator.
The simulated events were processed through the same reconstruction algorithms as the data.
For all samples containing a SM Higgs boson, its mass was fixed to 125~\GeV.
The same mass value is used in the calculation of the Higgs boson decay branching fractions
and in the calculation of the single-Higgs-boson and SM non-resonant $HH$ production
cross-sections.
Unless otherwise specified, the order of the cross-section calculation
refers to the expansion in the strong coupling constant ($\alphas$).
A summary of the event samples used for the simulation of the signal and background
processes is shown in Table~\ref{tab:samples}.

\begin{table}[tb!]
\caption{The generators used to simulate the signal and background
processes. If not otherwise specified, the order of the cross-section calculation refers
to the expansion in the strong coupling constant ($\alphas$).
The acronyms ME, PS and UE are used for matrix element, parton shower and underlying event, respectively.
Details of the simulation of the signal and background samples are described in the text.
$(\dagger)$ The NNLO(QCD)+NLO(EW) cross-section calculation for the $pp \rightarrow ZH$ process already includes the $gg\rightarrow ZH$ contribution.
The $qq\rightarrow ZH$ process is normalised to the NNLO(QCD)+NLO(EW) cross-section for the $pp \rightarrow ZH$ process,
after subtracting the $gg\rightarrow ZH$ contribution.
\protect\label{tab:samples}}
 
\begin{center}{\small
\resizebox{\textwidth}{!}{\begin{tabular}{llllllll}
\toprule
\hspace*{0.3cm} Process & ME generator & ME QCD & ME PDF & PS and & UE model & Cross-section \hspace{2.5cm}\\
& & order & & hadronisation & tune & order\\
\midrule
\multicolumn{7}{l}{\BF{Signal}} \\
\hspace*{0.3cm} non-resonant $gg\rightarrow HH$ (ggF)   &\POWHEGBOX[v2] & NLO & \PDFforLHC[15nlo] & \Pythia[8.244] & A14 & NNLO FTApprox  \\
\hspace*{0.3cm} non-resonant $qq\rightarrow qqHH$ (VBF)   & \MGNLO[2.7.3] & LO & \NNPDF[3.0nlo] & \Pythia[8.244] & A14 & N$^{3}$LO(QCD)  \\
\hspace*{0.3cm} resonant $gg\rightarrow X \rightarrow HH$ & \MGNLO[2.6.1] & LO & \NNPDF[2.3lo] & \Herwig[7.1.3] & H7.1-Default & --  \\
\midrule
\multicolumn{7}{l}{\BF{Top-quark}} \\
\hspace*{0.3cm} $t\bar{t}$  &\POWHEGBOX[v2] & NLO & \NNPDF[3.0nlo] & \Pythia[8.230] & A14 & NNLO+NNLL \\
\hspace*{0.3cm} $t$-channel &\POWHEGBOX[v2] & NLO & \NNPDF[3.0nlo] & \Pythia[8.230] & A14 & NLO \\
\hspace*{0.3cm} $s$-channel &\POWHEGBOX[v2] & NLO & \NNPDF[3.0nlo] & \Pythia[8.230] & A14 & NLO \\
\hspace*{0.3cm} $Wt$        &\POWHEGBOX[v2] & NLO & \NNPDF[3.0nlo] & \Pythia[8.230] & A14 & NLO \\
\hspace*{0.3cm} $t\bar{t}Z$ & \Sherpa[2.2.1]  & NLO & \NNPDF[3.0nnlo] &  \Sherpa[2.2.1] & Default & NLO  \\
\hspace*{0.3cm} $t\bar{t}W$ & \Sherpa[2.2.8]  & NLO & \NNPDF[3.0nnlo] &  \Sherpa[2.2.8] & Default & NLO  \\
\midrule
\multicolumn{7}{l}{\BF{Vector boson + jets}} \\
\hspace*{0.3cm} $W/Z+$jets & \Sherpa[2.2.1] & NLO ($\leq 2$ jets) & \NNPDF[3.0nnlo] &  \Sherpa[2.2.1] & Default & NNLO  \\
&                       & LO  (3,4 jets)     &              &                        &                 \\
\midrule
\multicolumn{7}{l}{\BF{Diboson}} \\
\hspace*{0.3cm}$WW,WZ,ZZ$ & \Sherpa[2.2.1] & NLO ($\leq 1$ jet)  & \NNPDF[3.0nnlo] &  \Sherpa[2.2.1] & Default & NLO  \\
&                       & LO (2,3 jets)       &              &                        &         &                     \\
\midrule
\multicolumn{7}{l}{\BF{Single Higgs boson}} \\
\hspace*{0.3cm} ggF         &\POWHEGBOX[v2] & NNLO & \NNPDF[3.0nlo] &  \Pythia[8.212] & AZNLO & N$^{3}$LO(QCD)+NLO(EW)  \\
\hspace*{0.3cm} VBF         &\POWHEGBOX[v2] & NLO  & \NNPDF[3.0nlo] &  \Pythia[8.212] & AZNLO & NNLO(QCD)+NLO(EW)  \\
\hspace*{0.3cm} $qq\rightarrow WH$  &\POWHEGBOX[v2] & NLO  & \NNPDF[3.0nlo] &  \Pythia[8.212] & AZNLO & NNLO(QCD)+NLO(EW)  \\
\hspace*{0.3cm} $qq\rightarrow ZH$  &\POWHEGBOX[v2] & NLO  & \NNPDF[3.0nlo] &  \Pythia[8.212] & AZNLO & NNLO(QCD)+NLO(EW)$^{(\dagger)}$  \\
\hspace*{0.3cm} $gg\rightarrow ZH$  &\POWHEGBOX[v2] & NLO  & \NNPDF[3.0nlo] &  \Pythia[8.212] & AZNLO & NLO+NLL \\
\hspace*{0.3cm} $t\bar{t}H$ &\POWHEGBOX[v2] & NLO  & \NNPDF[3.0nlo] &  \Pythia[8.230] & A14   & NLO  \\
\bottomrule
\end{tabular}}}
\end{center}
\end{table}


Simulated SM non-resonant $HH$ signal production includes the contributions
from the ggF and VBF processes.
The simulated ggF events were generated with the \POWHEGBOX[v2] generator~\cite{Alioli:2010xd}
at next-to-leading order (NLO) with finite top-quark mass,
and using the \PDFforLHC[15\_nlo\_30\_pdfas] (code 90400 in the LHAPDF database~\cite{LHAPDF6})
PDF set~\cite{Butterworth:2015oua}.
Parton showers and hadronisation were simulated using \Pythia[8.244]~\cite{Sjostrand:2007gs}
with the A14 tune~\cite{ATL-PHYS-PUB-2014-021,ATLAS:2012uec} and the \NNPDF[2.3lo] PDF set.
The cross-section for ggF non-resonant $HH$ production is calculated at next-to-next-to-leading order (NNLO)
using FTApprox~\cite{Grazzini:2018bsd}.
The VBF non-resonant $HH$ signal events were generated at LO using the
\MGNLO[2.7.3]~\cite{Alwall:2014hca} generator with the
\NNPDF[3.0nlo]~\cite{Ball:2014uwa} PDF set.
Parton showering and hadronisation were performed using \Pythia[8.244] with the
A14 tune and the \NNPDF[2.3lo] PDF set.
The cross-section for VBF non-resonant $HH$ production is calculated at
next-to-next-to-next-to-leading order (N$^{3}$LO) in QCD
in the limit of no partonic exchange between the two protons~\cite{Dreyer:2018qbw}.
The calculated cross-section values for ggF and VBF non-resonant $HH$ production
at $\sqrt{s}=13$~\TeV\ and $\mH=125$~\GeV\ are given in Section~\ref{sec:intro}.
Other non-resonant $HH$ production modes are not considered because their contributions
to the analysis sensitivity are expected to be negligible.
 
The BSM resonant $HH$ signal from the ggF production of a heavy spin-0 resonance
and its decay into a pair of SM Higgs bosons, $X \rightarrow HH$, was simulated with the
\MGNLO[2.6.1] generator using the \NNPDF[2.3lo] PDF
set at LO accuracy in QCD.
The simulated events were interfaced to \Herwig[7.1.3]~\cite{Bahr:2008pv,Bellm:2015jjp}
to model the parton shower, hadronisation and underlying event, using the H7.1-Default tune~\cite{Gieseke:2012ft}
and the \NNPDF[2.3lo] PDF set.
The resonant $HH$ signal was simulated for 19 values of the resonance mass, $m_{X}$,
between 251~\GeV\ and 1.6~\TeV.
The width of the heavy scalar $X$ was fixed to 10~\MeV.
 
The production of $t\bar{t}$ events, and of single-top-quark events in the $Wt$-, $s$- and $t$-channels,
was simulated using the \POWHEGBOX[v2] generator together with the \NNPDF[3.0nlo] PDF set.
The simulated events were interfaced to \Pythia[8.230] for parton showering and hadronisation
using the A14 tune together with the \NNPDF[2.3lo] PDF set.
The top-quark spin correlations were preserved for all these simulated top-quark processes.
The top-quark mass was set to 172.5~\GeV.
The $t\bar{t}$ production cross-section is calculated
at next-to-next-to-leading-order and next-to-next-to-leading-logarithm (NNLO+NNLL) accuracy~\cite{Czakon:2011xx}.
The cross-sections for the three single-top-quark production channels are calculated at
NLO~\cite{stop_sch_Xsec,stop_tch_Xsec,Kidonakis:2010ux}.
The $t\bar{t}$--$Wt$ interference was handled using the diagram removal scheme.

Events containing a $W$ or $Z$ boson produced in association with jets,
diboson ($WW$, $WZ$ and $ZZ$) production processes, and the $t\bar tZ$ production process
were simulated with the \Sherpa[2.2.1] generator~\cite{Bothmann:2019yzt},
whereas the $t\bar tW$ production process was simulated with the \Sherpa[2.2.8] generator.
These samples used the \NNPDF[3.0nnlo]~\cite{Ball:2014uwa} PDF set
with dedicated parton shower tuning developed by the \Sherpa authors.
For the simulation of $W$/$Z+$jets events, the matrix elements were calculated
for up to two partons at NLO and up to four partons at LO using the \OPENLOOPS~\cite{Cascioli:2011va}
and \textsc{Comix}~\cite{Gleisberg:2008fv} matrix-element generators.
The expected number of $W$/$Z+$jets events is normalised to the NNLO cross-sections~\cite{Catani:2009sm}.
Diboson production was simulated for up to one additional parton at NLO and up to three additional
partons at LO using the \OPENLOOPS and \textsc{Comix} programs.
The NLO cross-sections from \Sherpa are used to normalise the diboson
and the $t\bar tW/Z$ events.

SM single Higgs boson production is considered as part of
the background in this search, and its production modes were simulated using the
\POWHEGBOX[v2] generator and the \NNPDF[3.0nlo] PDF set.
Single Higgs boson production via ggF was simulated at NNLO accuracy in QCD using the
\POWHEG NNLOPS program~\cite{Hamilton:2013fea,Hamilton:2015nsa},
whereas VBF single Higgs boson production was simulated at NLO accuracy in QCD~\cite{Nason:2009ai}.
Events from both of these production modes were interfaced to \Pythia[8.212] for parton showering and hadronisation
using the AZNLO tune~\cite{AZNLOtune} together with the \CTEQ[6L1] PDF set~\cite{CTEQ6L1}.
The cross-section for ggF production of single Higgs bosons is based on a computation with
N$^{3}$LO accuracy in QCD, and NLO accuracy in the electroweak (EW)
expansion~\cite{deFlorian:2016spz_edited,Anastasiou:2016cez,Anastasiou:2015ema,Dulat:2018rbf,Actis:2008ug},
whereas the cross-section for VBF production of single Higgs bosons is taken from the NNLO(QCD)+NLO(EW)
calculation~\cite{deFlorian:2016spz_edited,Ciccolini:2007jr,Ciccolini:2007ec,Bolzoni:2010xr}.
The $qq \rightarrow WH$, $qq \rightarrow ZH$ and $gg \rightarrow ZH$ simulated events
were interfaced to \Pythia[8.212] for parton showering
and hadronisation using the AZNLO tune together with the
\CTEQ[6L1] PDF set.
The cross-sections are taken from the NNLO(QCD)+NLO(EW) calculations for
$qq \rightarrow WH$ and $qq \rightarrow ZH$~\cite{Ciccolini:2003jy,Brein:2003wg,Ferrera:2011bk,Brein:2011vx,Ferrera:2013yga,Ferrera:2014lca,Campbell:2016jau},
and from calculations at next-to-leading-order and next-to-leading-logarithm (NLO+NLL) accuracy in QCD for
$gg \rightarrow ZH$~\cite{Altenkamp:2012sx,Hespel:2015zea,ggzhnll,Harlander:2013mla,Brein:2012ne}.
For Higgs boson production in association with a pair of top quarks ($t\bar{t}H$),
the simulated events were interfaced to \Pythia[8.230] for parton showering
and hadronisation using the A14 tune and the \NNPDF[2.3lo] PDF set.
The cross-section for $t\bar{t}H$ production is taken from NLO calculations~\cite{deFlorian:2016spz_edited}.
The contributions from other single Higgs production modes are expected to be negligible
and thus not considered.
SM single Higgs boson production plays a more important role as a background
in the non-resonant $HH$ search than in the resonant $HH$ search,
due to more similar production kinematics.


\section{Object reconstruction}
\label{sec:objects}

Electrons, muons, $\tau_\text{had-vis}$, jets from the hadronisation of quarks and gluons, including $b$-tagged jets, and $\mathbf{p}_\text{T}^\text{miss}$ are used in this search. An anti-$\tau_\text{had-vis}$ object, defined as a $\tau_\text{had-vis}$ with modified identification requirements, is also used to estimate the backgrounds from hadronic jets misidentified as $\tau_\text{had-vis}$. Tracks are used in the reconstruction, identification, isolation and vertex compatibility requirements and calibration of many of the physics objects described below, and in vertex reconstruction; they are reconstructed from hits in the inner tracking detectors, and are required to have $p_\text{T}>500~\MeV$~\cite{PERF-2015-08,ATL-PHYS-PUB-2015-051}. Events are required to have at least one collision vertex reconstructed from two or more associated tracks. If multiple vertices are found, the one with the largest $\sum p_\text{T}^2$ of the associated tracks is selected as the primary vertex. Finally, an overlap removal procedure is applied to ensure that no detector signature is identified as multiple reconstructed objects.
 
Electron candidates are reconstructed by matching tracks reconstructed in the inner detector to topological energy clusters in the electromagnetic calorimeter, with a reconstruction efficiency of around 98\%~\cite{EGAM-2018-01}. Electron candidates are required to have $p_\text{T}>7$~\GeV\ and $\vert\eta\vert<2.47$, and to be outside the transition region between the calorimeter's barrel and endcaps, $1.37<\vert\eta\vert<1.52$. They must pass track-quality requirements, followed by a loose likelihood-based selection that requires the shower profile to be compatible with that of an electromagnetic shower. These requirements have an efficiency of around 93\%. Isolation requirements are applied; these are based on the presence of tracks in a cone of $\pT^e$-dependent size $\Delta R$ around the electron and of calorimetric energy deposits in a fixed-size cone. Lastly, the electron energy scale is calibrated in data, and the energy resolution is calibrated in simulation, using $Z\to ee$ events~\cite{EGAM-2018-01}.
 
Muon candidates are reconstructed from tracks in the muon spectrometer, matched to tracks in the inner detector where available~\cite{MUON-2018-03}. In the absence of full tracks in the muon spectrometer, muons with $\vert\eta\vert<0.1$ can be reconstructed from track segments in the muon spectrometer, or energy deposits compatible with that of a minimum-ionising particle in the calorimeters. If an inner-detector track is present, it must match the direction and momentum of the muon spectrometer track for it to be included. The muon momentum is defined by using information from both the muon spectrometer and the inner detector where available. Only muon candidates with $p_\text{T}>7$~\GeV\ and $\vert\eta\vert<2.7$, and passing loose quality requirements based on the number of hits used to reconstruct the tracks, with an efficiency of around 99\%, are considered for further analysis. Lastly, isolation requirements with an efficiency of around 95\% that are based on the presence of particle-flow objects~\cite{PERF-2015-09} in a cone of $\pT^\mu$-dependent size $\Delta R$ around the muon are applied, except for muons used in the $b$-tagged-jet energy correction described below.
 
Jets are reconstructed using the anti-$k_t$ jet clustering algorithm~\cite{Cacciari:2008gp,Cacciari:2011ma} with a radius parameter of 0.4, applied to noise-suppressed positive-energy topological energy clusters~\cite{PERF-2014-07,ATL-PHYS-PUB-2015-036} and charged-particle tracks, processed using a particle-flow algorithm~\cite{PERF-2015-09}. Jet energies are corrected for contributions from pile-up, calibrated using energy- and $\eta$-dependent correction factors determined from comparisons between particle-level objects and reconstructed physics objects in simulated events, and then corrections are applied to account for effects due to the initiating-parton type and hadron composition~\cite{JETM-2018-05}. In data, a residual in situ correction is applied in order to correct for differences relative to simulation. Jets are required to have $p_\text{T}>20$~\GeV\ and $\vert\eta\vert<2.5$, or $p_\text{T}>30$~\GeV\ and $2.5<\vert\eta\vert<4.5$.
To reject jets from pile-up, jets with $p_\text{T}<60$~\GeV\ and $\vert\eta\vert<2.5$ ($20<p_\text{T}<120$~\GeV\ and $2.5<\vert\eta\vert<4.5$) are required to pass a `jet vertex tagger'~\cite{PERF-2014-03} (`forward jet vertex tagger'~\cite{ATL-PHYS-PUB-2015-034}) requirement to determine if they originate from the primary vertex~\cite{ATL-PHYS-PUB-2015-026}.
Lastly, quality criteria~\cite{ATLAS-CONF-2015-029} are applied to jets reconstructed from topological clusters~\cite{PERF-2014-07,ATL-PHYS-PUB-2015-036} in the calorimeter in order to identify and remove events containing jets from non-collision backgrounds and calorimeter noise.
 
A multivariate classification algorithm based on a deep neural network, the DL1r tagger~\cite{FTAG-2018-01,ATL-PHYS-PUB-2017-013}, is used to distinguish $b$-jets from the background of light-flavour- and charm-quark-initiated jets using information about the jet kinematics, the impact parameters of tracks associated with the jet, and the presence of displaced vertices. The inputs to the DL1r network include variables based on a recurrent neural network (RNNIP)~\cite{ATL-PHYS-PUB-2017-003}, which can exploit the spatial and kinematic correlations between tracks that are initiated from the same $b$-hadrons. This analysis uses a $b$-jet working point that targets an efficiency of 77\%. A simulation-based $p_\text{T}$-dependent correction is applied to the $b$-tagged jet momentum~\cite{HIGG-2016-29}. Finally, for $b$-tagged jets containing a muon, the difference between the reconstructed muon momentum and the momentum corresponding to the energy of its associated cluster in the calorimeter is added to the $b$-tagged jet momentum~\cite{HIGG-2016-29}. This corrects for the fact that the muon typically deposits only a small fraction of its energy in the calorimeters. These two corrections improve the resolution of the invariant mass of the pair of $b$-tagged jets by approximately 16\% for the non-resonant $HH$ signal.
 
The reconstructed $\tau_\text{had-vis}$ candidates~\cite{PERF-2013-06} are seeded by jets. The $\tau_\text{had-vis}$ energy is calibrated using multivariate methods with information from tracks and calorimeter energy clusters~\cite{ATLAS-CONF-2017-029}. The $\tau_\text{had-vis}$ are required to have $p_\text{T}>20$~\GeV\ and $\vert\eta\vert<2.5$, excluding $1.37<\vert\eta\vert<1.52$. Boosted decision trees are used to determine if tracks near a cluster originate from a $\tau_\text{had}$, and one or three tracks with a total charge of $\pm 1$ are required. The true-$\tau_\text{had-vis}$ are discriminated from quark- and gluon-initiated-jet backgrounds by using recurrent neural networks trained to target signatures with either one or three associated tracks~\cite{ATL-PHYS-PUB-2019-033}, and a loose requirement with an efficiency of around 85\% (75\%) for one-track (three-track) $\tau_\text{had-vis}$ candidates is applied. A separate boosted decision tree is then used to reject $\tau_\text{had-vis}$ candidates with one associated track that originate from electrons, with an efficiency of about 95\% for $\tau_\text{had}$~\cite{PERF-2013-06}. 
 
Anti-$\tau_\text{had-vis}$ objects are defined in order to estimate the background from jets misidentified as $\tau_\text{had-vis}$, as described in Section~\ref{sec:modelling:background}. They are only used for the estimation of these backgrounds, and only one anti-$\tau_\text{had-vis}$ per event is considered. Anti-$\tau_\text{had-vis}$ objects are reconstructed, and their energy is calibrated, in the same way as for $\tau_\text{had-vis}$ candidates, and they must satisfy the nominal $\tau_\text{had-vis}$ kinematic and track selection criteria. They are required to pass a looser recurrent neural-network requirement, corresponding to an efficiency of approximately 99\% for true-$\tau_\text{had-vis}$, but to fail the nominal recurrent neural-network requirement applied to the $\tau_\text{had-vis}$ candidates.
 
Neutrinos are produced in the decay of $\tau$-leptons and in the semileptonic decay of $b$-hadrons, and contribute to the total $\mathbf{p}_\text{T}^\text{miss}$ in the event. The $\mathbf{p}_\text{T}^\text{miss}$ is defined as the negative vector sum of the transverse momenta of all reconstructed and calibrated leptons, jets, and $\tau_\text{had-vis}$, and of all tracks matched to the primary vertex but not to other reconstructed objects in the event~\cite{PERF-2016-07}. The magnitude of $\mathbf{p}_\text{T}^\text{miss}$ is denoted by $E_\text{T}^\text{miss}$.

A sequential overlap removal procedure is applied to resolve ambiguities in which multiple electron, muon or jet candidates would otherwise be reconstructed from the same detector signature. This procedure uses a definition of angular distance, $\Delta R_y = \sqrt{(\Delta y)^{2}+(\Delta\phi)^{2}}$, that is based on the rapidity $y$ of the objects. The following steps are applied in order:
\begin{enumerate}
\item if any electrons share a track, all but the highest-$p_\text{T}$ electron are removed;
\item any $\tau_\text{had-vis}$ within $\Delta R_y=0.2$ of an electron or a muon is removed, except if the $\tau_\text{had-vis}$ has $p_\text{T}>50$~\GeV, in which case the muon must also be associated with a track in the muon spectrometer for the $\tau_\text{had-vis}$ to be removed;
\item if an electron and a muon share a track, the electron is removed if the muon is associated with a signature in the muon spectrometer, otherwise the muon is removed;
\item any jet within $\Delta R_y=0.2$ of an electron is removed;
\item any electron within $\Delta R_y=0.4$ of a jet is removed;
\item any jet within $\Delta R_y=0.2$ of a muon, or which would have the inner-detector track of a muon matched to it using ghost association~\cite{Cacciari:2007fd}, is removed if it has fewer than three associated tracks;
\item any muon within $\Delta R_y=0.4$ of a jet is removed;
\item any jet within $\Delta R_y=0.2$ of a $\tau_\text{had-vis}$ is removed;
\item if an anti-$\tau_\text{had-vis}$ is within $\Delta R_y=0.2$ of a jet, it is removed if the jet is $b$-tagged, otherwise the jet is removed.
\end{enumerate}


\section{Event selections}
\label{sec:selection}

Events are selected in three separate signal categories. The $\tau_\text{had}\tau_\text{had}$ category targets events with two oppositely charged $\tau_\text{had-vis}$ and two $b$-jets, whereas two $\tau_\text{lep}\tau_\text{had}$ categories target events with an electron or muon, an oppositely charged $\tau_\text{had-vis}$ and two $b$-jets. A control region ($Z$\,+\,HF CR) is also defined in order to determine the normalisation of the background in which a $Z$~boson is produced in association with one or more jets initiated by heavy-flavour quarks ($Z$\,+\,HF background). Events in the signal regions (SRs) are analysed for the presence of signal using multivariate discriminants, and discrimination between the $Z$\,+\,HF and $t\bar t$ backgrounds is achieved using the dilepton invariant mass ($m_{\ell\ell}$) in the $Z$\,+\,HF CR, as described in Section~\ref{sec:selection:zcr}.

\subsection{Signal regions}
\label{sec:selection:srs}
 
Events in the $\tau_\text{had}\tau_\text{had}$ category were recorded using a combination of single-$\tau_\text{had-vis}$ triggers (STTs) and di-$\tau_\text{had-vis}$ triggers (DTTs). The STTs accept events with at least a single $\tau_\text{had-vis}$ at the HLT with a minimum $p_\text{T}$ between 80~\GeV\ and 160~\GeV, depending on the data-taking period. The DTTs select events with at least a pair of $\tau_\text{had-vis}$ reconstructed at the HLT, with minimum $p_\text{T}$ of 35~\GeV\ (25~\GeV) for the (sub-)leading $\tau_\text{had-vis}$, where the (sub-)leading $\tau_\text{had-vis}$ is defined as the $\tau_\text{had-vis}$ with the (second-)highest $p_\text{T}$ in the event. From the 2016 data-taking period onward, additional requirements were applied in the L1 trigger to reduce the DTT trigger rates. During 2016 data-taking, an additional jet with $E_\text{T}>25~\GeV$ was required. For 2017 (2018) data-taking, if two offline\footnote{In this paper, offline objects are objects which are reconstructed after the data were collected, as opposed to trigger-level objects.} jets with $p_\text{T}>45~\GeV$ are found, then a trigger is used that requires two additional jets with $E_\text{T}>12~\GeV$ (and $\vert\eta\vert<2.3$) at L1, otherwise another trigger is used that requires one additional jet with $E_\text{T}>25~\GeV$ and the $\tau_\text{had-vis}$ to be reconstructed within $\Delta R=2.8$ of each other\footnote{This dependence of the online selection on the offline requirements ensures that both triggers will not be used for the same events, which avoids the need to study their combined efficiency.}.
In order to select events near the trigger efficiency plateaus where the efficiencies are well modelled, the offline $\tau_\text{had-vis}$ are required to be within $\Delta R=0.2$ of the corresponding HLT $\tau_\text{had-vis}$ objects and a minimum offline $p_\text{T}$ requirement is applied to $\tau_\text{had-vis}$ and to the jets. The offline $p_\text{T}$ thresholds for the $\tau_\text{had-vis}$ range between 100~\GeV\ and 180~\GeV\ for the STTs, and are 40~\GeV\ (30~\GeV) for the (sub-)leading $\tau_\text{had-vis}$ for the DTTs. Additional offline requirements for the DTTs are either that two additional jets with $p_\text{T}>45$~\GeV\ are present in the event, or that a jet with $p_\text{T}>80~\GeV$ is present in the event and the $\tau_\text{had-vis}$ are reconstructed within $\Delta R=2.5$ of each other. For events that pass both the STTs and DTTs, the offline requirements used for the STTs are applied. Events passing the $\tau_\text{had}\tau_\text{had}$ event selection are analysed together.
 
Events in the $\tau_\text{lep}\tau_\text{had}$ categories were recorded using a combination of single-lepton triggers (SLTs) and lepton-plus-$\tau_\text{had-vis}$ triggers (LTTs). The SLTs require an electron or muon to be reconstructed at the HLT with an $E_\text{T}$ threshold that ranges from 24~\GeV\ to 26~\GeV\ for electrons and a $p_\text{T}$ threshold that ranges from 20~\GeV\ to 26~\GeV\ for muons, depending on the data-taking period. The LTTs require that an electron with $p_\text{T}>17$~\GeV\ or a muon with $p_\text{T}>14$~\GeV , and a $\tau_\text{had-vis}$ with $p_\text{T}>25$~\GeV , are reconstructed at the HLT. The LTTs used to collect lepton-plus-$\tau_\text{had-vis}$ events with $\tau_\text{had-vis}$ $p_\text{T}<35~\GeV$ had additional requirements at the L1 trigger, requiring the presence of either an additional jet with $E_\text{T}>25~\GeV$ or two additional jets with $E_\text{T}>12~\GeV$. The electron-plus-$\tau_\text{had-vis}$ triggers that require an additional jet with $E_\text{T}>25~\GeV$ was only used to select an event if the electron-plus-$\tau_\text{had-vis}$ triggers that require two additional jets with $E_\text{T}>12~\GeV$ did not select the event. For any trigger to select an event, based on the presence of a muon, the muon must have $\vert\eta\vert<2.5$. In order to select events near the trigger efficiency plateaus where the trigger efficiencies are well modelled, the offline electrons, muons and $\tau_\text{had-vis}$ objects are required to be within $\Delta R=0.07$, $\Delta R=0.1$ and $\Delta R=0.2$ of the corresponding objects at the HLT, respectively. Minimum $p_\text{T}$ requirements are applied to the offline objects, and these are 1~\GeV\ above the thresholds for electrons and muons at the HLT, 5~\GeV\ above the thresholds for $\tau_\text{had-vis}$ at the HLT, and 80~\GeV\ (45~\GeV) for jets with an L1-trigger $E_\text{T}$ threshold of 25~\GeV\ (12~\GeV).
Events which pass the offline SLT lepton $p_\text{T}$ requirements are not considered for the LTT. This ensures that there is no overlap between the SLT and LTT categories. These two categories are analysed separately.

In the $\tau_\text{had}\tau_\text{had}$ category, in addition to the trigger selection described above, events are required to contain exactly two $\tau_\text{had-vis}$ with opposite charges and exactly two $b$-tagged jets. Events with any additional leptons are vetoed, and the (sub-)leading $b$-tagged jet is required to have $p_\text{T}>45~(20)$~\GeV. The invariant mass of the $\tau$-lepton pair ($m_{\tau\tau}^\text{MMC}$) is estimated from the four-momenta of the $\tau_\text{had-vis}$ and the $\mathbf{p}_\text{T}^\text{miss}$ using the Missing Mass Calculator (MMC)~\cite{Elagin:2010aw}, which assumes that the $\mathbf{p}_\text{T}^\text{miss}$ is exclusively from the neutrinos produced in the $\tau$-lepton decays. To reject background from low-mass Drell--Yan events, $m_{\tau\tau}^\text{MMC}$ is required to be above 60~\GeV.

In the $\tau_\text{lep}\tau_\text{had}$ categories, exactly one `loose' electron or `loose' muon, an oppositely charged $\tau_\text{had-vis}$, and exactly two $b$-tagged jets are required. The selected electron (muon) must also pass a tight (medium) identification requirement with an efficiency of around 80\% (97\%). Events are required to have $m_{\tau\tau}^\text{MMC}>60~\GeV$, where $m_{\tau\tau}^\text{MMC}$ is calculated using the four-momenta of the electron or muon, the $\tau_\text{had-vis}$ and the $\mathbf{p}_\text{T}^\text{miss}$. The $b$-tagged jet pair invariant mass ($m_{bb}$) is required to be less than 150~\GeV\ to reject $t\bar t$ background events, and to allow for the definition of a $t\bar t$-enriched region which is used in the estimation of $t\bar t$ backgrounds, as described in Section~\ref{sec:modelling:background}. A $\tau_\text{had-vis}$ with $p_\text{T}>20$~\GeV\ and $\vert\eta\vert<2.3$ is required in the SLT category, and a $\tau_\text{had-vis}$ with $p_\text{T}>30$~\GeV, or higher if required by the trigger, and $\vert\eta\vert<2.3$ is required in the LTT category. In both categories, the (sub-)leading $b$-tagged jet must have $p_\text{T}>45~(20)$~\GeV, in addition to any trigger-dependent requirements.
 
The full event selections are summarised in Table~\ref{tab:selection:selectionsummary}. The acceptance times efficiency for the non-resonant ggF+VBF $HH\to b\bar b\tau^+\tau^-$ signal, evaluated with respect to the targeted $\tau$-lepton pair decay mode, is 4.0\%, 4.0\% and 1.0\%, in the $\tau_\text{had}\tau_\text{had}$ category, and the $\tau_\text{lep}\tau_\text{had}$ SLT and $\tau_\text{lep}\tau_\text{had}$ LTT categories, respectively. For the resonant $HH$ signal, the acceptance times efficiency is shown in Figure~\ref{fig:selection:acceptances} as a function of the resonance mass. The decrease in acceptance times efficiency for $m_X$ greater than about 1000~\GeV\ is due to the Lorentz boost of the Higgs bosons causing their decay products to become highly collimated more often.
 
\begin{table}[!htbp]
\caption{Summary of the event selections, shown separately for events that are selected by different triggers. In cases where pairs of reconstructed objects of the same type are required, thresholds for the (sub-)leading $p_\text{T}$ object are given outside (within) parentheses where different event selection thresholds are applied. When the selection depends on the year of data-taking, the possible values of the requirements are separated by commas, except for the jet selection in the LTT and DTT triggers, which use multiple selection criteria as described in Section~\ref{sec:selection:srs}. The trigger $p_\text{T}$ thresholds shown are applied to the offline physics objects that are matched to the corresponding trigger objects.}
\label{tab:selection:selectionsummary}
\centering
\resizebox{\textwidth}{!}{
\begin{tabular}{C{0.27\textwidth}C{0.27\textwidth}C{0.27\textwidth}C{0.27\textwidth}}
\toprule
\multicolumn{2}{c}{$\tau_\text{had}\tau_\text{had}$ category} & \multicolumn{2}{c}{$\tau_\text{lep}\tau_\text{had}$ categories}\\
STT & DTT & SLT & LTT\\
\midrule
\multicolumn{4}{c}{\BF{$e/\mu$ selection}}\\
\multicolumn{2}{c}{No loose $e/\mu$} & \multicolumn{2}{c}{Exactly one loose $e/\mu$}\\
& & \multicolumn{2}{c}{$e$ ($\mu$) must be tight (medium and have $\vert\eta\vert<2.5$)}\\
& & $p_\text{T}^e>25,27~\GeV$ & $18~\GeV<p_\text{T}^e<\text{SLT cut}$\\
& & $p_\text{T}^\mu>21,27~\GeV$ & $15~\GeV<p_\text{T}^\mu<\text{SLT cut}$\\
\midrule
\multicolumn{4}{c}{\BF{$\tau_\text{had-vis}$ selection}}\\
\multicolumn{2}{c}{Two loose $\tau_\text{had-vis}$} & \multicolumn{2}{c}{One loose $\tau_\text{had-vis}$}\\
& & \multicolumn{2}{c}{$\vert\eta\vert<2.3$}\\
$p_\text{T}>100,140,180~(25)~\GeV$ & $p_\text{T}>40~(30)~\GeV$ & & $p_\text{T}>30~\GeV$\\
\midrule
\multicolumn{4}{c}{\BF{Jet selection}}\\
\multicolumn{4}{c}{$\geq 2$ jets with $\vert\eta\vert<2.5$}\\
Leading jet $p_\text{T}>45~\GeV$ & Trigger dependent & Leading jet $p_\text{T}>45~\GeV$ & Trigger dependent\\
\midrule
\multicolumn{4}{c}{\BF{Event-level selection}}\\
\multicolumn{4}{c}{Trigger requirements passed}\\
\multicolumn{4}{c}{Collision vertex reconstructed}\\
\multicolumn{4}{c}{$m_{\tau\tau}^\text{MMC}>60~\GeV$}\\
\multicolumn{4}{c}{Opposite-sign electric charges of $e/\mu/\tau_\text{had-vis}$ and $\tau_\text{had-vis}$}\\
\multicolumn{4}{c}{Exactly two $b$-tagged jets}\\
& & \multicolumn{2}{c}{$m_{bb}<150~\GeV$}\\
\bottomrule
\end{tabular}
}
\end{table}
 
\begin{figure}[!htbp]
\centering
\includegraphics[width=0.65\linewidth]{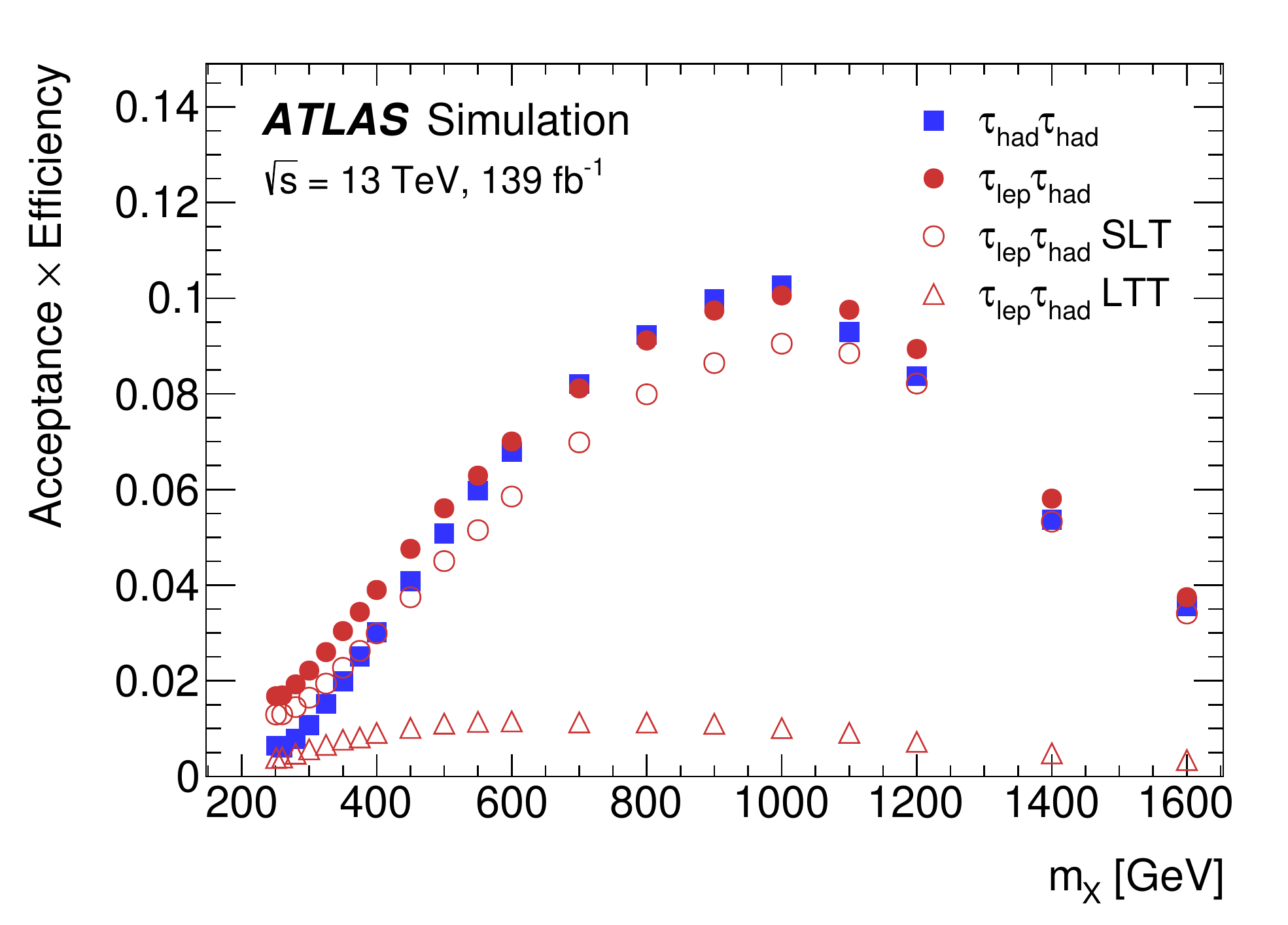}
\caption{Acceptance times efficiency for the full analysis selections as a function of the resonance mass $m_X$
in the $\tau_\text{had}\tau_\text{had}$, $\tau_\text{lep}\tau_\text{had}$ SLT and
$\tau_\text{lep}\tau_\text{had}$ LTT trigger categories, shown in solid square markers,
empty circle markers and empty triangle markers, respectively.
The solid circle markers are the acceptance times efficiency values for the
combined $\tau_\text{lep}\tau_\text{had}$ category.
The acceptance times efficiency for $X\to HH\to b\bar{b}\tau^+\tau^-$ decays is evaluated with
respect to the targeted $\tau$-lepton pair decay mode ($\tau_\text{lep}\tau_\text{had}$ or
$\tau_\text{had}\tau_\text{had}$).}
\label{fig:selection:acceptances}
\end{figure}
 
\subsection{Multivariate signal extraction}
\label{sec:selection:mvas}
 
Multivariate discriminants (MVAs) evaluated for events passing the above selections are used to extract possible signals. Parameterised neural networks (PNNs)~\cite{Baldi:2016fzo} are used in the search for resonant $HH$ production, and a boosted decision tree (BDT) and neural networks (NNs) are used in the $\tau_\text{had}\tau_\text{had}$ category and $\tau_\text{lep}\tau_\text{had}$ categories of the non-resonant $HH$ search, respectively. The (P)NNs are trained using Keras~\cite{chollet2015keras} with the Tensorflow~\cite{tensorflow2015-whitepaper} backend, and the BDT is trained using TMVA~\cite{TMVA}. During training, the sum of all backgrounds normalised to their respective cross-sections is used. The backgrounds containing one or more $\tau_\text{had-vis}$ from a misidentified quark- or gluon-initiated jet are modelled using simulation, except for the multi-jet background in the $\tau_\text{had}\tau_\text{had}$ category, where the data-driven estimate is used. The PNNs are parameterised in the mass of the heavy resonance, providing near-optimal sensitivity and continuity over the range of signal masses considered.

The same choice of MVA input variables is used for the resonant and non-resonant production modes, although different input variables are used in the different analysis categories. These variables are listed in Table~\ref{tab:selection:mvas:HHinputs}, and are defined as follows:
\begin{itemize}
\item $m_{HH}$ is the invariant mass of the $HH$ system as reconstructed from the $\tau$-lepton pair (calculated using the MMC) and the $b$-tagged jet pair. In the $\tau_\text{lep}\tau_\text{had}$ categories the four-momenta of the $b\bar{b}$ and $\tau^{+}\tau^{-}$ systems are multiplied by correction factors (125 GeV/$m_{bb}$ and 125 GeV/$m_{\tau\tau}^\text{MMC}$, respectively) in order to improve the $m_{HH}$ mass resolution;
\item $\Delta  R(\tau, \tau)$ is evaluated between the two $\tau_\text{had-vis}$ (the electron or muon and the $\tau_\text{had-vis}$) in the $\tau_\text{had}\tau_\text{had}$ category ($\tau_\text{lep}\tau_\text{had}$ categories);
\item $\Delta  R(b, b)$ is evaluated between the $b$-tagged jets;
\item $\Delta p_\text{T}(\ell, \tau)$ is the difference between the transverse momenta of the lepton and the $\tau_\text{had-vis}$;
\item $m_\text{T}^W =\sqrt{2p_\text{T}^\ell E_\text{T}^\text{miss}(1-\cos\Delta\phi_{\ell,\mathbf{p}_\text{T}^\text{miss}})}$ is the transverse mass of the lepton and the $\mathbf{p}_\text{T}^\text{miss}$;
 
\item the $\mathbf{p}_\text{T}^\text{miss}~\phi$ centrality specifies the angular position of the $\mathbf{p}_\text{T}^{\text{miss}}$ relative to the $\tau_\text{had-vis}$ in the transverse plane~\cite{HIGG-2013-32} and is defined as $(A+B)/\sqrt{A^2+B^2}$, where $A=\sin(\phi_{\mathbf{p}_\text{T}^\text{miss}}-\phi_{\tau_2})/\sin(\phi_{\tau_1}-\phi_{\tau_2})$, $B=\sin(\phi_{\tau_1}-\phi_{\mathbf{p}_\text{T}^\text{miss}})/\sin(\phi_{\tau_1}-\phi_{\tau_2})$, and $\tau_1$ and $\tau_2$ represent the two $\tau_\text{had-vis}$ (electron or muon and $\tau_\text{had-vis}$) in the case of the $\tau_\text{had}\tau_\text{had}$ category ($\tau_\text{lep}\tau_\text{had}$ categories);
 
\item $\Delta\phi(\ell\tau, bb)$ is the azimuthal angle between the $\ell+\tau_\text{had-vis}$ system and the $b$-tagged jet pair;
\item $\Delta\phi(\ell, \mathbf{p}_\text{T}^\text{miss})$ is the azimuthal angle between the lepton and the $\mathbf{p}_\text{T}^\text{miss}$;
\item $\Delta\phi(\tau\tau, \mathbf{p}_\text{T}^\text{miss})$ is the azimuthal angle between the $\tau$-lepton pair system (estimated using the MMC) and the $\mathbf{p}_\text{T}^\text{miss}$;
\item $S_\text{T}$ is the total transverse energy in the event, summed over all jets, $\tau_\text{had-vis}$ and leptons in the event and $E_\text{T}^\text{miss}$.
\end{itemize}
The $m_{HH}$, $m_{\tau\tau}^\text{MMC}$ and $m_{bb}$ distributions are shown in Figure~\ref{fig:selection:mvas:mHHmMMCmbb}. For all categories of the non-resonant search, $m_{\tau\tau}^\text{MMC}$ and $m_{bb}$ are among the three most important MVA input variables. For the resonant search, five values of $m_X$ were tested in all categories, and $m_{HH}$ was found to be the most important MVA input variable in all cases except at lower values of $m_X$ in the $\tau_\text{had}\tau_\text{had}$ category. The $m_{\tau\tau}^\text{MMC}$ and $m_{bb}$ variables separate resonant $H\to\tau\tau$ and $H\to bb$ signals (respectively) from backgrounds that do not contain these processes, and the $m_{HH}$ variable targets resonant $X\to HH$ decays in the resonant search.
 
\begin{table}[!htbp]
\caption{Variables used as inputs to the MVAs in the three analysis categories. The same choice of input variables is used for the resonant and non-resonant production modes. The variables are defined in the main text.}
\label{tab:selection:mvas:HHinputs}
\centering
\begin{tabular}{lccc}
\toprule
Variable & $\tau_\text{had}\tau_\text{had}$ & $\tau_\text{lep}\tau_\text{had}$ SLT & $\tau_\text{lep}\tau_\text{had}$ LTT\\
\midrule
$m_{HH}$ & \ding{51} & \ding{51} & \ding{51} \\
$m_{\tau\tau}^\text{MMC}$ & \ding{51} & \ding{51} & \ding{51} \\
$m_{bb}$ & \ding{51} & \ding{51} & \ding{51} \\
$\Delta R(\tau, \tau)$ & \ding{51} & \ding{51} & \ding{51} \\
$\Delta R(b, b)$ & \ding{51} & \ding{51} & \\
$\Delta p_\text{T}(\ell, \tau)$ & & \ding{51} & \ding{51} \\
Sub-leading $b$-tagged jet $p_\text{T}$ & & \ding{51} & \\
$m_\text{T}^W$ & & \ding{51} & \\
$E_\text{T}^\text{miss}$ & & \ding{51} & \\
$\mathbf{p}_\text{T}^\text{miss}~\phi$ centrality & & \ding{51} & \\
$\Delta\phi(\ell\tau, bb)$ & & \ding{51} & \\
$\Delta\phi(\ell, \mathbf{p}_\text{T}^\text{miss})$ & & & \ding{51} \\
$\Delta\phi(\tau\tau, \mathbf{p}_\text{T}^\text{miss})$ & & & \ding{51} \\
$S_\text{T}$ & & & \ding{51} \\
\bottomrule
\end{tabular}
\end{table}

\begin{figure}[!htbp]
\centering
\subfigure[]{\includegraphics[width=0.32\textwidth]{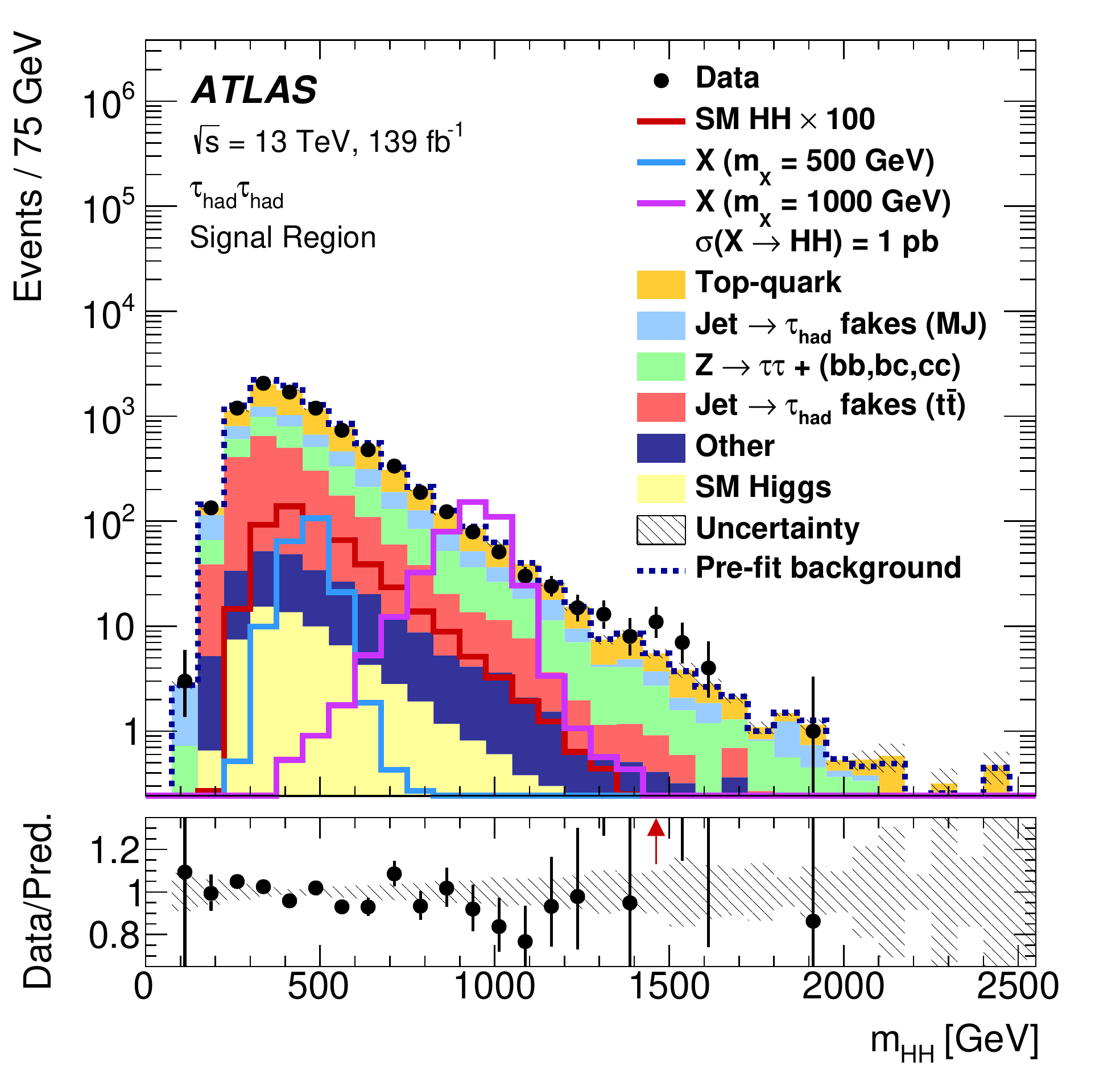}}
\subfigure[]{\includegraphics[width=0.32\textwidth]{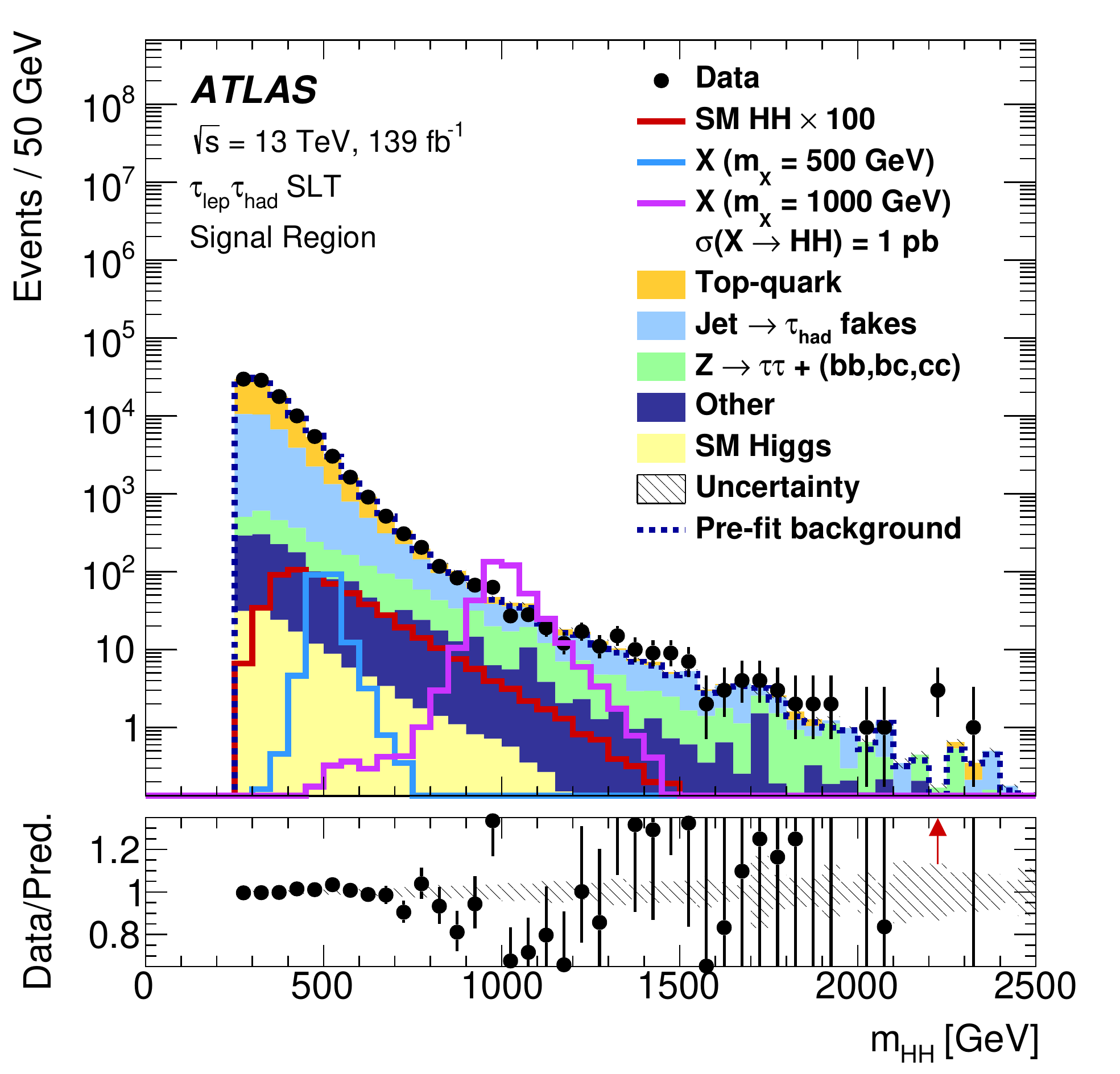}}
\subfigure[]{\includegraphics[width=0.32\textwidth]{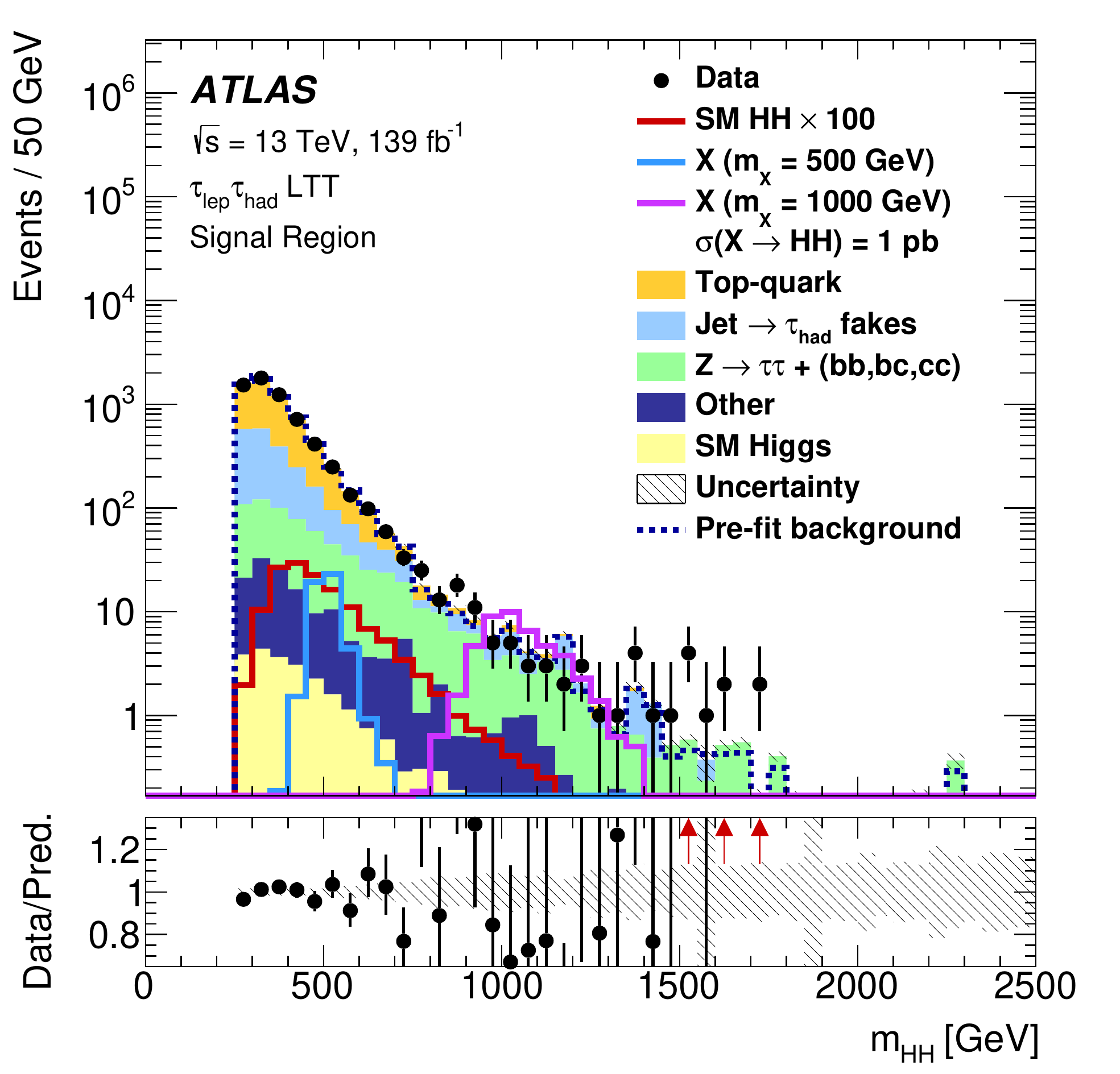}}\\
\subfigure[]{\includegraphics[width=0.32\textwidth]{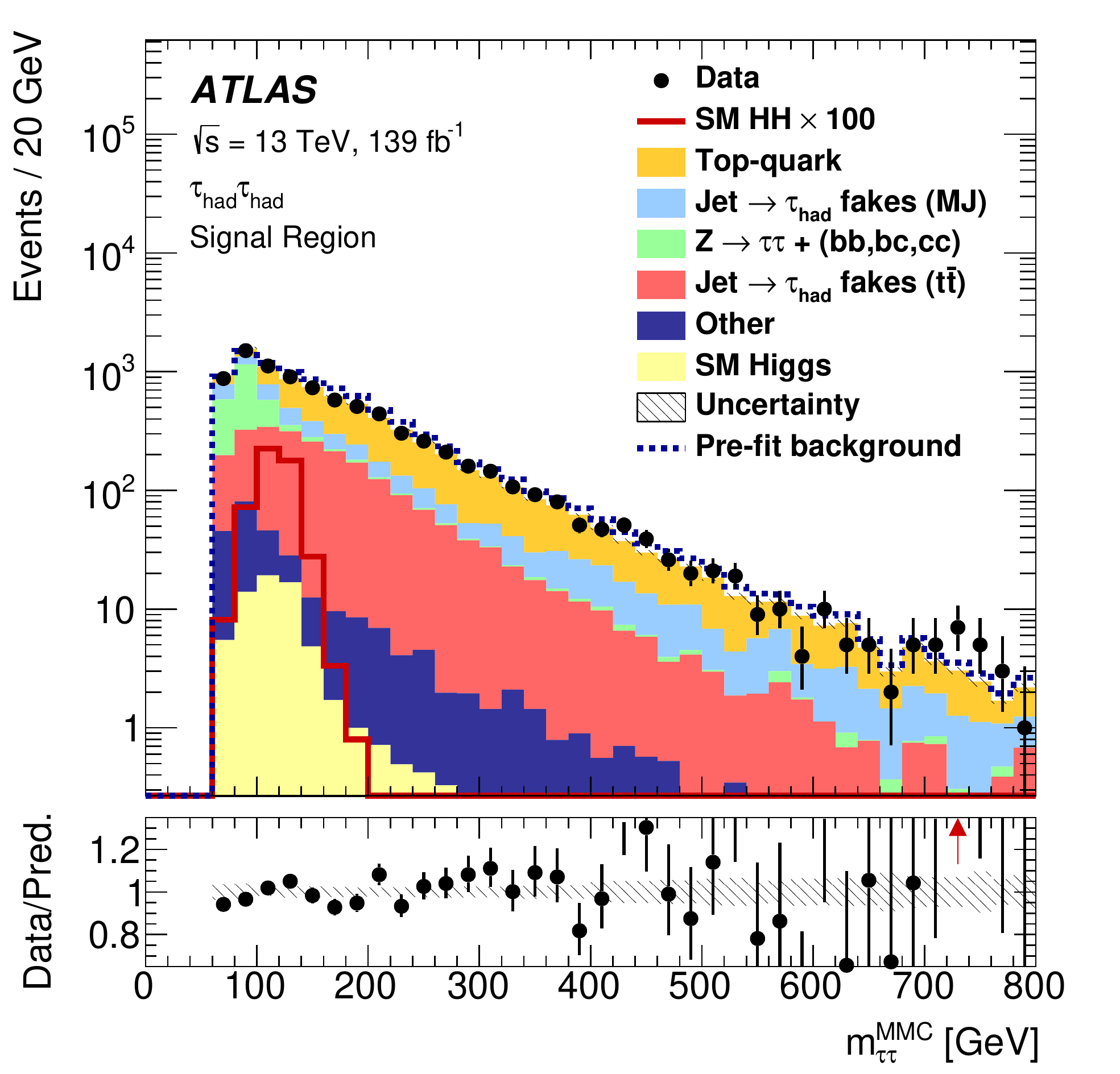}}
\subfigure[]{\includegraphics[width=0.32\textwidth]{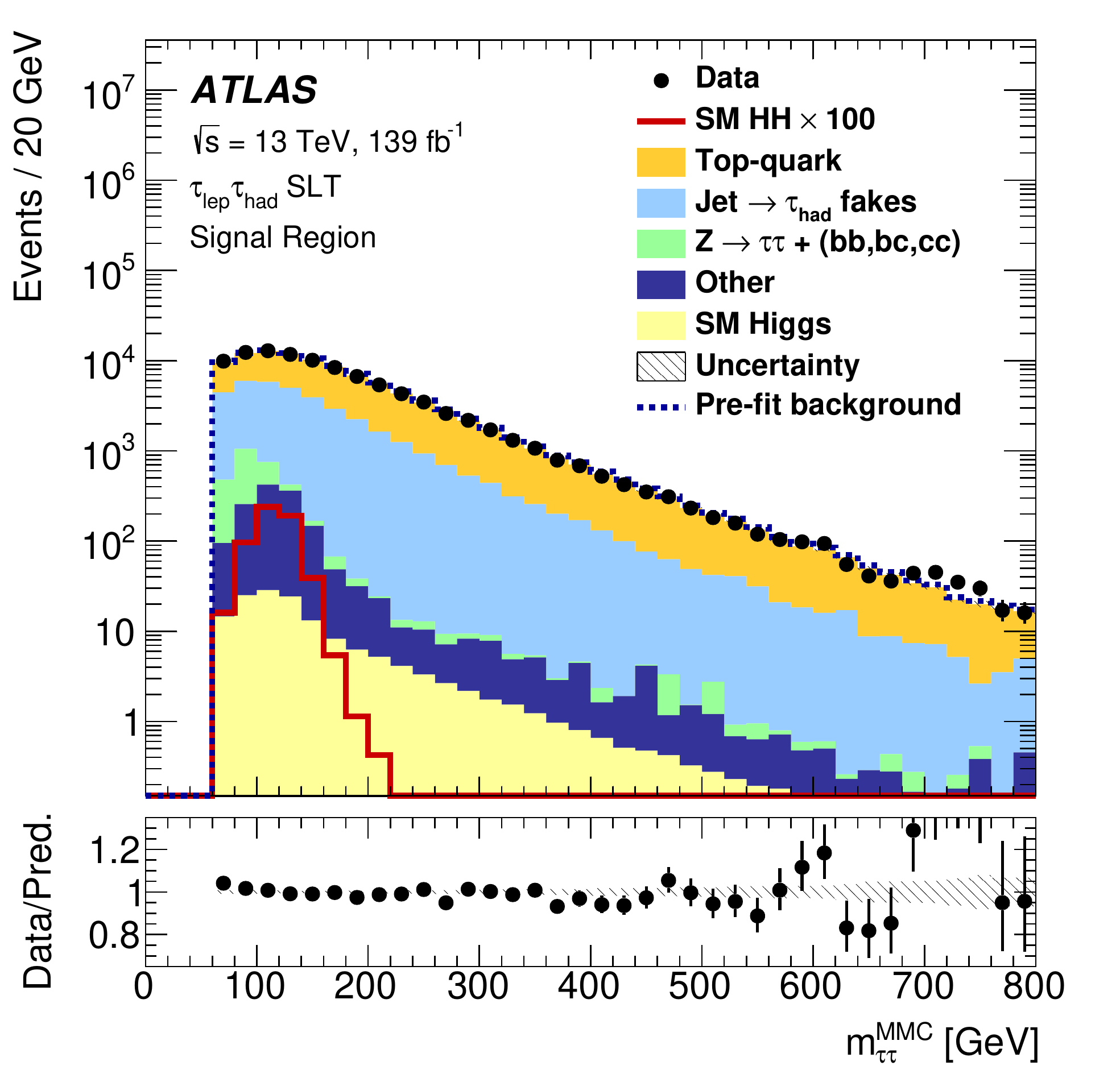}}
\subfigure[]{\includegraphics[width=0.32\textwidth]{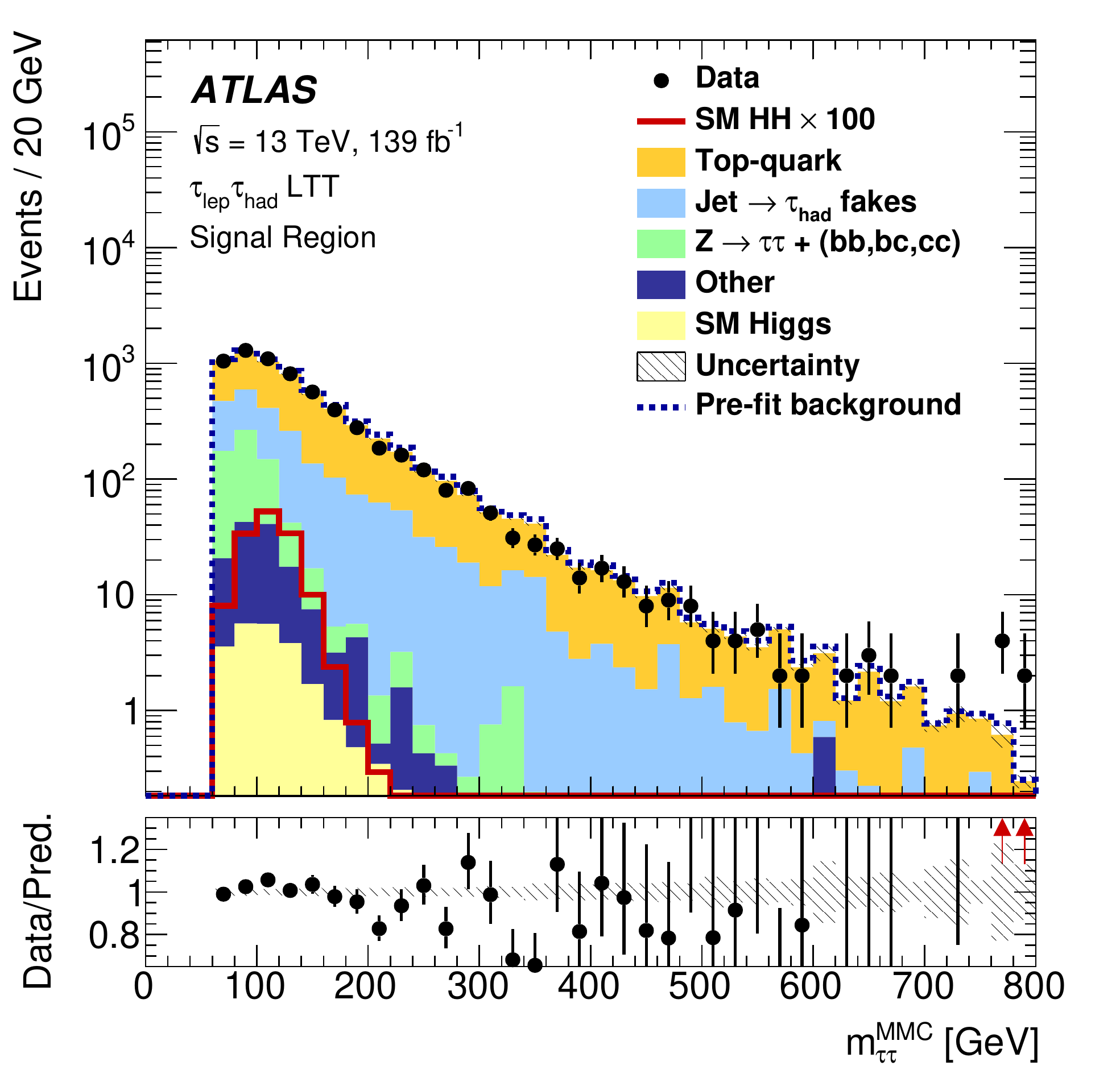}}\\
\subfigure[]{\includegraphics[width=0.32\textwidth]{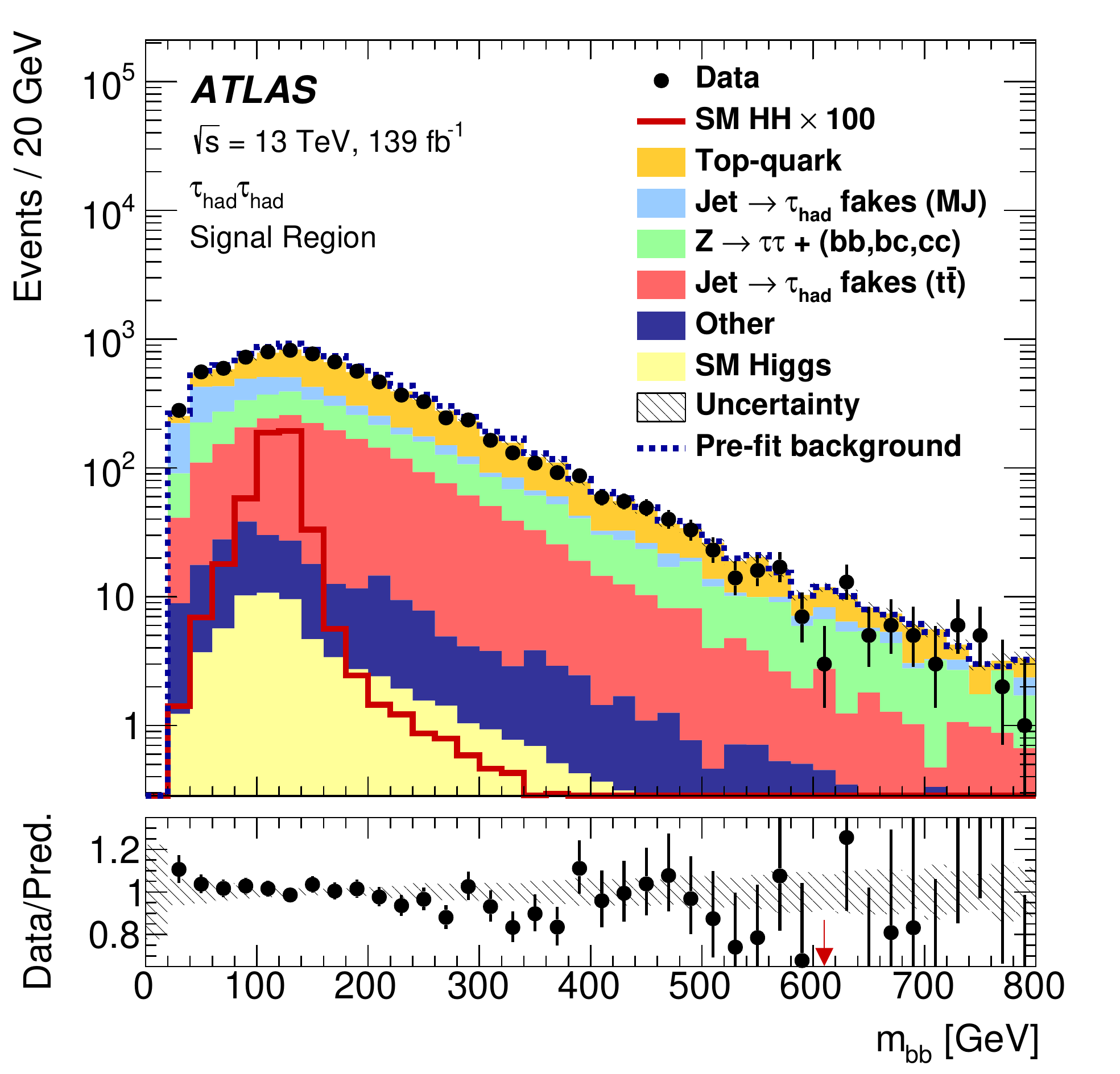}}
\subfigure[]{\includegraphics[width=0.32\textwidth]{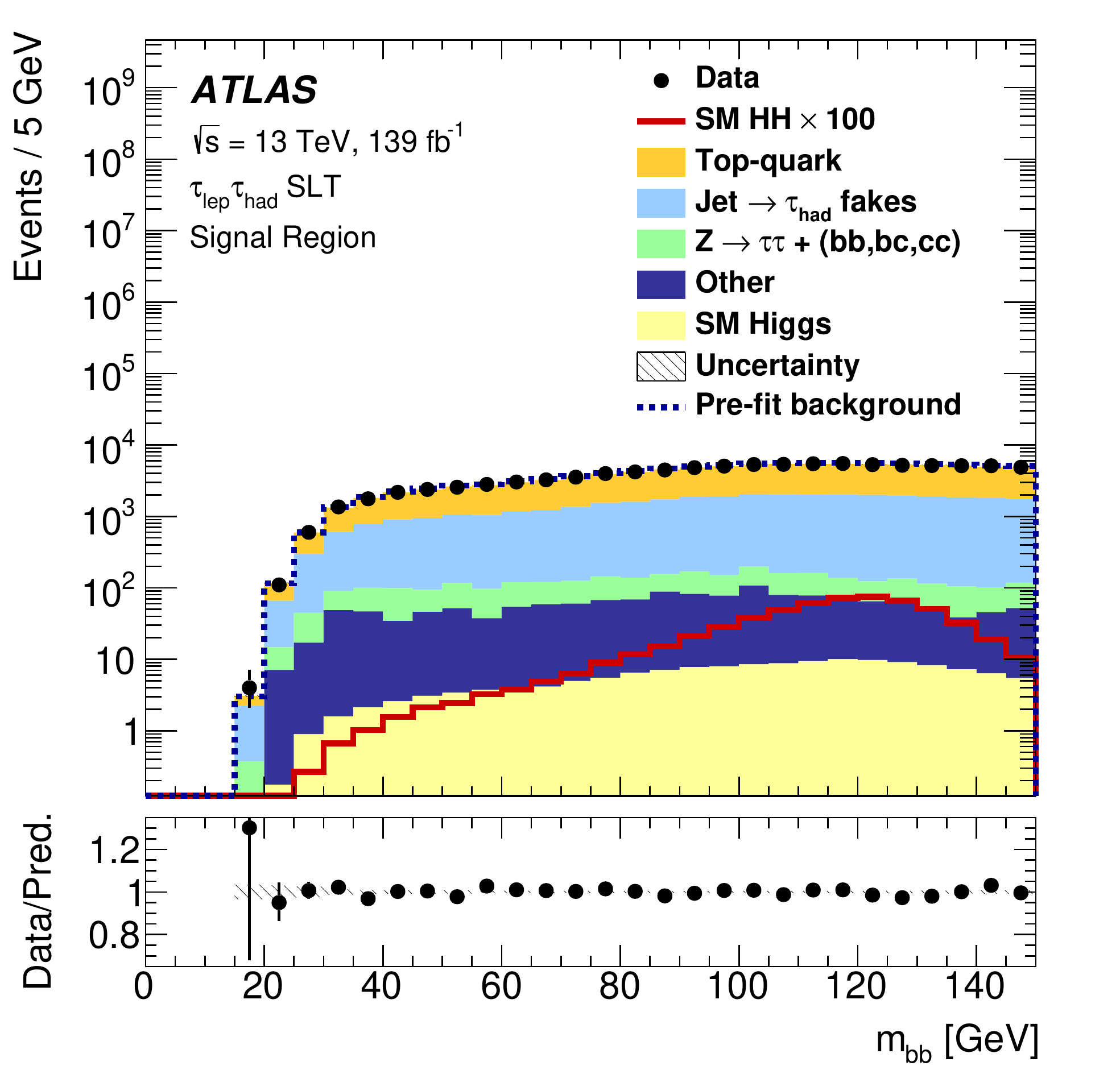}}
\subfigure[]{\includegraphics[width=0.32\textwidth]{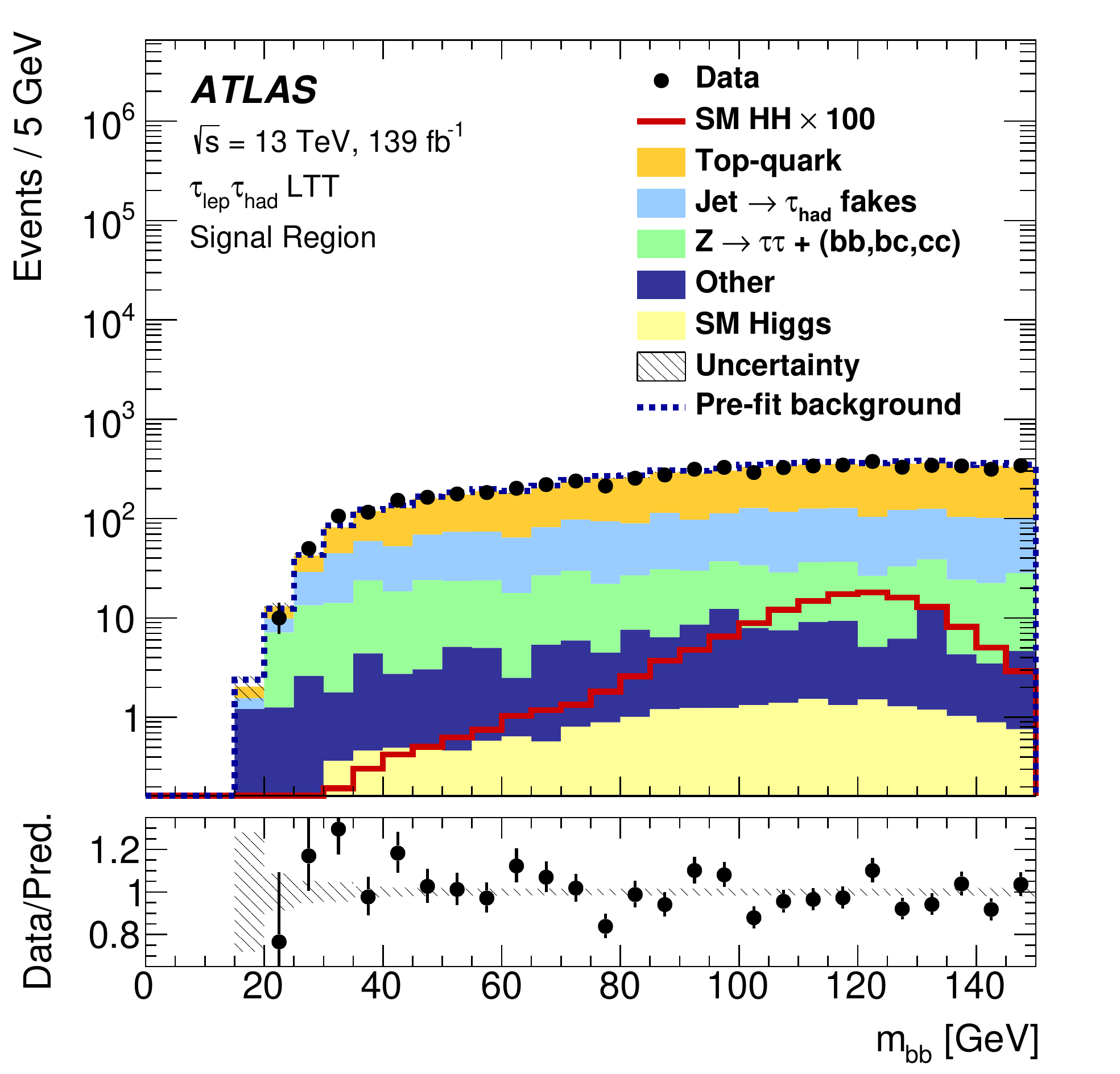}}
 
\caption{Signal (solid lines), background (filled histograms) and data (dots with error bars) distributions of $m_{HH}$ (top), $m_{\tau\tau}^\text{MMC}$ (middle row) and $m_{bb}$ (bottom) for events in the $\tau_\text{had}\tau_\text{had}$ (left), $\tau_\text{lep}\tau_\text{had}$ SLT (middle column) and $\tau_\text{lep}\tau_\text{had}$ LTT (right) categories. The normalisation and shape of the backgrounds and the uncertainty in the total background shown are determined from the likelihood fit (described in Section~\ref{sec:stats}) to data in the non-resonant $HH$ search. The expected non-resonant signal is overlaid with its normalisation scaled by a factor of 100, and the $m_X=500~\GeV$ and $m_X=1000~\GeV$ resonant signals are overlaid in the $m_{HH}$ distributions with their cross-section set to 1~pb.
The dashed histogram shows the total pre-fit background.
The size of the combined statistical and systematic uncertainty of the background is indicated by the hatched band.
The ratio of the data to the sum of the backgrounds is shown in the lower panels.}
\label{fig:selection:mvas:mHHmMMCmbb}
\end{figure}
 
The (P)NNs use rectified linear unit and sigmoidal activation functions for the hidden and output layers, respectively, binary cross entropy as the loss function, and stochastic (mini-batch) gradient descent as the optimiser~\cite{Goodfellow-et-al-2016}. The PNN used in the $\tau_\text{had}\tau_\text{had}$ category has 3 hidden layers of 128 nodes, followed by 1 hidden layer of 16 nodes. The (P)NN used in the $\tau_\text{lep}\tau_\text{had}$ SLT category has 2 hidden layers of 512 nodes, and the (P)NN used in the $\tau_\text{lep}\tau_\text{had}$ LTT category has 3 hidden layers of 512 (256) nodes. The (P)NN input variables are standardised by subtracting the median value and dividing by the interquartile range. Nesterov momentum and learning-rate decay were used in the training of all (P)NNs, and in the $\tau_\text{lep}\tau_\text{had}$ categories they used an L2 regularisation (`ridge regression') term in the loss function~\cite{Goodfellow-et-al-2016}. The BDT uses 1500 trees with a maximum depth of 2 and a minimum node size of 1\% of the training events. Gradient boosting is used with a shrinkage of 0.2.

\subsection{$Z$\,+\,HF control region}
\label{sec:selection:zcr}
 
The normalisation of the $Z$\,+\,HF background is determined from data by fitting the $m_{\ell\ell}$ distribution in the $Z$\,+\,HF CR in the likelihood fit (described in Section~\ref{sec:stats}). This is to account for a known discrepancy between the $Z$\,+\,HF production cross-section provided at NLO by \Sherpa and the cross-section observed in data. The $Z$\,+\,HF CR targets events containing $Z$~boson decays into electron or muon pairs by using triggers that require either a lepton or a pair of same-flavour leptons. Exactly two oppositely charged same-flavour leptons and exactly two $b$-tagged jets must be reconstructed offline. The leptons are required to have $p_\text{T}>9~\GeV$, pass offline $p_\text{T}$ thresholds based on the trigger thresholds, be compatible with originating from the primary vertex, and pass medium identification and loose isolation requirements. Lastly, the invariant mass of the lepton pair is required to be between 75~\GeV\ and 110~\GeV\ to select events which include a $Z$~boson decay, and $m_{bb}$ is required to be less than 40~\GeV\ or greater than 210~\GeV\ to ensure orthogonality to the event selection of another ATLAS search. This region also provides constraints on the normalisation of the $t\bar t$ background. Figure~\ref{fig:zcr} shows the post-fit background and data $m_{\ell\ell}$ distributions in the $Z$\,+\,HF control region.

\begin{figure}[!htbp]
\centering
\subfigure[]{\includegraphics[width=0.50\textwidth]{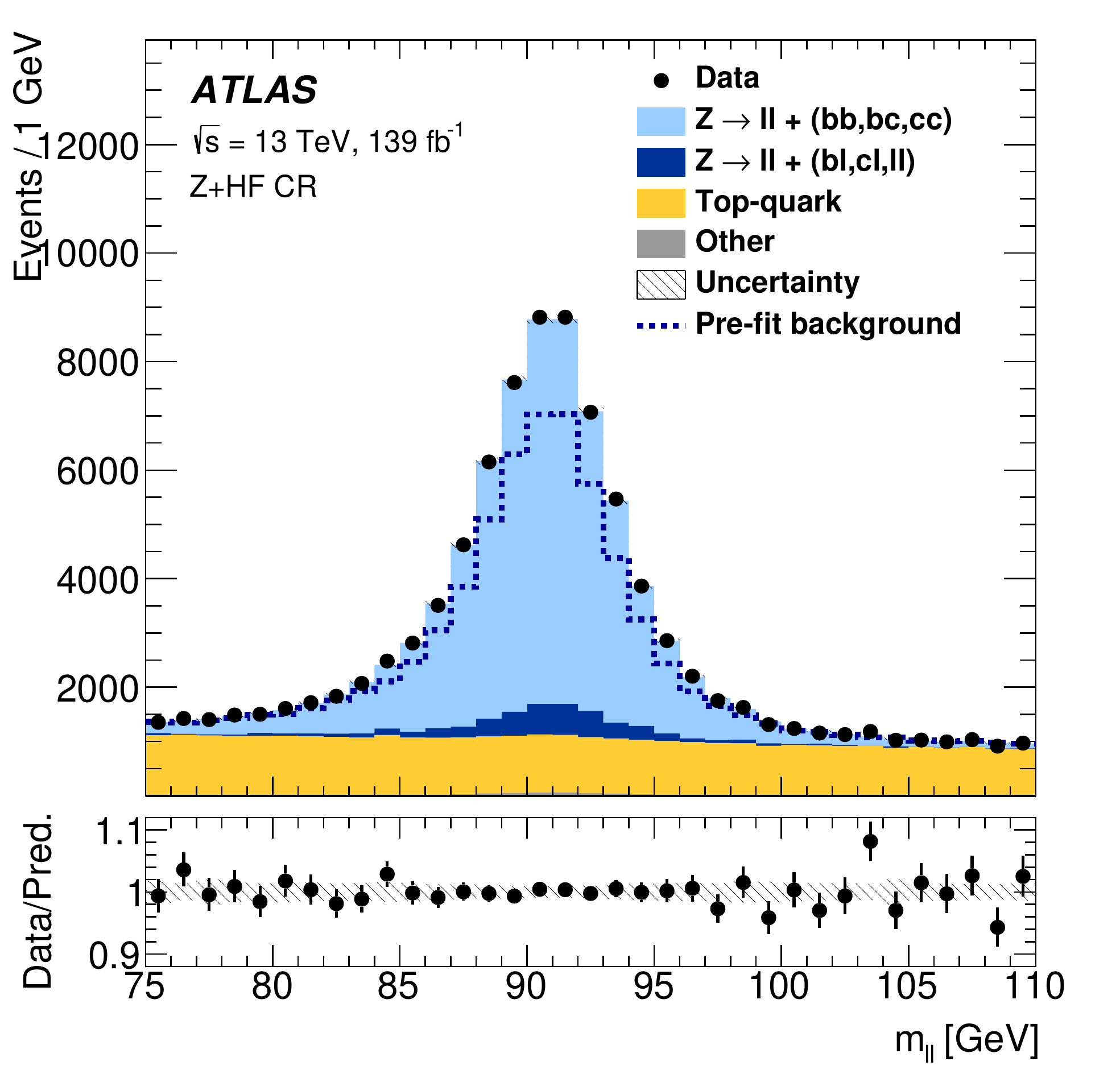}}
\caption{Post-fit background and data $m_{\ell\ell}$ distributions in the $Z$\,+\,HF control region. The normalisation and shape of the backgrounds and the uncertainty in the total background are shown as determined from the likelihood fit (described in Section~\ref{sec:stats}) to data in the non-resonant $HH$ search. The uncertainty band includes statistical and systematic uncertainties of the total background. The dashed histogram shows the total pre-fit background.}
\label{fig:zcr}
\end{figure}


\section{Background modelling}
\label{sec:modelling:background}


Backgrounds in this search are estimated using a combination of simulation-based and data-driven techniques. The main sources of background are top-quark, $Z$+jets, $W$+jets, diboson,
single Higgs boson and multi-jet production. A reconstructed $\tau_\text{had-vis}$, in these background events, can originate either from a $\tau_\text{had}$ decay (true-$\tau_\text{had-vis}$), or from a misidentified quark- or gluon-initiated jet (fake-$\tau_\text{had-vis}$).
Events in which an electron or a muon is misidentified as a $\tau_\text{had-vis}$
represent a small additional background.

Most of the background events with fake-$\tau_\text{had-vis}$ are from $t\bar{t}$ or multi-jet production.
In $t\bar{t}$ events, fake-$\tau_\text{had-vis}$ typically originate from quark-initiated jets from the top-quark decay.
In multi-jet events, both quark- and gluon-initiated jets may be misidentified as $\tau_\text{had-vis}$.

The simulated event samples, summarised in Section~\ref{sec:samples:simulation}, are used to model
background events containing true-$\tau_\text{had-vis}$ and events with an electron or a muon misidentified as a $\tau_\text{had-vis}$. Events with fake-$\tau_\text{had-vis}$ in $t\bar{t}$ or multi-jet production are estimated
using techniques relying on both simulated events and data, as detailed in the following
subsections. Smaller backgrounds with fake-$\tau_\text{had-vis}$ from other production processes
are estimated from simulation.
 
The normalisations of simulated $t\bar{t}$, for both the true- and fake-$\tau_\text{had-vis}$ components,
and $Z$\,+\,HF backgrounds are determined from data in the likelihood fits of signal and control regions,
as outlined in Section~\ref{sec:stats}.

\subsection{Fake-$\tau_\text{had-vis}$ background in the $\tau_\text{lep}\tau_\text{had}$ channel}
\label{sec:faketau_lephad}
 
In the $\tau_\text{lep}\tau_\text{had}$ channel, a combined fake-factor method similar to that
described in Ref.~\cite{HIGG-2016-16-witherratum} is used to estimate multi-jet and $t\bar{t}$
backgrounds with fake-$\tau_{\text{had-vis}}$.
A schematic depiction of this method is shown in Figure~\ref{fig:lephad_BG}.
This method employs events in two groups of regions.
The events in the identification (ID) regions require one identified $\tau_\text{had-vis}$,
whereas events in the anti-identification (anti-ID) regions contain one anti-$\tau_{\text{had-vis}}$ candidate.
 
Prior to the last step of the overlap removal procedure outlined in Section~\ref{sec:objects},
if an event does not contain an identified $\tau_{\text{had-vis}}$, it is checked for reconstructed
$\tau_{\text{had-vis}}$ candidates satisfying the anti-$\tau_{\text{had-vis}}$ requirements,
as defined in Section~\ref{sec:objects}. If an event contains multiple anti-$\tau_{\text{had-vis}}$ candidates,
one is chosen randomly.
In the LTT category, however, only the anti-$\tau_\text{had-vis}$ candidate that is within $\Delta R=0.2$
of the HLT $\tau_\text{had-vis}$ object is considered.
 
In order to define a fake-$\tau_\text{had-vis}$ background template,
an anti-ID region is defined using event selection criteria equivalent to the SR selection,
with one anti-$\tau_{\text{had-vis}}$ candidate instead of one identified $\tau_{\text{had-vis}}$.
This region, which is enriched with fake-$\tau_\text{had-vis}$, is defined as the SR Template region.
The template for estimating fake-$\tau_\text{had-vis}$ in the SR
is obtained by subtracting from the data distribution in the SR Template region
the distribution of simulated background events in which the $\tau_\text{had-vis}$ candidate is not
a fake-$\tau_\text{had-vis}$ originating from jets.
This subtraction is referred to as the true-$\tau_{\text{had-vis}}$ subtraction given
that the number of events in which an electron or muon is misidentified as a $\tau_{\text{had-vis}}$,
which are also subtracted, is very small.
The data and simulated events that are used to build the template are scaled with event weights, referred
to as fake-factors (FFs), to estimate the fake-$\tau_\text{had-vis}$ background in the SR.

\begin{figure}
\centering
\includegraphics[width=0.70\linewidth]{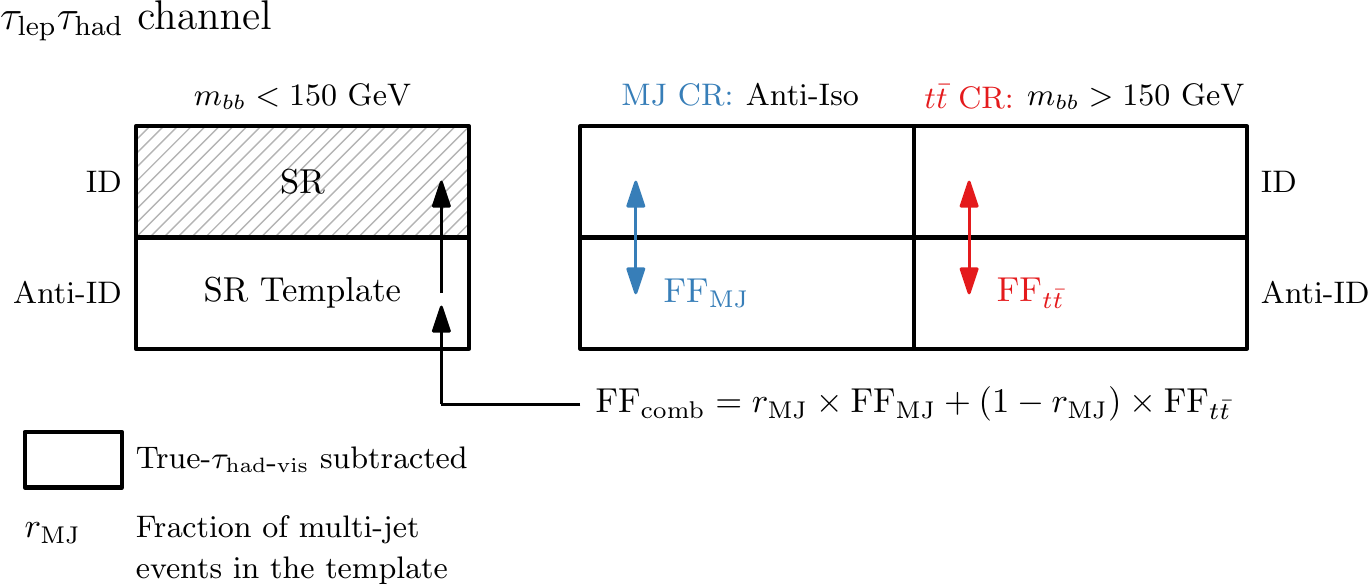}
\\
\caption{Schematic depiction of the combined fake-factor method used to estimate multi-jet and
$t\bar{t}$ backgrounds with fake-$\tau_\text{had-vis}$ in the $\tau_\text{lep}\tau_\text{had}$ channel.
Backgrounds which are not from  events with fake-$\tau_\text{had-vis}$ originating from jets
are  estimated from simulation and are subtracted from data in all control regions.
Events in which an electron or a muon  is misidentified as a $\tau_{\text{had-vis}}$ are also subtracted,
but their contribution is very small. Both sources are indicated by
`True-$\tau_\text{had-vis}$ subtracted' in the legend.
\label{fig:lephad_BG}}
\end{figure}

The FFs are derived separately for multi-jet ($\text{FF}_\text{MJ}$) and
$t\bar{t}$ ($\text{FF}_{t\bar{t}}$) events in dedicated control regions.
The multi-jet control regions (MJ CRs) and $t\bar{t}$ control regions ($t\bar{t}$ CRs)
are defined separately for the ID and the anti-ID regions, depending on whether they contain
one identified $\tau_{\text{had-vis}}$ or one anti-$\tau_{\text{had-vis}}$ candidate, respectively.
Besides the $\tau_{\text{had-vis}}$ selection, the MJ CRs are defined using the SR selection with
an inverted electron or muon isolation requirement (anti-Iso) and without the $m_{bb}<150$~\GeV\ requirement.
The MJ CR's purity in multi-jet production events varies between 65\% and 90\% depending on the
trigger category type (SLT or LTT) and whether the MJ CR is in the ID or anti-ID region.
Similarly, the $t\bar{t}$ CRs are defined using the SR selection with
an inverted $m_{bb}$ requirement ($m_{bb}>150$~\GeV).
The $t\bar{t}$ CR is about 95\% pure in the events from $t\bar{t}$ production.
Contamination from either non-resonant or resonant $HH$ signal in these CRs is estimated
to be negligible.
Backgrounds which are not from events with fake-$\tau_\text{had-vis}$ originating from jets,
are estimated from simulation and are subtracted from the distribution of the data in
all the control regions used for the FF measurement.
After the subtraction the FFs are derived as the ratio
of the number of events in the ID region to the number of events in the anti-ID region.
They are parameterised in terms of the $\tau_\text{had-vis}$ $p_\text{T}$,
independently for 1- and 3-prong $\tau_\text{had-vis}$
(`1- and 3-prong' refers to the number of tracks associated with a reconstructed $\tau_\text{had-vis}$),
and separately for the SLT and LTT categories.
The FFs corresponding to the individual background processes are combined as
\begin{equation*}
\text{FF}_\text{comb} = r_\text{MJ} \times \text{FF}_\text{MJ}  + (1 - r_\text{MJ}) \times \text{FF}_{t\bar{t}} \text{,}
\end{equation*}
where $r_\text{MJ}$ is the expected fraction of multi-jet events in the SR Template.
The number of multi-jet events in the SR Template is estimated by taking the number of
data events in the SR Template and subtracting the expected number of non-multi-jet
background events with both true- and fake-$\tau_\text{had-vis}$ as estimated from simulation.
The non-multi-jet background events are dominated by $t\bar{t}$ production.
The $r_\text{MJ}$ is parameterised as a function of the $\tau_\text{had-vis}$ $p_\text{T}$,
and it is measured separately for the $\tau_{e}\tau_\text{had}$ and $\tau_{\mu}\tau_\text{had}$ events,
for 1- and 3-prong $\tau_\text{had-vis}$ categories, and for the SLT and LTT categories.
The $\text{FF}_\text{comb}$ is used to scale the events used for the SR Template in order
to obtain the fake-$\tau_\text{had-vis}$ background prediction in the SR.

The determination of the fake-$\tau_\text{had-vis}$ background using the combined FF method
is sensitive to the modelling of simulated $t\bar{t}$ events with true-$\tau_\text{had-vis}$
given that this is the dominant background that is subtracted from data in the derivation of
the FF and $r_\text{MJ}$, and when obtaining the SR Template. Additionally, the derivation
of $r_\text{MJ}$ is sensitive to the modelling of simulated $t\bar{t}$ events with fake-$\tau_\text{had-vis}$.
To improve predictions, the simulated events from $t\bar{t}$ production are differentially reweighted
to data distributions depending on the jet multiplicity and the scalar sum of the transverse momentum
of all visible final-state objects in the event.
These reweighting factors are determined from another $t\bar{t}$ control region ($t\bar{t}$ CR2),
which is about 93\% pure in events from $t\bar{t}$ production,
and is defined using a selection identical to the SR selection,
but with an inverted $m_{bb}$ requirement ($m_{bb}>150$~\GeV)
and with an additional $m^{W}_\text{T}>40$~\GeV\ requirement.
Furthermore, events in this control region are required
to have a reconstructed $\tau_\text{had-vis}$ candidate, but this candidate is not required
to meet the recurrent neural-network identification criteria defined in Section~\ref{sec:objects}.
The $m^{W}_\text{T}$ requirement is introduced to reduce any potential contamination from multi-jet events.

Statistical uncertainties in $\text{FF}_{t\bar{t}}$, $\text{FF}_\text{MJ}$ and $r_\text{MJ}$
are evaluated and propagated to the final result.
The difference between the fake-$\tau_\text{had-vis}$ background estimates obtained with and without
the aforementioned $t\bar{t}$ modelling correction is taken as an uncertainty in the background estimate.
A conservative 30\% modelling uncertainty is assigned to simulated non-$t\bar t$ backgrounds
which are subtracted from data.
Due to its large dependence on the modelling of simulated $t\bar{t}$ events with fake-$\tau_\text{had-vis}$,
the obtained values of $r_\text{MJ}$ are varied by $\pm 0.5$, with the constraint $0\leq r_\text{MJ}\leq 1$.
The impact of such a conservative uncertainty is small since the FFs in multi-jet and $t\bar{t}$
events are found to be similar.
The total uncertainty in the $\text{FF}_\text{comb}$ value for the SLT category is at most 10\%, and
at most 25\% for the LTT category.
The combined FF method is checked for closure in the $t\bar{t}$ CR and it is validated in
the 0-$b$-tagged and 1-$b$-tagged regions, which are the same as the $\tau_\text{lep}\tau_\text{had}$ SR except
for the requirement on the number of $b$-tagged jets.
The signal contamination in the 0-$b$-tagged and 1-$b$-tagged regions is negligible.
The estimated background distributions agree well with the observed distributions in all validation regions.

\subsection{Fake-$\tau_\text{had-vis}$ background in the $\tau_\text{had}\tau_\text{had}$ channel}
\label{sec:faketau_hadhad}
 
In the $\tau_\text{had}\tau_\text{had}$ channel, two separate methods are used to estimate the
backgrounds with fake-$\tau_\text{had-vis}$ from $t\bar{t}$ and multi-jet production.
Multi-jet events can only enter the signal selection when both $\tau_\text{had-vis}$
are fake, whereas for $t\bar{t}$ production, usually no more than one
reconstructed $\tau_\text{had-vis}$ is fake.

\subsubsection{Fake-$\tau_\text{had-vis}$ background from multi-jet production}
\label{sec:faketau_hadhad_MJ}
In the $\tau_\text{had}\tau_\text{had}$ channel, the fake-$\tau_\text{had-vis}$ background
from multi-jet production is estimated using a fake-factor method.
A schematic depiction of this method is shown in Figure~\ref{fig:hadhad_BG1}. The ID region selection refers to the selection of events with two identified $\tau_\text{had-vis}$. In order to define an anti-ID region selection, prior to the last step of the overlap removal procedure outlined in Section~\ref{sec:objects}, events that have only one identified $\tau_\text{had-vis}$ are checked for a reconstructed $\tau_{\text{had-vis}}$ candidate satisfying the anti-$\tau_{\text{had-vis}}$ requirements. The selected anti-$\tau_{\text{had-vis}}$ candidate is required to be within $\Delta R=0.2$ of an HLT $\tau_\text{had-vis}$ object, except in the STT category for events in which the identified $\tau_\text{had-vis}$ is already trigger-matched. If multiple anti-$\tau_\text{had-vis}$ candidates fulfil the defined criteria, one is selected randomly.
 
In order to define a multi-jet background template, an anti-ID region is defined by using
a selection equivalent to the SR selection, with one identified $\tau_{\text{had-vis}}$
and one anti-$\tau_{\text{had-vis}}$ candidate, instead of two identified $\tau_{\text{had-vis}}$.
The template for estimating the multi-jet background in the SR (SR Template) is obtained by
subtracting simulated non-multi-jet events from data in the template region.
A large fraction of the subtracted non-multi-jet events are from $t\bar{t}$ production,
and these simulated $t\bar{t}$ events with fake-$\tau_\text{had-vis}$
are corrected with scale factors of the fake-$\tau_\text{had-vis}$ misidentification efficiencies
in the anti-ID region, which are described in Section~\ref{sec:faketau_hadhad_tt}.
Similarly to the procedure used in the $\tau_\text{lep}\tau_\text{had}$ channel,
the events that are used to build the template are further scaled with FFs to estimate
the multi-jet background in the SR.
 
A multi-jet-enriched control region is defined by using the $\tau_\text{had}\tau_\text{had}$ SR selection,
but requiring the two $\tau_\text{had-vis}$ to have same-sign (SS) charges,
as opposed to the SR selection that requires two $\tau_\text{had-vis}$ with opposite-sign (OS) charges.
Additionally, events in the control region are required to have exactly one $b$-tagged jet per event
(SS CR with 1 $b$-tagged jet). This control region, with its corresponding anti-ID counterpart,
is used for FF measurements.
This control region's purity in multi-jet production events varies from 80\% to 90\%
depending on the trigger type (STT or DTT).
The FFs are measured as the ratio of the number of events in the
ID region to the number of events in the anti-ID region after subtracting all simulated non-multi-jet
backgrounds from data.
The FFs are determined separately for the STT and DTT categories, and for the different years of data-taking
to account for the changes to the $\tau_\text{had-vis}$ identification algorithms and event selection topologies
used in the trigger.
The FFs are derived in the SS CR with 1 $b$-tagged jet, due to the limited number of selected events and large
$t\bar t$ background contamination in the SS region with 2 $b$-tagged jets.
For that reason, transfer factors (TFs) are defined to account for the extrapolation from 1-$b$-tagged-jet events to 2-$b$-tagged-jet events.
In the DTT category, the FFs are binned in $p_\text{T}$ and $\eta$ of the anti-$\tau_\text{had-vis}$.
In the STT category, due to the smaller number of available events, the FFs are measured inclusively in $p_\text{T}$
and $\eta$, but separately according to whether the selected anti-$\tau_\text{had-vis}$ is the
leading or sub-leading $\tau_\text{had-vis}$ candidate in $p_\text{T}$.
In both categories, the FFs are measured separately for events with 1- or 3-prong anti-$\tau_\text{had-vis}$
candidates.

The TFs are defined as ratios of the FFs measured in the SS CR with 2 $b$-tagged jets to
the FFs measured in the SS CR with 1 $b$-tagged jet, inclusively for the STT and DTT categories.
The large contamination from $t\bar{t}$ background in the SS CR with 2 $b$-tagged jets
is removed in the subtraction of all simulated non-multi-jet backgrounds from data.
The TFs are also measured inclusively in $p_\text{T}$ and $\eta$ of the $\tau_\text{had-vis}$,
but separately for events with 1- and 3-prong anti-$\tau_\text{had-vis}$ candidates,
separately according to whether the selected anti-$\tau_\text{had-vis}$ is the leading
or sub-leading $\tau_\text{had-vis}$ candidate in $p_\text{T}$, and separately for the different years of data-taking.
Their values are compatible with unity within the statistical uncertainty.
 
\begin{figure}
\centering
\includegraphics[width=0.70\linewidth]{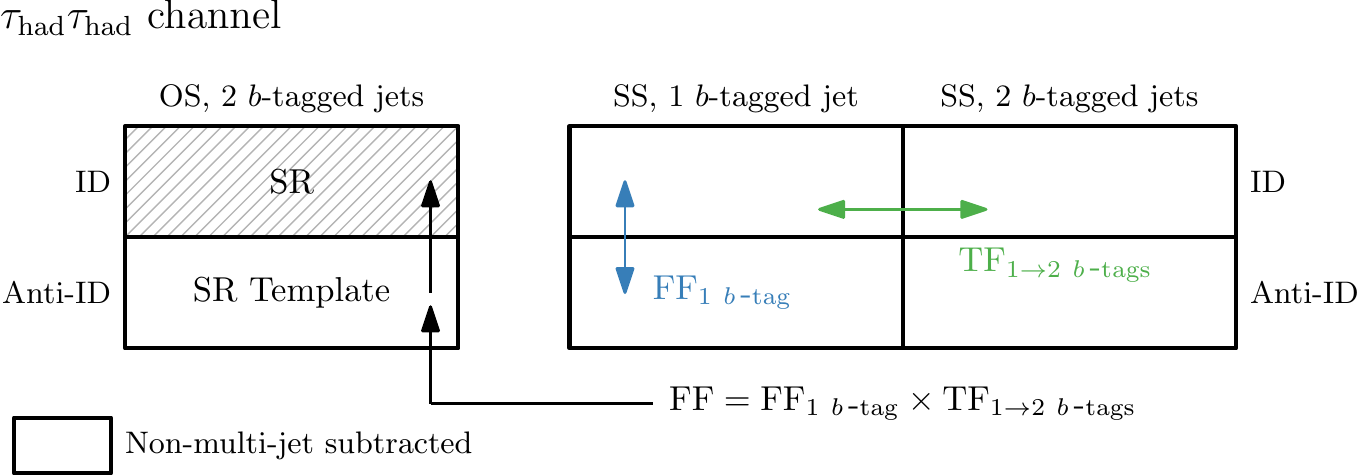}
\\
\caption{Schematic depiction of the fake-factor method to estimate the multi-jet
background with fake-$\tau_\text{had-vis}$ in the $\tau_\text{had}\tau_\text{had}$ channel.
Backgrounds that are not from multi-jet events are simulated
and subtracted from data in all the control regions.
This is indicated by `Non-multi-jet subtracted' in the legend.
\label{fig:hadhad_BG1}}
\end{figure}

The sources of uncertainty in the estimate of the fake-$\tau_\text{had-vis}$ background
from multi-jet production include the statistical uncertainties in the measured FFs and TFs,
and uncertainties in the normalisation and shape of the non-multi-jet backgrounds that are subtracted
from data when deriving the SR Template.
An uncertainty is also introduced to account for the extrapolation from the SS events to the OS events.
The associated systematic uncertainty is estimated by comparing the FFs derived from SS 1-$b$-tagged-jet events with those derived from OS 1-$b$-tagged-jet events.
To ensure that the sample is dominated by multi-jet events,
the OS events are additionally required to satisfy $m_{\tau\tau}^\text{MMC}>110~\GeV$
and $E_\text{T}^\text{miss}/\sigma(E_\text{T}^\text{miss})<3$, where $\sigma(E_\text{T}^\text{miss})$
is the event-based approximation to the resolution of the $E_\text{T}^\text{miss}$~\cite{ATLAS-CONF-2018-038}.
The modelling of the multi-jet background is checked for closure in the SS CRs with 1 and 2 $b$-tagged jets
and good agreement between data and prediction is observed.
The FFs are measured in the SS CR with 1 $b$-tagged jet,
and the TFs between the SS CR with 1 $b$-tagged jet and SS CR with 2 $b$-tagged jets
only correct for the normalisation of the fake-$\tau_\text{had-vis}$ prediction.
Thus the validation performed on the SS CR with 2 $b$-tagged jets provides a check of the
shape extrapolation from 1 $b$-tagged jet to 2 $b$-tagged jets.

\subsubsection{Fake-$\tau_\text{had-vis}$ background from $t\bar{t}$ production}
\label{sec:faketau_hadhad_tt}
 
Background events with fake-$\tau_\text{had-vis}$ from $t\bar{t}$ production in the
$\tau_\text{had}\tau_\text{had}$ channel are estimated using simulation. However, the fake-$\tau_\text{had-vis}$
misidentification efficiencies are corrected by scale factors (SFs) derived from data.
A schematic depiction of this method is shown in Figure~\ref{fig:hadhad_BG2}.
 
\begin{figure}
\centering
\includegraphics[width=0.55\linewidth]{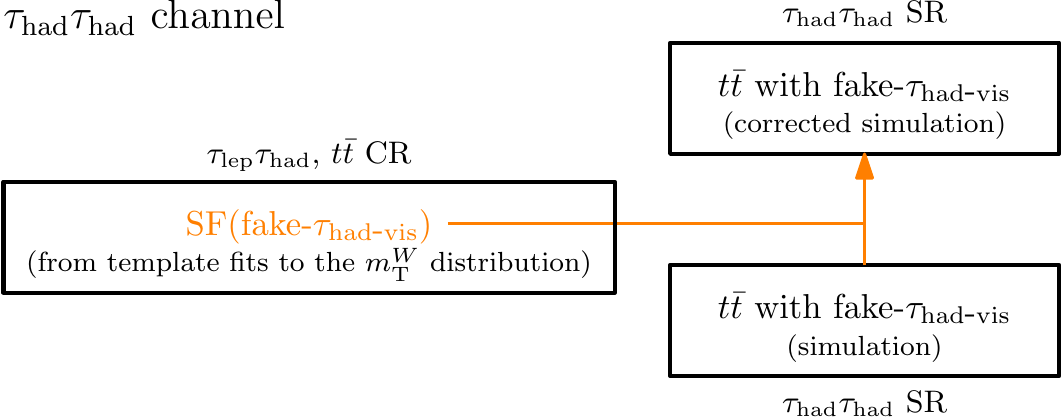}
\\
\caption{Schematic depiction of the fake-$\tau_\text{had-vis}$ scale-factor method to
estimate the $t\bar{t}$ background with fake-$\tau_\text{had-vis}$ in the $\tau_\text{had}\tau_\text{had}$ channel.
\label{fig:hadhad_BG2}}
\end{figure}

The SFs are determined in the $t\bar{t}$ CR defined within the $\tau_\text{lep}\tau_\text{had}$ SLT category,
as described in Section~\ref{sec:faketau_lephad}.
However, in order to harmonise it with the $\tau_\text{had}\tau_\text{had}$ SR selection,
the $t\bar{t}$ CR is redefined to select events with $\tau_\text{had-vis}$ $|\eta|<2.5$.
The SFs are extracted as a function of the fake-$\tau_\text{had-vis}$ $p_\text{T}$,
separately for 1- and 3-prong fake-$\tau_\text{had-vis}$ objects, by fitting the $m^{W}_\text{T}$
distribution of simulated events to data using a profile-likelihood fit.
The fit of the $m^{W}_\text{T}$ distribution allows the contributions of the
$t\bar{t}$ events with true- and fake-$\tau_\text{had-vis}$ to be determined while correcting for the modelling
of the $t\bar{t}$ simulation that is common to both contributions.
Separate fits are performed for different trigger categories.
For 1-prong fake-$\tau_\text{had-vis}$ the SFs are close to unity at fake-$\tau_\text{had-vis}$ $p_\text{T}$
below 40~\GeV\ and decrease to $\text{SF} \sim 0.6$ for fake-$\tau_\text{had-vis}$ $p_\text{T}$ above 70~\GeV.
The SFs for the 3-prong fake-$\tau_\text{had-vis}$ are generally about $\sim 20\%$ larger than for the 1-prong fake-$\tau_\text{had-vis}$ objects.
 
The $t\bar{t}$ background with fake-$\tau_\text{had-vis}$ in the $\tau_\text{had}\tau_\text{had}$
SR is estimated from the simulated events that pass the SR selection,
weighted by the corresponding SFs for each fake-$\tau_\text{had-vis}$ in the event.
 
Uncertainties in the detector response and in the modelling of the $t\bar{t}$ events,
as well as of the other minor contributing processes, are taken into account in the likelihood fit
when extracting the SFs.
The covariance matrix of the measured SFs, which contains all statistical and systematic uncertainties
of the measurement, is diagonalised and the resulting eigenvectors are used to define independent
nuisance parameters which are propagated to the final signal extraction fit.
Theoretical modelling uncertainties in simulated $t\bar{t}$ events to which the SFs
are applied are estimated as described in Section~\ref{sec:systematics} and they are
also propagated to the final signal extraction fit.
 
When estimating the fake-$\tau_\text{had-vis}$ background from multi-jet production
using the fake-factor method (cf.\ Section~\ref{sec:faketau_hadhad_MJ}),
a large fraction of $t\bar{t}$ events containing at least one fake-$\tau_\text{had-vis}$
needs to be subtracted from data in the OS 2-$b$-tagged-jet anti-ID region (SR Template)
to estimate the multi-jet contribution in the $\tau_\text{had}\tau_\text{had}$ SR.
The modelling of the simulated $t\bar{t}$ events with fake-$\tau_\text{had-vis}$ in the anti-ID region
is corrected with SFs that are measured using the same method as described above.
These SFs are measured in a control region similar to the $t\bar{t}$ CR except that
the $\tau_\text{had-vis}$ candidate has to satisfy the anti-$\tau_\text{had-vis}$ requirements.
The measured SFs for 1-prong fake-$\tau_\text{had-vis}$ in the anti-ID region are
close to unity at fake-$\tau_\text{had-vis}$ $p_\text{T}$ below 40~\GeV,
and follow the same trend of decreasing in value with increasing fake-$\tau_\text{had-vis}$ $p_\text{T}$,
as observed in the ID region.
The SFs for the 3-prong fake-$\tau_\text{had-vis}$ are generally about 10\%--20\% larger than
for the 1-prong fake-$\tau_\text{had-vis}$ objects.


\section{Systematic uncertainties}
\label{sec:systematics}

The most significant uncertainty in this search is the statistical uncertainty of the data in the signal region. Nonetheless, experimental, theoretical, and modelling uncertainties in the normalisation and shape of the signal and background estimates give a non-negligible contribution to the total uncertainty and are thus evaluated for use in the statistical model (described in Section~\ref{sec:stats}); these are described below. Statistical uncertainties in the predicted background processes are modelled using a simplified version of the Beeston--Barlow technique~\cite{Barlow:1993dm}, in which only the uncertainty in the total background content in each bin is considered.

Systematic uncertainties in the detector response are considered in this search. Where relevant, uncertainties in the trigger, reconstruction, identification, and isolation efficiencies, and in the momentum of electrons~\cite{EGAM-2018-01}, muons~\cite{MUON-2018-03} and $\tau_\text{had-vis}$~\cite{ATLAS-CONF-2017-029} are estimated.
Additional uncertainties are estimated for the efficiencies of the electron veto for $\tau_\text{had-vis}$ and the track-to-vertex-matching requirements for muons. Jet energy scale and resolution uncertainties~\cite{JETM-2018-05} and the uncertainty in the efficiency of matching jets to the primary vertex~\cite{PERF-2014-03} are estimated. These energy scale and resolution uncertainties, in addition to an uncertainty in the tracks matched to the primary vertex but not associated with other reconstructed objects in the event, are propagated to the $E_\text{T}^\text{miss}$ calculation~\cite{PERF-2016-07,ATLAS-CONF-2018-023}. Uncertainties in the $b$-jet tagging efficiency are considered, and these are evaluated as a function of $p_\text{T}$ for $b$-tagged jets and $c$-tagged jets by using $t\bar t$ events~\cite{FTAG-2018-01,Aad:2781747}, and as a function of $p_\text{T}$ for light-flavour jets by using $Z+\text{jets}$ events~\cite{ATLAS-CONF-2018-006}. An uncertainty of 1.7\% in the total integrated luminosity~\cite{ATLAS-CONF-2019-021}, obtained using the LUCID-2 detector~\cite{ATLAS-Lumi2}, is assigned to physics processes whose normalisations are taken from simulation. An uncertainty arising from the correction of the pile-up distribution in simulation to that in data is also estimated.

The effects of uncertainties in the parton shower on the acceptances of the resonant (non-resonant) $HH$ signals are assessed by comparing the nominal MC samples with alternative samples which use \Pythia[8] (\Herwig[7]) to model the parton shower. Alternative samples were not generated for some of the resonant $HH$ signals, and uncertainties for these signals are derived by interpolating or extrapolating the uncertainties from signals nearby in $m_X$ for which alternative samples were produced. Independent parton-shower acceptance uncertainties are accounted for in the normalisation of the non-resonant ggF, non-resonant VBF and resonant $HH$ signals, and these range between 0.1\% and 19\% of the nominal acceptance values. The effects of the renormalisation and factorisation scale, PDF and $\alphas$ uncertainties on the ggF and VBF $HH$ signal acceptances are included when they are found to be non-negligible.
Separate normalisation uncertainties are applied in the $\tau_\text{had}\tau_\text{had}$
and $\tau_\text{lep}\tau_\text{had}$ LTT categories to account for observed differences in
the $\tau_\text{had-vis}$ trigger acceptances between fast-simulation and full-simulation
resonant $HH$ signal MC samples.
The effects of renormalisation and factorisation scale, PDF and $\alphas$ uncertainties on the SM ggF and VBF $HH$ cross-sections are considered, and an uncertainty arising from the scheme and value used for the top-quark mass in calculations of the virtual top-quark loop contributions in the SM ggF $HH$ cross-section~\cite{Baglio:2020wgt} is included. Lastly, uncertainties in the $H\to b\bar b$ and $H\to\tau^+\tau^-$ branching fractions~\cite{deFlorian:2016spz_edited} are considered in the $HH$ signal and single-Higgs-boson background samples.

The $t\bar t$ normalisation is determined in the likelihood fits, so the analysis is not sensitive to uncertainties in its expected cross-section. However, the relative acceptances of the analysis categories and the distribution shapes of the discriminating variables are still sensitive to modelling uncertainties, and these are evaluated using a method that closely follows Ref.~\cite{TOPQ-2018-18}. The $t\bar t$ hard-scatter and parton-shower uncertainties are evaluated by comparing the nominal sample with samples generated using $\MGNLO+\Pythia[8]$ and $\POWHEGBOX[\text{v2}]+\Herwig[7]$, respectively. The hard-scatter uncertainty also accounts for the matrix-element--parton-shower matching and merging uncertainty (henceforth referred to as the matching uncertainty). The impact of the uncertainty in the amount of initial-state QCD radiation is evaluated by co-varying the $h_\text{damp}$ parameter, the renormalisation and factorisation scales, and the showering tune. The effects of uncertainties in the amount of final-state QCD radiation, the PDF, the $\alphas$ value, and the renormalisation and factorisation scales are all evaluated by varying generator parameter values or the PDF. All of these uncertainties affect the relative acceptances of the analysis categories. Additionally, the effects of the hard-scatter and parton-shower uncertainties on the distribution shapes of the MVA output variables are evaluated directly in all categories, but are found to be negligible in the $\tau_\text{lep}\tau_\text{had}$ LTT category. The impact of the uncertainty in the amount of initial-state QCD radiation on the MVA output variables is considered in the $\tau_\text{had}\tau_\text{had}$ category.
 
The $Z$\,+\,HF normalisation is determined in the likelihood fits. The effects of uncertainties in the hard scatter, matching, parton shower, the renormalisation, factorisation and resummation scales, PDF and $\alphas$ on the relative acceptances of the analysis categories are all considered. For the renormalisation, factorisation and resummation scales, these uncertainties are evaluated by varying the relevant scales upwards and downwards from their nominal values by a factor of two; both simultaneous and independent variations are considered for the renormalisation and factorisation scales, while the resummation scale is varied independently. In the $\tau_\text{had}\tau_\text{had}$ category, the impact of the hard-scatter and parton-shower uncertainty is evaluated by comparing the nominal sample with a $\MGNLO +\Pythia[8]$ sample, and the impact on distribution shapes is parameterised in $m_{bb}$ and propagated to the MVA output variables. This uncertainty is not applied in the $\tau_\text{lep}\tau_\text{had}$ category and the $Z$\,+\,HF CR. In the $\tau_\text{had}\tau_\text{had}$ category, the impact of the renormalisation and factorisation scale uncertainties on the shape of the MVA output variables is evaluated directly, and in the $\tau_\text{lep}\tau_\text{had}$ categories, their impact is parameterised in the $p_\text{T}$ of the $b$-tagged jet pair and propagated to the MVA output variables.

The effects of uncertainties in the PDF and the amount of initial- and final-state QCD radiation on the acceptances of the single-top-quark backgrounds are estimated. Cross-section uncertainties are also considered. Despite these backgrounds being small inclusively, their total yields in the most sensitive MVA bins are significant, and the acceptance uncertainties in the different analysis categories range between 14\% and 34\% of the nominal values. The impact of the uncertainty in the amount of initial-state QCD radiation is evaluated by co-varying the renormalisation and factorisation scales, and the showering tunes. The impact of the uncertainty in the amount of final-state QCD radiation is evaluated by varying the \textsc{Var2} parameter~\cite{ATL-PHYS-PUB-2014-021}. The impact of uncertainties in the combined hard scatter and matching (parton shower) of the $Wt$ background is estimated by comparing the nominal sample with a sample generated using $\MGNLO+\Pythia[8]$ ($\POWHEGBOX[\text{v2}]+\Herwig[7]$). The impact of the uncertainty in the interference between the $Wt$ and $t\bar t$ backgrounds on the acceptance is evaluated by comparing simulated $Wt$
samples produced with the diagram removal and diagram subtraction schemes~\cite{Frixione:2008yi}, parameterising this difference in the $p_\text{T}$ of the $b$-tagged jet pair, and propagating it to the normalisation and shape of the MVA output distributions.

The effects of uncertainties in the renormalisation and factorisation scales, PDF and $\alphas$ on the normalisation of the single-Higgs-boson backgrounds are evaluated. Uncertainties of 100\% are applied separately to the normalisations of the ggF, VBF and $WH$ single-Higgs-boson backgrounds where the Higgs boson decays into $\tau$-leptons, to account for difficulties in the modelling of these processes in association with heavy-flavour jets~\cite{TOPQ-2012-16,STDM-2012-11}. The impact of uncertainties in the parton shower on the acceptances are evaluated for the $ZH$ and $t\bar tH$ backgrounds by comparing the nominal samples with alternative samples generated using \Herwig[7] and \EVTGEN. Uncertainties in the matching and the amount of initial- and final-state QCD radiation are also considered for the $t\bar tH$ backgrounds, and the impact of QCD radiation uncertainties is evaluated by varying the \textsc{Var3c} parameter~\cite{ATL-PHYS-PUB-2014-021} and the renormalisation scale. Uncertainties in the shape of the MVA output distributions of all single-Higgs-boson processes are found to be negligible. Cross-section uncertainties are applied to the normalisation of the simulated single-Higgs-boson~\cite{deFlorian:2016spz_edited}, single-top-quark, $Z$+light-flavour jets, $W$+jets, and diboson backgrounds, and acceptance uncertainties are applied to the normalisation of various small backgrounds.

Systematic uncertainties affecting the data-driven background estimates are described in Section~\ref{sec:modelling:background}.
 
Table~\ref{tab:systematics:breakdown} shows the breakdown of the relative contributions to the uncertainty in the non-resonant and resonant extracted signal cross-sections, which is obtained from the likelihood fit described in Section~\ref{sec:stats}. Systematic uncertainties in the single-Higgs-boson backgrounds affect the non-resonant search more than these resonant searches, due to the more significant contribution of ggF single-Higgs-boson background in the most signal-like regions of the MVA distributions in the non-resonant search.
 
\begin{table}
\caption{Breakdown of the relative contributions to the uncertainties in the extracted
signal cross-sections, as determined in the likelihood fit (described in Section~\ref{sec:stats}) to data.
They are obtained by fixing the relevant nuisance parameters in the likelihood fit,
subtracting the square of the obtained uncertainty in the fitted signal cross-section
from the square of the total uncertainty, taking the square root, and then dividing by the total uncertainty.
The sum in quadrature of the individual components differs from the total uncertainty due to
correlations between uncertainties in the different groups.
}
\label{tab:systematics:breakdown}
\centering
\resizebox{\textwidth}{!}{\begin{tabular}{lcccc}
\toprule
\multirow{2}{*}{Uncertainty source} & \multirow{2}{*}{Non-resonant $HH$} & & Resonant $X\to HH$ & \\
& & 300~\GeV & 500~\GeV & 1000~\GeV \\
\midrule
\BF{Data statistical + floating normalisation}    &  81\%  &  76\%  &  90\%  &  93\%  \\
\hline
~~~{Data statistical}                             &  81\%  &  76\%  &  90\%  &  93\%  \\
~~~$t\bar t$ and $Z+\text{HF}$ normalisations     &   ~~4\%  &   ~~8\%  &   ~~3\%  &   ~~5\%  \\
\hline
\BF{Systematic}                                   &  58\%  &  65\%  &  43\%  &  37\%  \\
\hline
~~~MC statistical                                 &  28\%  &  44\%  &  33\%  &  18\%  \\
~~~\BF{Experimental}                              &  12\%  &  31\%  &   ~~8\%  &  12\%  \\
\hline
~~~~~~Jet and $E_\text{T}^\text{miss}$           &   ~~8\%  &  27\%  &   ~~5\%  &   ~~4\%  \\
~~~~~~$b$-jet tagging                            &   ~~5\%  &   ~~5\%  &   ~~3\%  &   ~~7\%  \\
~~~~~~$\tau_\text{had-vis}$                      &   ~~6\%  &  12\%  &   ~~3\%  &   ~~8\%  \\
~~~~~~Electrons and muons                        &   ~~3\%  &   ~~3\%  &   ~~2\%  &   ~~2\%  \\
~~~~~~Luminosity and pile-up                      &   ~~3\%  &   ~~2\%  &   ~~2\%  &   ~~5\%  \\
\hline
~~~\BF{Background and signal and modelling}       &  42\%  &  39\%  &  26\%  &  30\%  \\
\hline
~~~~~~Fake-$\tau_\text{had-vis}$                  &   ~~8\%  &  19\%  &   ~~4\%  &   ~~8\%  \\
~~~~~~Top-quark                                   &  24\%  &  17\%  &  12\%  &   ~~8\%  \\
~~~~~~$Z(\to\tau\tau)+\text{HF}$                  &   ~~9\%  &  17\%  &   ~~9\%  &  15\%  \\
~~~~~~Single Higgs boson                          &  29\%  &   ~~2\%  &  14\%  &  15\%  \\
~~~~~~Other backgrounds                           &   ~~3\%  &   ~~2\%  &   ~~5\%  &   ~~3\%  \\
~~~~~~Signal                                      &   ~~5\%  &  14\%  &   ~~7\%  &  15\%  \\
\bottomrule
\end{tabular}}
\end{table}


\section{Statistical interpretation}
\label{sec:stats}


The $HH$ signal yields are each estimated from data using a simultaneous binned maximum-likelihood fit to the MVA output distributions in the $\tau_\text{had}\tau_\text{had}$, $\tau_\text{lep}\tau_\text{had}$ SLT, and $\tau_\text{lep}\tau_\text{had}$ LTT event categories, and to the $m_{\ell\ell}$ distribution in the $Z$\,+\,HF CR. The $t\bar t$ and $Z$\,+\,HF background normalisations are free parameters in the fit, and are primarily constrained by the background-like MVA output bins and the $Z$\,+\,HF CR. Upper limits are set on the signal normalisations at the 95\% confidence level (CL) using the profile-likelihood-ratio test statistic~\cite{Cowan:2010js-witherratum} and the modified frequentist $\text{CL}_\text{s}$ technique~\cite{Read:2002hq} in the asymptotic approximation~\cite{Cowan:2010js-witherratum}.
 
The systematic uncertainties described in Sections~\ref{sec:modelling:background} and \ref{sec:systematics} are represented in the fit as Gaussian- or Poisson-constrained nuisance parameters, which modify the normalisation, relative normalisation between event categories, and/or distribution shape of the discriminating variable for the signal and background processes. Systematic uncertainties are symmetrised and shape uncertainties are smoothed where physically motivated, and then those with a negligible impact are removed from the likelihood fit. Experimental, cross-section, and acceptance uncertainties are correlated across the event categories, except for the parton-shower uncertainty of the $t\bar t$ background, since the kinematic and topological properties of this background differ between event categories.
Modelling uncertainties in data-driven background estimates are not correlated across different estimation strategies, because different sources of fake-$\tau_\text{had-vis}$ are estimated using different procedures.

The binning schemes for the MVA output distributions used in the likelihood fit were chosen to minimise the number of bins, while also maximising the retained expected sensitivity, and ensuring the stability of the fit and the validity of the asymptotic approximation. The binning schemes start from finely binned histograms, and bins are iteratively merged beginning from the most signal-like MVA bins until the following channel-dependent criteria are fulfilled. In the $\tau_\text{had}\tau_\text{had}$ channel, the bins are required to satisfy $\sigma^\text{MC}_\text{b}<0.5f_\text{s}+1\%$, where $\sigma^\text{MC}_\text{b}$ is the relative MC statistical uncertainty of the background estimate and $f_\text{s}$ is the fraction of the total signal in the bin. In the $\tau_\text{lep}\tau_\text{had}$ channel, the bins are required to satisfy $10f_\text{s}+5f_\text{b}>1$, where $f_\text{s}$ and $f_\text{b}$ are the fractions of the total signal and background in the bin, respectively. Bins in all channels must be expected to contain at least five background events to ensure that the asymptotic approximation is valid.


\section{Results}
\label{sec:results}

 
The post-fit normalisation factors of the unconstrained $t\bar{t}$ and $Z$\,+\,HF
backgrounds for the combined $\tau_\text{had}\tau_\text{had}$ and $\tau_\text{lep}\tau_\text{had}$
channels, as obtained in the search for non-resonant $HH$ production,
are $0.97 \pm 0.04$ and $1.40 \pm 0.11$, respectively.
These post-fit background normalisation factors are compatible within uncertainties
with those obtained for the resonant $HH$ production searches.

The MVA output distributions in the search for non-resonant $HH$ production and
for two resonance mass hypotheses ($m_{X}=500$~\GeV\ and 1~\TeV) are shown in Figure~\ref{fig:RHH_mva_output},
after performing the fits to the data and assuming a background-only hypothesis.
The data, expected signal and post-fit background yields for events entering the most signal-like bin
in the non-resonant MVA output histograms in the three event categories are shown in Table~\ref{tab:yieldstopmvabin}.

\begin{figure}
\centering
\subfigure[]{\includegraphics[width=0.32\linewidth]{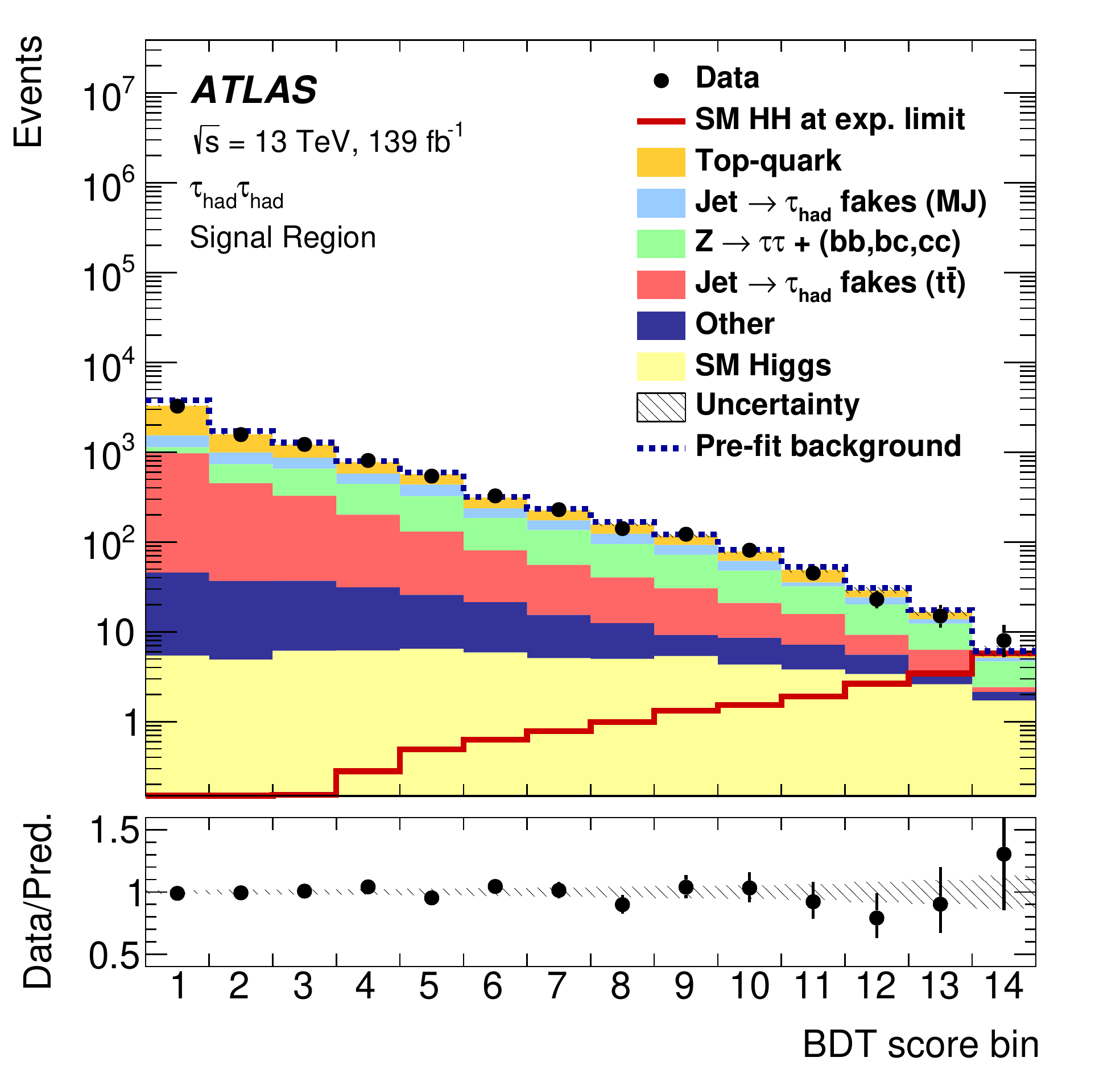}}
\subfigure[]{\includegraphics[width=0.32\linewidth]{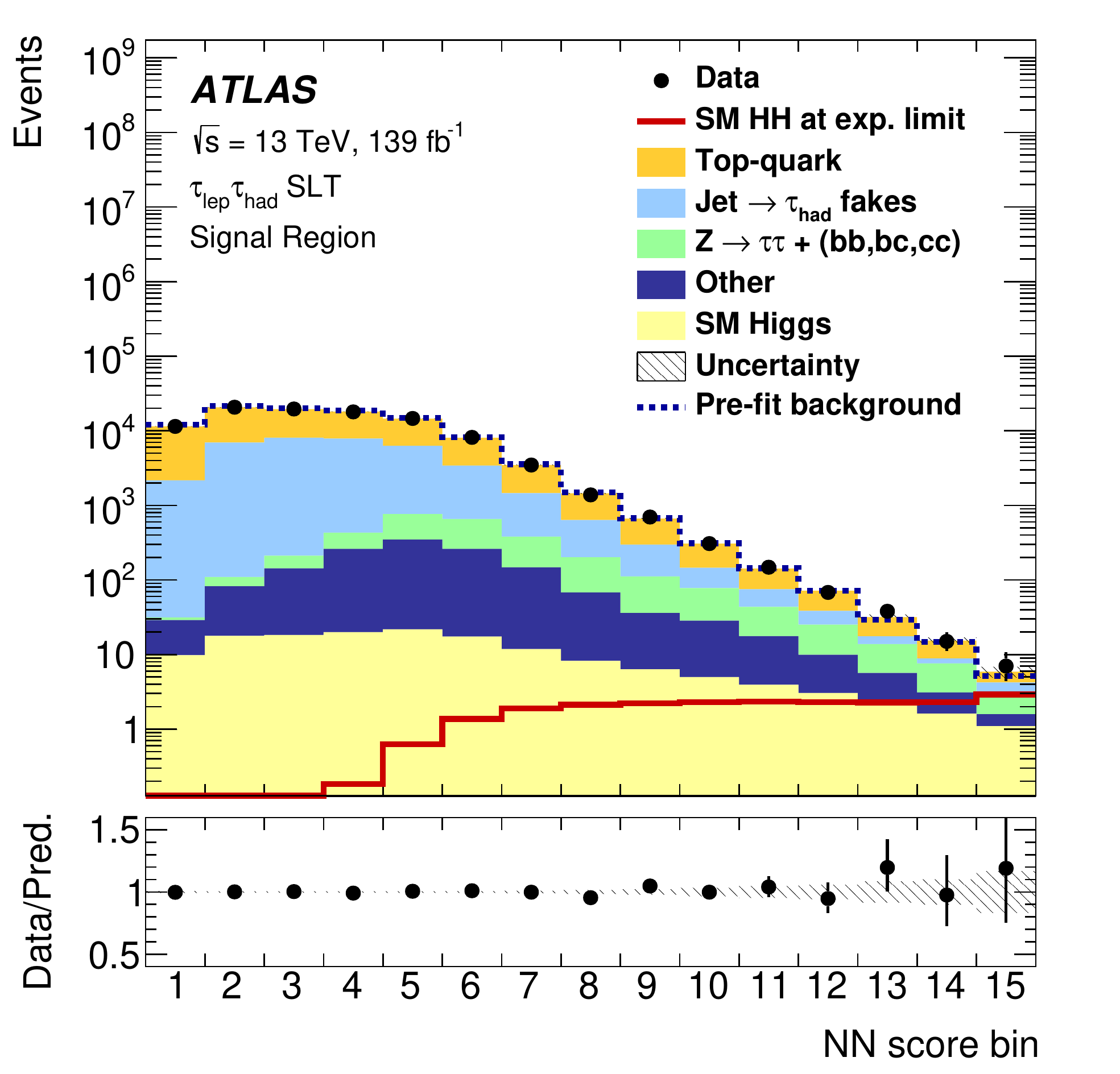}}
\subfigure[]{\includegraphics[width=0.32\linewidth]{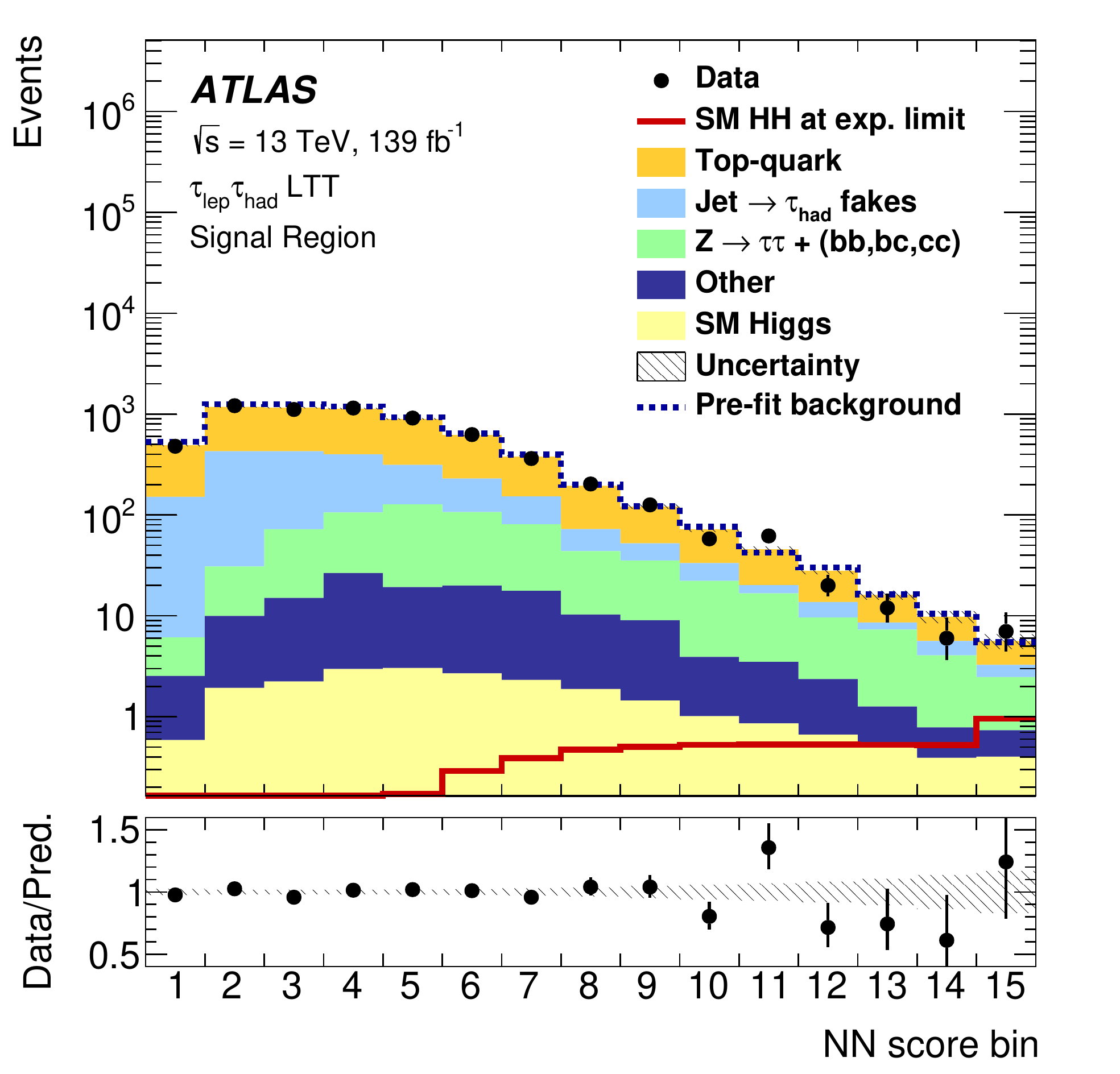}}\\
\subfigure[]{\includegraphics[width=0.32\linewidth]{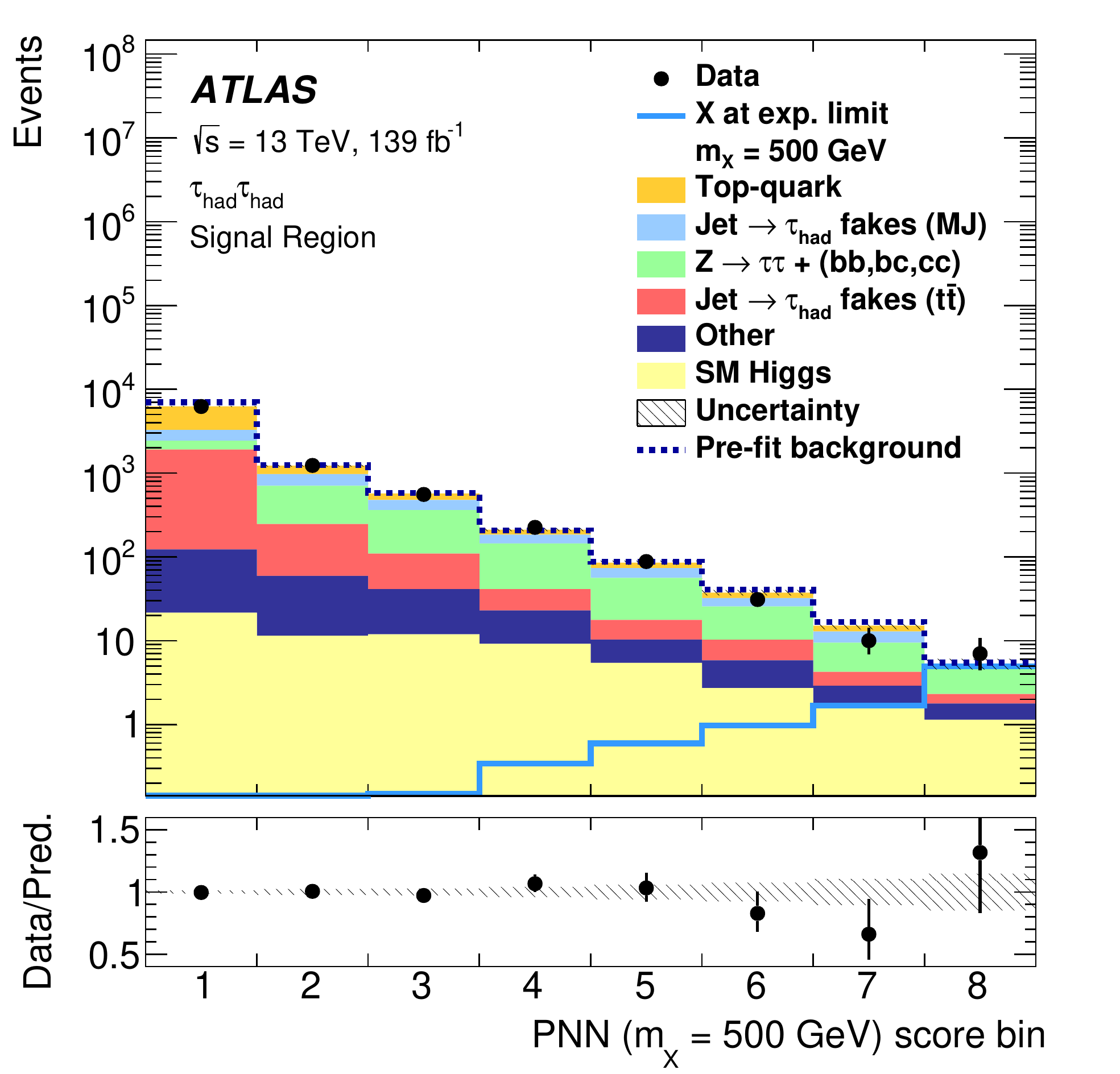}}
\subfigure[]{\includegraphics[width=0.32\linewidth]{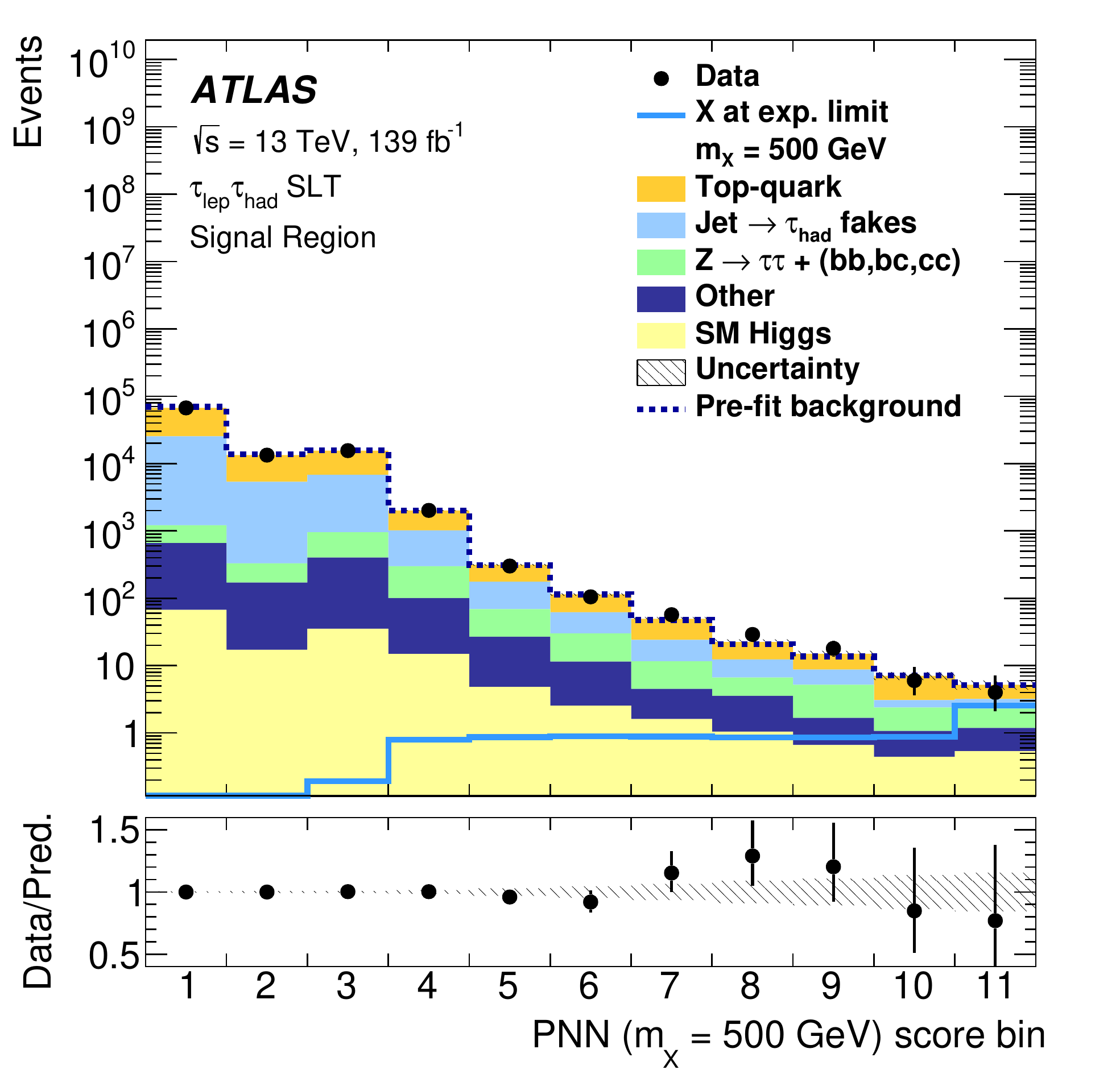}}
\subfigure[]{\includegraphics[width=0.32\linewidth]{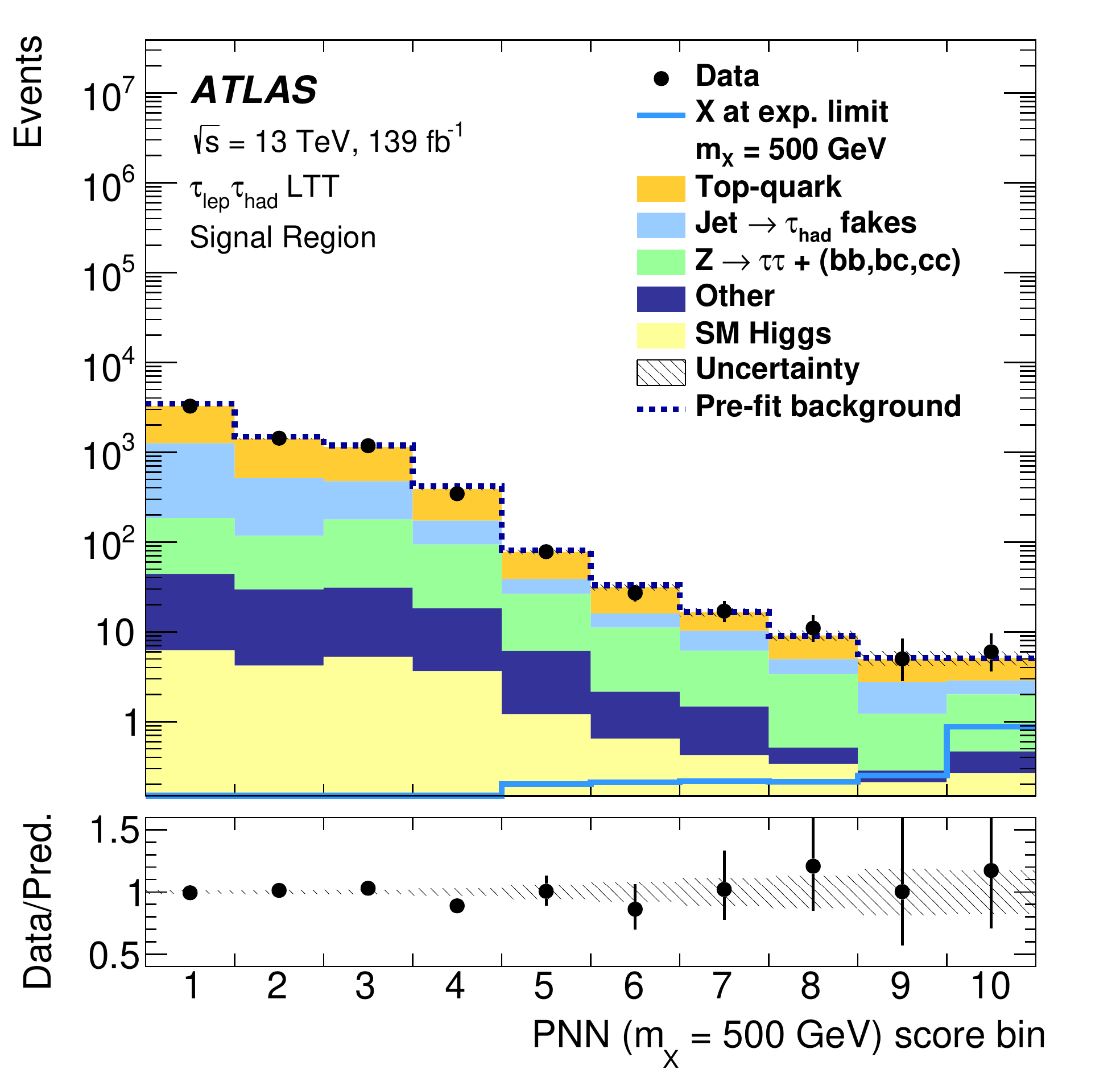}}\\
\subfigure[]{\includegraphics[width=0.32\linewidth]{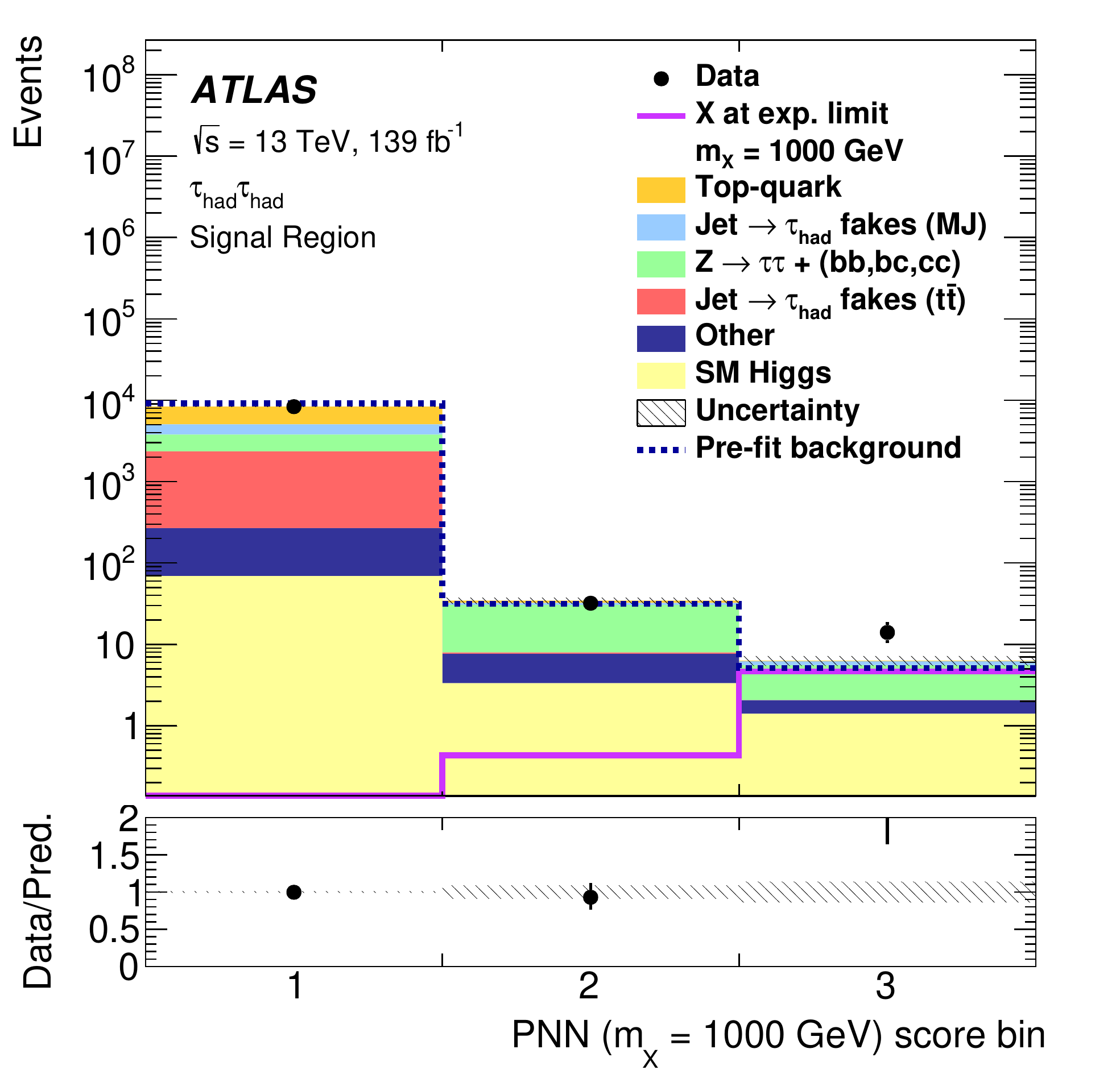}}
\subfigure[]{\includegraphics[width=0.32\linewidth]{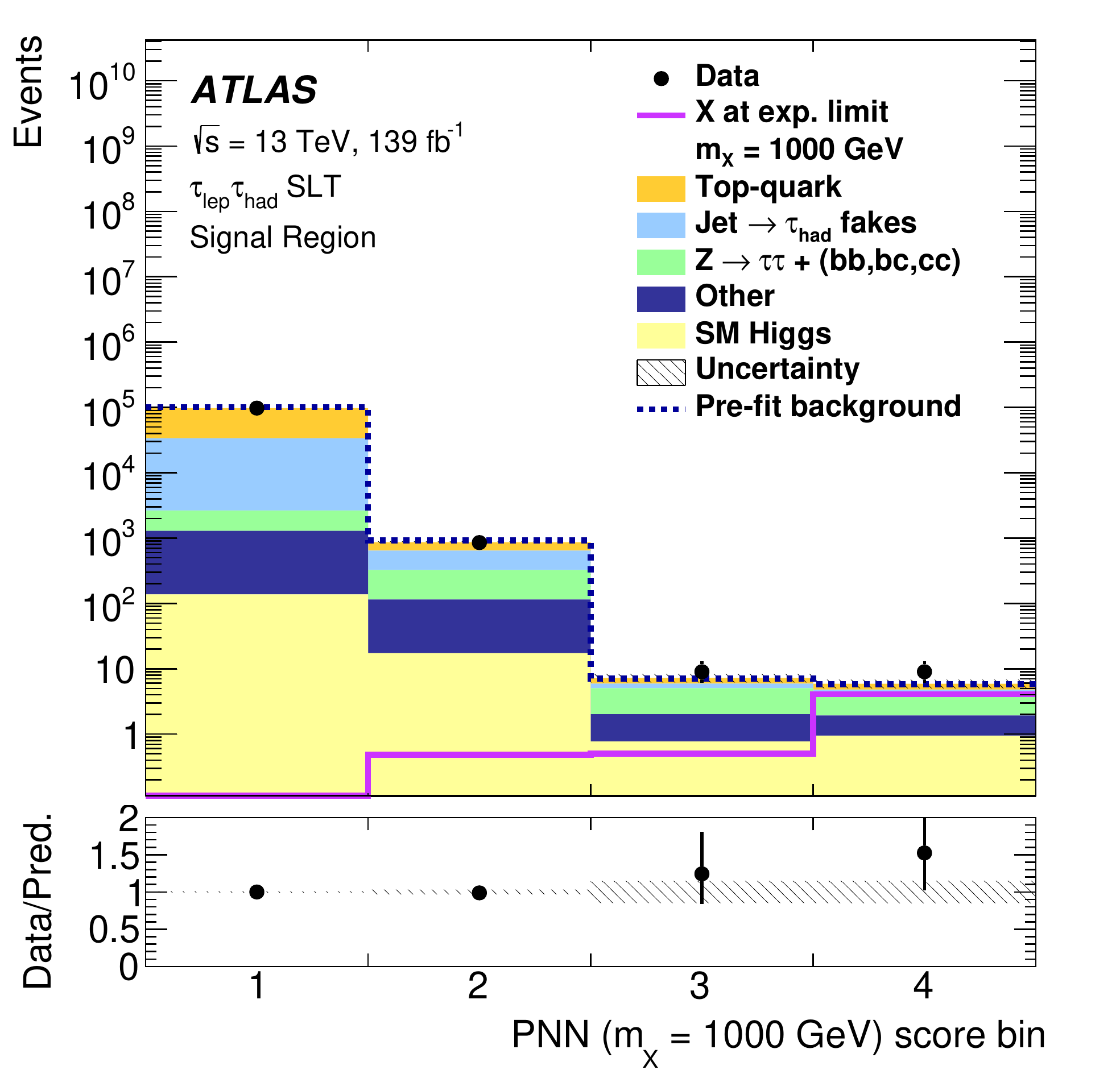}}
\subfigure[]{\includegraphics[width=0.32\linewidth]{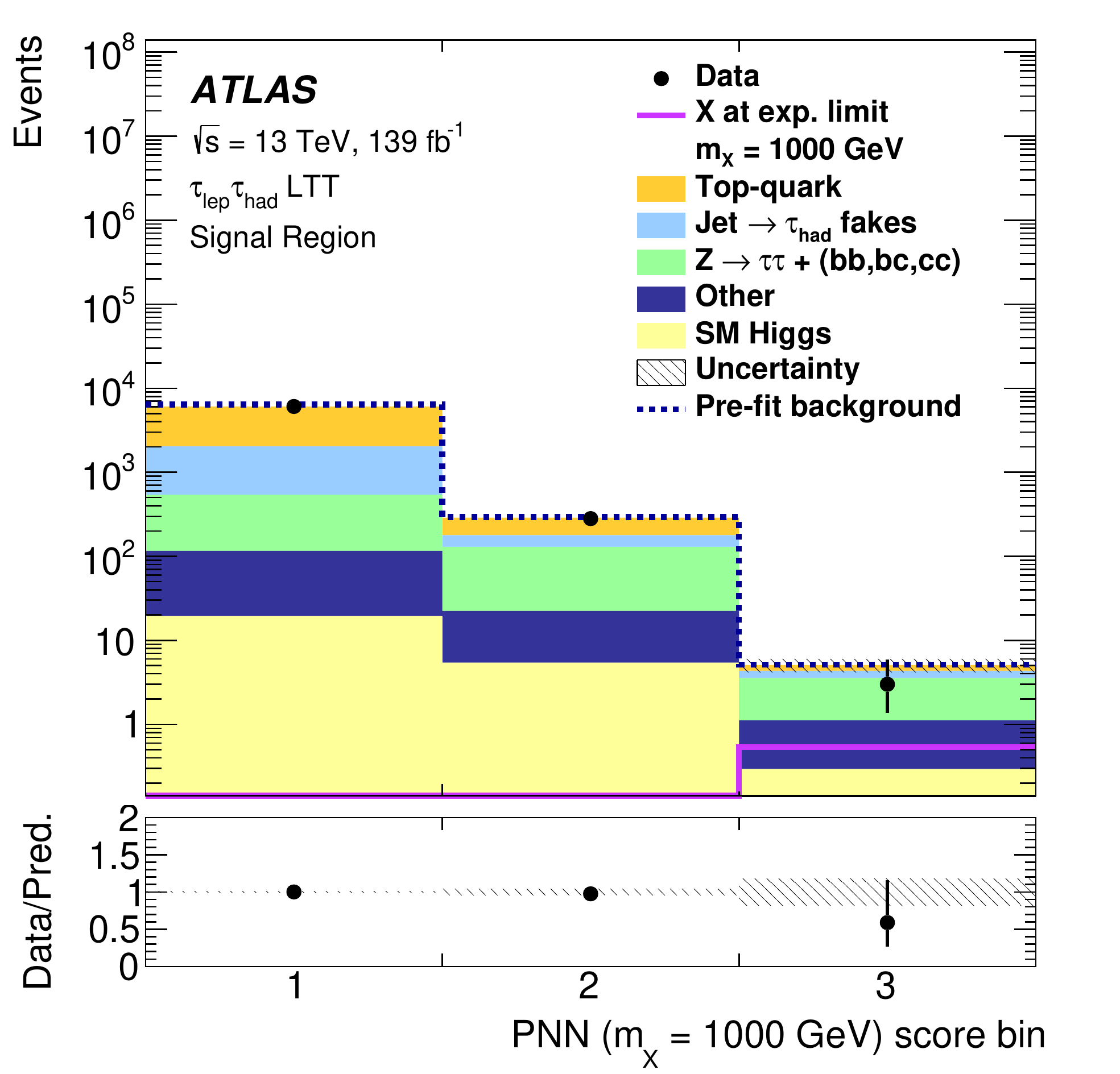}}
\caption{The MVA output distributions in the search for non-resonant $HH$ signal (top)
and in the search for resonant $HH$ signal with $m_{X}=500$~\GeV\ (middle row) and $m_{X}=1000$~\GeV\ (bottom),
in the $\tau_\text{had}\tau_\text{had}$ (left),
$\tau_\text{lep}\tau_\text{had}$ SLT (middle column) and
$\tau_\text{lep}\tau_\text{had}$ LTT (right) categories.
The distributions are shown after performing the fits to data and assuming the background-only hypothesis.
The signal is overlaid and scaled to the combined expected limit.
The dashed histogram shows the total pre-fit background.
The lower panels show the ratio of data to the total post-fit background,
where the hatched band shows the statistical and systematic uncertainties of that background.
For visualisation purposes, these histograms are displayed using uniform bin widths instead of the bin edges used in the fit, although the bin contents correspond to those used in the fit. Indices are used to label the bins.
\label{fig:RHH_mva_output}}
\end{figure}
 
\begin{table}[!htbp]
\caption{Data, expected signal and background yields for events passing the event selection and entering the most signal-like bin in the non-resonant MVA output histograms in the three event categories.
The background yields are calculated following a likelihood fit of the background model to data in the non-resonant $HH$ search. The non-resonant $HH$ signal yields are normalised using the SM cross-section. The `multi-jet fakes' and `$t\bar t$ fakes' estimates are only used in the $\tau_\text{had}\tau_\text{had}$ category, and the `combined fakes' estimate is only used in the $\tau_\text{lep}\tau_\text{had}$ categories. All systematic uncertainties are included in the indicated uncertainties.}
\label{tab:yieldstopmvabin}
\centering
\begin{tabular}{lccc}
\toprule
& $\tau_\text{had}\tau_\text{had}$ & $\tau_\text{lep}\tau_\text{had}$ SLT & $\tau_\text{lep}\tau_\text{had}$ LTT \\
\midrule
Data                  & 8                  & 7                    & 7                    \\
\midrule
Non-resonant ggF $HH$ & $1.58\pm 0.27$     & $0.77\pm 0.13$       & $0.25\pm 0.05$       \\
Non-resonant VBF $HH$ & $0.0227\pm 0.0019$ & $0.0075\pm 0.0007$   & $0.00455\pm 0.00035$ \\
\midrule
$t\bar t$             & $0.5\pm 0.1$       & $0.44\pm 0.12$       & $1.76\pm 0.32$       \\
Single top-quark      & $0.47\pm 0.20$     & $1.2\pm 0.7$         & $0.61\pm 0.35$       \\
$Z+\text{HF}$         & $2.3\pm 0.4$       & $1.55\pm 0.33$       & $1.7\pm 0.4$         \\
Combined fakes        & --                 & $1.09\pm 0.22$       & $0.8\pm 0.5$         \\
Multi-jet fakes       & $0.47\pm 0.17$     & --                   & --                   \\
$t\bar t$ fakes       & $0.29\pm 0.07$     & --                   & --                   \\
Single Higgs boson    & $1.7\pm 0.5$       & $1.10\pm 0.28$       & $0.4\pm 0.1$         \\
Other backgrounds     & $0.4\pm 0.1$       & $0.48\pm 0.09$       & $0.33\pm 0.07$       \\
Total background      & $6.1\pm 0.8$       & $6\pm 1$             & $6\pm 1$             \\
\bottomrule
\end{tabular}
\end{table}

Table~\ref{tab:NRHH_limit} shows the 95\% CL upper limits on the cross-sections for non-resonant $HH$ production,
and on the ratios of these cross-sections to their SM predictions.
The combined observed (expected) limit on the SM non-resonant $HH$ production cross-section via the ggF and VBF processes
is 140~fb ($110^{+40}_{-30}$~fb), while the limit on the ratio to the SM prediction is 4.7 ($3.9^{+1.5}_{-1.1}$).
The expected limit is calculated assuming no $HH$ production.

Figure~\ref{fig:NRHH_logSovB} shows the data, background and non-resonant $HH$ signal yields,
where final-discriminant bins from the $\tau_\text{had}\tau_\text{had}$, $\tau_\text{lep}\tau_\text{had}$ SLT
and $\tau_\text{lep}\tau_\text{had}$ LTT categories are combined into bins of log$_{10}(S/B)$.
The fitted background yield $B$ assumes the background-only hypothesis.
The signal $S$ that is shown in the overlaid histograms is scaled either to the SM expected cross-section
or to the combined expected limit cross-section.

\begin{table}
\begin{center}
\caption{Observed and expected 95\% CL upper limits on the cross-section for non-resonant $HH$ production,
assuming SM kinematics,
and on the cross-section for non-resonant $HH$ production divided by the SM prediction.
The $\pm 1\sigma$ and $\pm 2\sigma$ variations around the expected limit are also shown.
\protect\label{tab:NRHH_limit}}
\begin{tabular}{lccccccc}
\toprule
&                                                         & Observed & $-2\sigma$ & $-1\sigma$ & Expected & $+1\sigma$ & $+2\sigma$ \\ \midrule
\multirow{2}{*}{$\tau_\text{had}\tau_\text{had}$} & $\sigma_\text{ggF+VBF}$ [fb]                  & 150~~~~  & 70~~        & 95~~       & 130~~~~ & 180~~~~ & 240~~~~ \\
& $\sigma_\text{ggF+VBF}/\sigma^\text{SM}_\text{ggF+VBF}$ & ~~~5.0     & ~~~2.4     & ~~~3.2    & ~~~4.4    & ~~~6.1    & ~~~8.2    \\ \midrule
\multirow{2}{*}{$\tau_\text{lep}\tau_\text{had}$} & $\sigma_\text{ggF+VBF}$ [fb]                   & 280~~~~  & 120~~~~  & 170~~~~ & 230~~~~ & 320~~~~ & 430~~~~ \\
& $\sigma_\text{ggF+VBF}/\sigma^\text{SM}_\text{ggF+VBF}$ & ~~~9.7     & ~~~4.2     & ~~~5.6    & ~~~7.8    & 11~~        & 15~~       \\ \midrule
\multirow{2}{*}{Combined }                        & $\sigma_\text{ggF+VBF}$ [fb]                            & 140~~~    & 62~~        & 83~~       & 110~~~~ & 160~~~~ & 210~~~~ \\
& $\sigma_\text{ggF+VBF}/\sigma^\text{SM}_\text{ggF+VBF}$ & ~~~4.7     & ~~~2.1     & ~~~2.8    & ~~~3.9    & ~~~5.4    & ~~~7.2    \\
\bottomrule
\end{tabular}
\end{center}
\end{table}

\begin{figure}
\centering
\includegraphics[width=0.55\linewidth]{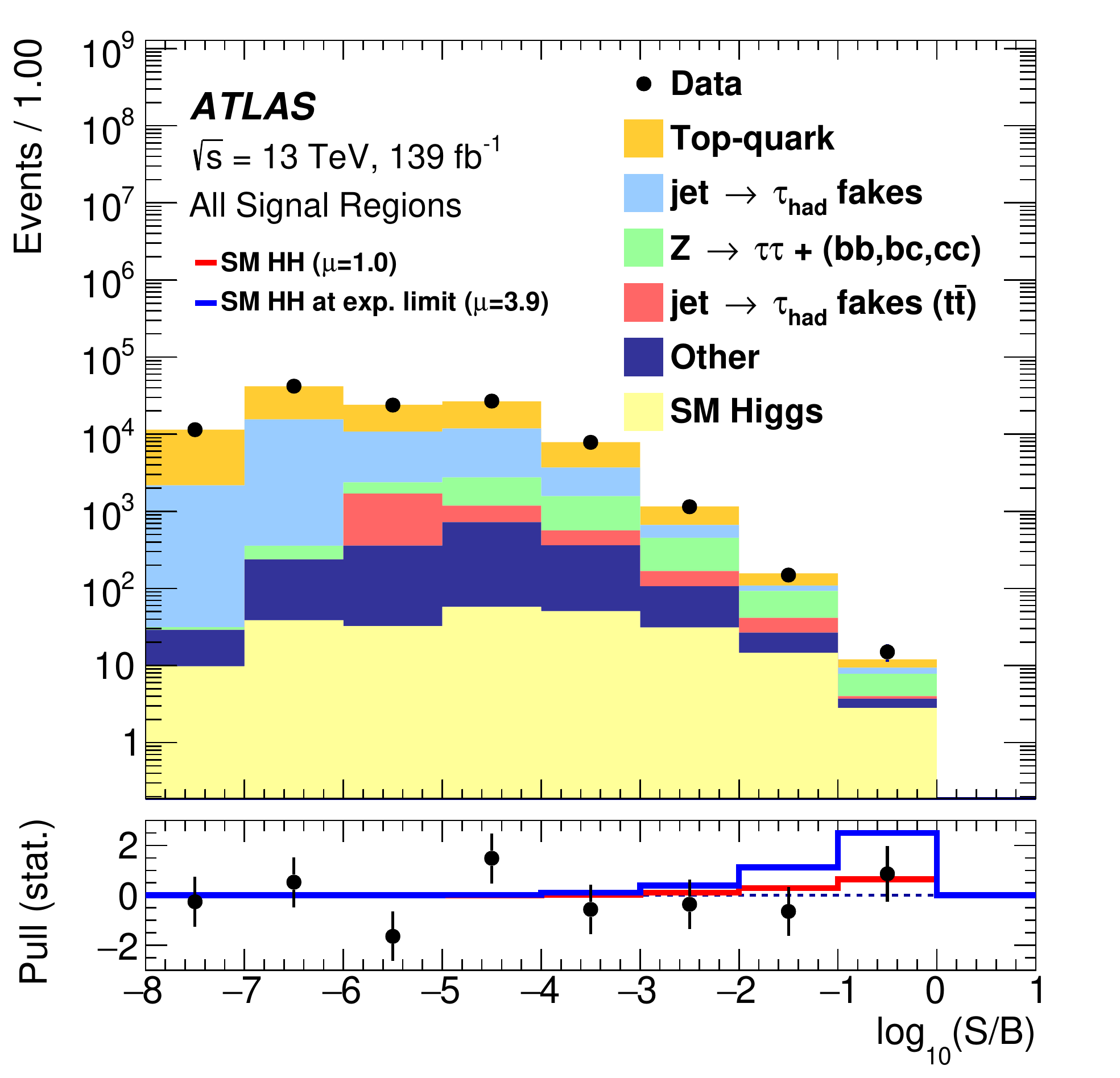}
\caption{Event yields as a function of log$_{10}(S/B)$ for data, background and non-resonant $HH$ signal.
Final-discriminant bins from the $\tau_\text{had}\tau_\text{had}$, $\tau_\text{lep}\tau_\text{had}$ SLT
and $\tau_\text{lep}\tau_\text{had}$ LTT categories are combined into bins of log$_{10}(S/B)$.
The fitted background yield $B$ assumes the background-only hypothesis.
In the lower panel, the pull of the data relative to the background
(the statistical significance of the difference between data and fitted background)
is shown with statistical uncertainties only.
The full lines indicate the pull expected from the sum of the fitted background and the signal,
which is scaled to either the SM expected cross-section (red) or the combined expected limit cross-section (blue),
relative to the fitted background.
\label{fig:NRHH_logSovB}}
\end{figure}

For resonant $HH$ production, the exclusion limits on the cross-section
are shown as a function of $m_{X}$ in Figure~\ref{fig:RHH_limit}.
The $\tau_\text{had}\tau_\text{had}$ channel has better sensitivity than the
$\tau_\text{lep}\tau_\text{had}$ channel below a resonance mass of 1~\TeV,
while their sensitivities become comparable at higher masses.
A broad excess is observed in both channels in the mass range $700~\GeV < m_{X} < 1.2~\TeV$.
The most significant excess for the $\tau_\text{had}\tau_\text{had}$ ($\tau_\text{lep}\tau_\text{had}$) channel
is found for the signal mass hypothesis of 1~\TeV\ (1.1~\TeV) with a local significance of $2.8\sigma$ ($1.6\sigma$).
The most significant combined excess is at a signal mass hypothesis of 1~\TeV\ with a local significance of $3.1\sigma$
and a global significance of $2.0\sigma$.
 
The $p$-value defining the global significance is given by the probability of finding a
maximum local excess larger than the observed value under the background-only hypothesis,
regardless of the resonance mass value $m_{X}$ where the largest excess is found.
It is estimated using fits of `toy' experiments drawn from a model providing a joint description of
all observables in the statistical interpretation.
The set of observables comprises the number of events observed in a bin entering either
a signal or control region, separately for all bins, as well as the central values of auxiliary measurements
used to propagate systematic uncertainties to the results.
This model differs from the probabilistic model used to define the likelihood functions
since these describe only one PNN discriminant at a time,
targeting a single signal mass hypothesis.
However, non-trivial correlations between observables are expected since events are subject
to the same selection criteria regardless of the resonance mass value $m_{X}$ to be probed,
but different multivariate discriminants are used for each of these signal hypotheses.
Observables describing the number of events in a given bin are generated from a multivariate
Poisson distribution with the dependence structure between observables given by the copula~\cite{Nelsen:2006}
of a centred multivariate normal distribution with covariance matrix given by the expected
linear correlations estimated from a simulation- and data-driven background model.
Observables describing the statistical precision of the background estimate are generated
using a resampling approach~\cite{ATL-PHYS-PUB-2021-011}.
Observables associated with all other experimental and theoretical uncertainties are varied coherently
for all hypothesis tests.

\begin{figure}
\centering
\includegraphics[width=0.70\linewidth]{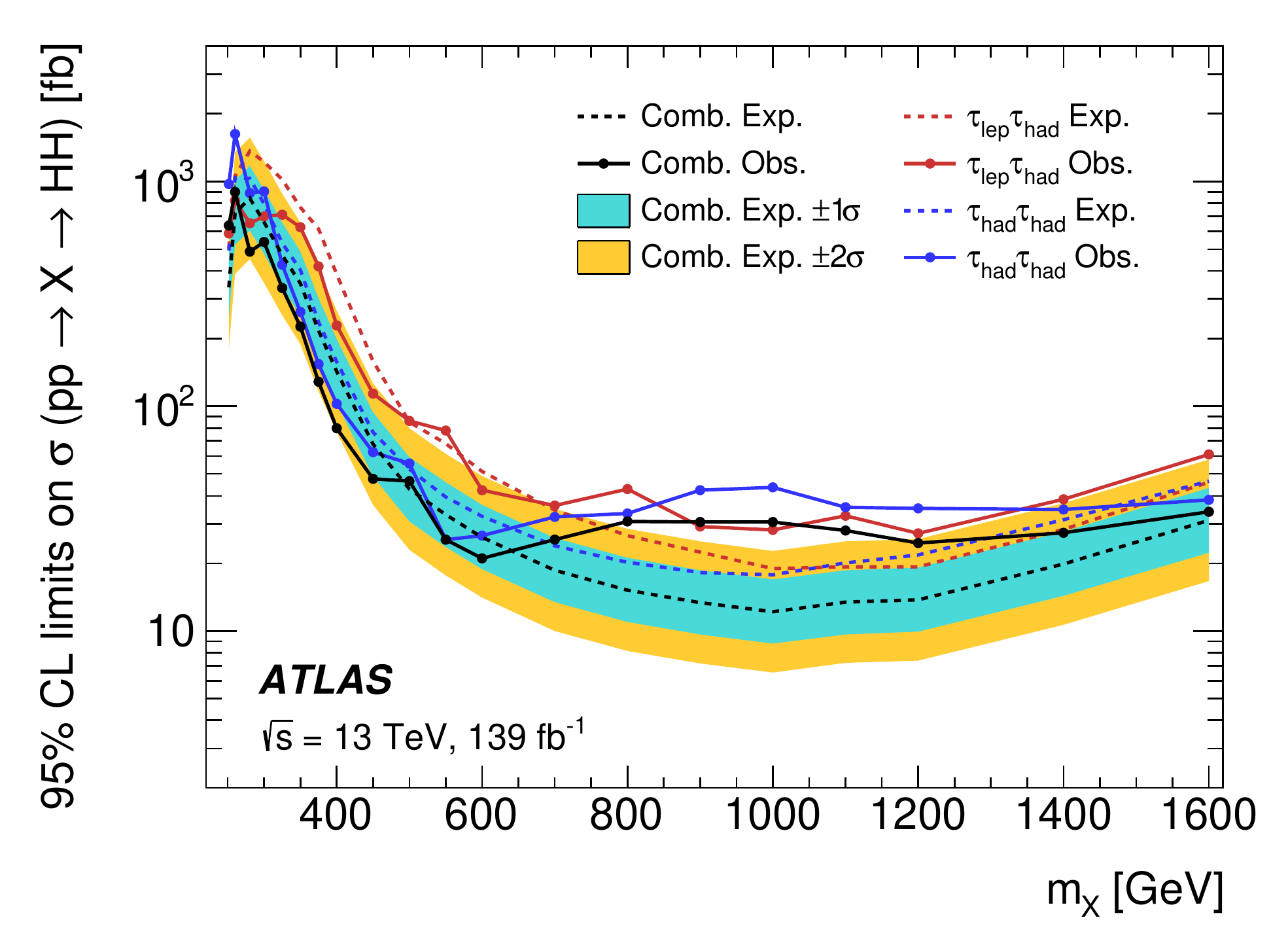}
\caption{Observed and expected limits at 95\% CL on the cross-section for resonant $HH$ production
as a function of the scalar resonance mass $m_{X}$.
The dashed lines show the expected limits while the solid lines show the observed limits.
The blue and red lines are the limits for the $\tau_\text{had}\tau_\text{had}$ channel and
$\tau_\text{lep}\tau_\text{had}$ channel, respectively.
The black lines are the combined limits of the two channels.
The $\pm 1\sigma$ and $\pm 2\sigma$ variations around the expected combined limit are indicated by the
turquoise and yellow bands, respectively.
The limits are obtained using the profile-likelihood test statistic and the modified frequentist
$\text{CL}_\text{s}$ technique.
\label{fig:RHH_limit}}
\end{figure}


\section{Conclusion}
\label{sec:conclusions}

 
A search for non-resonant and resonant Higgs boson pair production in $b\bar b\tau^+\tau^-$ events is conducted, where the non-resonant signal is assumed to be produced with SM kinematics, and the resonant signal corresponds to a narrow scalar resonance with a mass $m_X$ in the range 251 to 1600~\GeV. The 13~\TeV\ $pp$ collision dataset used was collected at the LHC by the ATLAS experiment between 2015 and 2018, and corresponds to an integrated luminosity of $139~\text{fb}^{-1}$. The sensitivity of this search to the non-resonant signal hypothesis improves on the previous ATLAS search in this channel by around a factor of four. Roughly half of this improvement is due to the larger dataset, while most of the remaining sensitivity gain is due to significant improvements in the $\tau_\text{had-vis}$ and $b$-jet reconstruction and identification. Analysis-level improvements include the use of more sophisticated multivariate techniques to target the resonant signal hypotheses, and new fake-$\tau_\text{had-vis}$ estimation methods. The data are found to be compatible with the background-only hypothesis, with the largest deviation being found in the search for resonant $HH$ production at a mass of 1~\TeV\ and corresponding to a local (global) significance of $3.1\sigma$ ($2.0\sigma$). The observed (expected) upper limit on the non-resonant Higgs boson pair-production cross-section, set at the 95\% confidence level, is 4.7 (3.9) times the SM expectation. Observed (expected) upper limits are placed at the 95\% confidence level on resonant Higgs boson production and exclude cross-sections above 21--900~fb (12--840~fb), depending on the mass of the resonance. This search provides the highest expected sensitivity to non-resonant $HH$ production of any individual search to date, and provides limits on resonant $HH$ production that are more stringent than, or competitive with, the most recently published ATLAS and CMS $HH$ resonant search combinations over much of the $m_X$ range explored.


\section*{Acknowledgements}


We thank CERN for the very successful operation of the LHC, as well as the
support staff from our institutions without whom ATLAS could not be
operated efficiently.
 
We acknowledge the support of
ANPCyT, Argentina;
YerPhI, Armenia;
ARC, Australia;
BMWFW and FWF, Austria;
ANAS, Azerbaijan;
CNPq and FAPESP, Brazil;
NSERC, NRC and CFI, Canada;
CERN;
ANID, Chile;
CAS, MOST and NSFC, China;
Minciencias, Colombia;
MEYS CR, Czech Republic;
DNRF and DNSRC, Denmark;
IN2P3-CNRS and CEA-DRF/IRFU, France;
SRNSFG, Georgia;
BMBF, HGF and MPG, Germany;
GSRI, Greece;
RGC and Hong Kong SAR, China;
ISF and Benoziyo Center, Israel;
INFN, Italy;
MEXT and JSPS, Japan;
CNRST, Morocco;
NWO, Netherlands;
RCN, Norway;
MEiN, Poland;
FCT, Portugal;
MNE/IFA, Romania;
MESTD, Serbia;
MSSR, Slovakia;
ARRS and MIZ\v{S}, Slovenia;
DSI/NRF, South Africa;
MICINN, Spain;
SRC and Wallenberg Foundation, Sweden;
SERI, SNSF and Cantons of Bern and Geneva, Switzerland;
MOST, Taiwan;
TENMAK, T\"urkiye;
STFC, United Kingdom;
DOE and NSF, United States of America.
In addition, individual groups and members have received support from
BCKDF, CANARIE, Compute Canada and CRC, Canada;
PRIMUS 21/SCI/017 and UNCE SCI/013, Czech Republic;
COST, ERC, ERDF, Horizon 2020 and Marie Sk{\l}odowska-Curie Actions, European Union;
Investissements d'Avenir Labex, Investissements d'Avenir Idex and ANR, France;
DFG and AvH Foundation, Germany;
Herakleitos, Thales and Aristeia programmes co-financed by EU-ESF and the Greek NSRF, Greece;
BSF-NSF and MINERVA, Israel;
Norwegian Financial Mechanism 2014-2021, Norway;
NCN and NAWA, Poland;
La Caixa Banking Foundation, CERCA Programme Generalitat de Catalunya and PROMETEO and GenT Programmes Generalitat Valenciana, Spain;
G\"{o}ran Gustafssons Stiftelse, Sweden;
The Royal Society and Leverhulme Trust, United Kingdom.
 
The crucial computing support from all WLCG partners is acknowledged gratefully, in particular from CERN, the ATLAS Tier-1 facilities at TRIUMF (Canada), NDGF (Denmark, Norway, Sweden), CC-IN2P3 (France), KIT/GridKA (Germany), INFN-CNAF (Italy), NL-T1 (Netherlands), PIC (Spain), ASGC (Taiwan), RAL (UK) and BNL (USA), the Tier-2 facilities worldwide and large non-WLCG resource providers. Major contributors of computing resources are listed in Ref.~\cite{ATL-SOFT-PUB-2021-003}.


\printbibliography
\clearpage
 
\begin{flushleft}
\hypersetup{urlcolor=black}
{\Large The ATLAS Collaboration}

\bigskip

\AtlasOrcid[0000-0002-6665-4934]{G.~Aad}$^\textrm{\scriptsize 99}$,
\AtlasOrcid[0000-0002-5888-2734]{B.~Abbott}$^\textrm{\scriptsize 117}$,
\AtlasOrcid[0000-0002-7248-3203]{D.C.~Abbott}$^\textrm{\scriptsize 100}$,
\AtlasOrcid[0000-0002-2788-3822]{A.~Abed~Abud}$^\textrm{\scriptsize 34}$,
\AtlasOrcid[0000-0002-1002-1652]{K.~Abeling}$^\textrm{\scriptsize 53}$,
\AtlasOrcid[0000-0002-2987-4006]{D.K.~Abhayasinghe}$^\textrm{\scriptsize 92}$,
\AtlasOrcid[0000-0002-8496-9294]{S.H.~Abidi}$^\textrm{\scriptsize 27}$,
\AtlasOrcid[0000-0002-9987-2292]{A.~Aboulhorma}$^\textrm{\scriptsize 33e}$,
\AtlasOrcid[0000-0001-5329-6640]{H.~Abramowicz}$^\textrm{\scriptsize 148}$,
\AtlasOrcid[0000-0002-1599-2896]{H.~Abreu}$^\textrm{\scriptsize 147}$
% ...
\bigskip
\\

% ...
$^{27}$Physics Department, Brookhaven National Laboratory, Upton NY; United States of America.\\
% ...
$^{33}$$^{(a)}$Facult\'e des Sciences Ain Chock, R\'eseau Universitaire de Physique des Hautes Energies - Universit\'e Hassan II, Casablanca;$^{(b)}$Facult\'{e} des Sciences, Universit\'{e} Ibn-Tofail, K\'{e}nitra;$^{(c)}$Facult\'e des Sciences Semlalia, Universit\'e Cadi Ayyad, LPHEA-Marrakech;$^{(d)}$LPMR, Facult\'e des Sciences, Universit\'e Mohamed Premier, Oujda;$^{(e)}$Facult\'e des sciences, Universit\'e Mohammed V, Rabat;$^{(f)}$Institute of Applied Physics, Mohammed VI Polytechnic University, Ben Guerir; Morocco.\\
$^{34}$CERN, Geneva; Switzerland.\\
% ...
$^{53}$II. Physikalisches Institut, Georg-August-Universit\"{a}t G\"ottingen, G\"ottingen; Germany.\\
% ...
$^{92}$Department of Physics, Royal Holloway University of London, Egham; United Kingdom.\\
% ...
$^{99}$CPPM, Aix-Marseille Universit\'e, CNRS/IN2P3, Marseille; France.\\
$^{100}$Department of Physics, University of Massachusetts, Amherst MA; United States of America.\\
% ...
$^{117}$Homer L. Dodge Department of Physics and Astronomy, University of Oklahoma, Norman OK; United States of America.\\
% ...
$^{147}$Department of Physics, Technion, Israel Institute of Technology, Haifa; Israel.\\
$^{148}$Raymond and Beverly Sackler School of Physics and Astronomy, Tel Aviv University, Tel Aviv; Israel.\\
% ...
$^{a}$ Also Affiliated with an institute covered by a cooperation agreement with CERN.\\
$^{b}$ Also at Borough of Manhattan Community College, City University of New York, New York NY; United States of America.\\
$^{c}$ Also at Bruno Kessler Foundation, Trento; Italy.\\
$^{d}$ Also at Center for High Energy Physics, Peking University; China.\\
$^{e}$ Also at Centro Studi e Ricerche Enrico Fermi; Italy.\\
$^{f}$ Also at CERN, Geneva; Switzerland.\\
% ...

\end{flushleft}


\end{document}